\documentclass[superscriptaddress,groupedaddress,nofootnoteinbib,11pt]{article}
\pdfoutput=1 
\usepackage{setspace}
\usepackage{jheppub}

\usepackage[utf8x]{inputenc}
\usepackage{mathtools,slashed,mathrsfs}
\usepackage[caption=false]{subfig}
\usepackage{dcolumn}
\usepackage{multirow}
\usepackage{tabularx}
\usepackage{booktabs}
\usepackage{bm}
\usepackage{comment}
\usepackage[dvipsnames]{xcolor}
\usepackage[normalem]{ulem} 
\usepackage{enumerate}
\usepackage{siunitx}
\usepackage{xspace}
\usepackage{adjustbox}
\allowdisplaybreaks

\newcommand{\Sec}[1]{Sec.~\ref{#1}}
\newcommand{\Subsec}[1]{Subsec.~\ref{#1}}
\newcommand{\App}[1]{Appendix~\ref{#1}}
\newcommand{\Fig}[1]{Fig.~\ref{#1}}
\newcommand{\Eq}[1]{Eq.~(\ref{#1})}

\newcommand{\beq}{\begin{equation}}
\newcommand{\eeq}{\end{equation}}
\newcommand{\ba}{\begin{array}}
\newcommand{\ea}{\end{array}}
\newcommand{\bea}{\begin{eqnarray}}
\newcommand{\eea}{\end{eqnarray} }
\newcommand{\be}{\begin{eqnarray}}
\newcommand{\ee}{\end{eqnarray}}
\newcommand{\bal}{\begin{align}}
\newcommand{\eal}{\end{align}}
\newcommand{\bi}{\begin{itemize}}
\newcommand{\ei}{\end{itemize}}
\newcommand{\ben}{\begin{enumerate}}
\newcommand{\een}{\end{enumerate}}
\newcommand{\bc}{\begin{center}}
\newcommand{\ec}{\end{center}}
\newcommand{\bt}{\begin{table}}
\newcommand{\et}{\end{table}}
\newcommand{\btb}{\begin{tabular}}
\newcommand{\etb}{\end{tabular}}

\newcommand{\bl}{\left}
\newcommand{\br}{\right}
\newcommand{\eg}{\textit{e.g.}}
\newcommand{\ie}{\textit{i.e.}}

\newcommand{\eV}{\mathrm{eV}}

\newcommand{\MeV}{\mathrm{MeV}}
\newcommand{\GeV}{\mathrm{GeV}}
\newcommand{\TeV}{\mathrm{TeV}}

\newcommand{\Mpc}{\mathrm{Mpc}}

\newcommand{\cm}{\mathrm{cm}}

\newcommand{\km}{\mathrm{km}}
\newcommand{\seg}{\mathrm{s}}

\newcommand{\dd}{\ensuremath{\mathrm{d}}}
\newcommand{\mO}{\mathcal{O}}

\newcommand{\mH}{\mathcal{H}}
\newcommand{\mF}{\mathcal{F}}

\newcommand{\mD}{\mathcal{D}}
\newcommand{\mDH}{\mathcal{DH}}

\newcommand{\mDF}{\mathcal{DF}}
\newcommand{\mDHF}{\mathcal{DHF}}

\newcommand{\pl}{\ensuremath{\mathrm{Pl}}}
\newcommand{\eq}{\ensuremath{\mathrm{eq}}}
\newcommand{\dec}{\ensuremath{\mathrm{dec}}}

\newcommand{\lcdm}{\ensuremath{\Lambda\mathrm{CDM}}\xspace}

\newcommand{\dm}{\mathrm{dm}}
\newcommand{\cdm}{\mathrm{cdm}}

\newcommand{\adm}{\mathrm{adm}}
\newcommand{\dr}{\mathrm{dr}}

\newcommand{\anti}[1]{\ensuremath{\overline{#1}}}
\newcommand{\logmm}{\ensuremath{\log_{10}\bl( m_{e'} / m_{p'} \br)}}

\newcommand{\Neff}{\ensuremath{N_{\mathrm{eff}}}\xspace}
\newcommand{\DNeff}{\ensuremath{\Delta N_\mathrm{eff}}\xspace}

\newcommand{\HO}{\ensuremath{H_0}\xspace}

\newcommand{\nuadam}{\ensuremath{\text{nuADaM}}\xspace}

\newcommand{\Ap}{\ensuremath{A'}}

\newcommand{\ignore}[1]{}

\usepackage{tikz,xcolor,hyperref}
\definecolor{lime}{HTML}{A6CE39}
\DeclareRobustCommand{\orcidicon}{%
	\begin{tikzpicture}
	\draw[lime, fill=lime] (0,0) 
	circle [radius=0.16] 
	node[white] {{\fontfamily{qag}\selectfont \tiny ID}};	\draw[white, fill=white] (-0.0625,0.095) 
	circle [radius=0.007];	\end{tikzpicture}
	\hspace{-2mm}}
\foreach \x in {A,...,Z}{%
	\expandafter\xdef\csname orcid\x\endcsname{\noexpand\href{https://orcid.org/\csname orcidauthor\x\endcsname}{\noexpand\orcidicon}}
}

\begin{document}
\preprint{\begin{flushright}
    UTWI-37-2024\\
\end{flushright}}

\title{Atomic Dark Matter, Interacting Dark Radiation, and the Hubble Tension}

\date{\today}
\author[a,b]{Manuel A. Buen-Abad,\orcidA{}}
\author[a]{Zackaria Chacko,\orcidB{}}
\author[a]{Ina Flood,\orcidC{}}
\author[c]{Can Kilic,\orcidD{}}
\author[d]{Gustavo Marques-Tavares,\orcidE{}}
\author[e,f]{Taewook Youn,\orcidF{}}

\affiliation[a]{Maryland Center for Fundamental Physics, Department of Physics, University of Maryland, College Park, MD 20742, U.S.A.}
\affiliation[b]{Dual CP Institute of High Energy Physics, C.P. 28045, Colima, M\'{e}xico}
\affiliation[c]{Theory Group, Weinberg Institute for Theoretical Physics, University of Texas at Austin, Austin, TX 78712, U.S.A.}
\affiliation[d]{Department of Physics and Astronomy, University of Utah, Salt Lake City, UT 84112, U.S.A.}
\affiliation[e]{Laboratory for Elementary Particle Physics, Cornell University, Ithaca, NY 14853, USA}
\affiliation[f]{School of Physics, Korea Institute for Advanced Study,  Seoul 02455, Republic of Korea}
\emailAdd{buenabad@umd.edu}
\emailAdd{zchacko@umd.edu}
\emailAdd{iflood@umd.edu}
\emailAdd{kilic@physics.utexas.edu}
\emailAdd{g.marques@utah.edu}
\emailAdd{taewook.youn@cornell.edu}

\abstract{
We present a new class of interacting dark sector models that can address the Hubble tension.
Interacting dark radiation (DR) has previously been put forward as a solution to the problem, but this proposal is disfavored by the high-$\ell$ cosmic microwave background (CMB) data.
We modify this basic framework by introducing a subcomponent of dark matter (DM) that interacts strongly with the DR, so that together they constitute a tightly coupled fluid at early times.
We show that if this subcomponent decouples from the interacting DR during the CMB epoch, the $\ell$ modes of the CMB that entered the horizon before decoupling are impacted differently from those that entered after, allowing a solution to the problem.
We present a model that realizes this framework, which we dub ``New Atomic Dark Matter'', or \nuadam, in which the interacting dark matter (iDM) subcomponent is composed of dark atoms, and dark ``neutrinos'' with long-range interactions contribute to the DR, hence the name of the model.
This iDM subcomponent is acoustic at early times but decouples from the DR following dark recombination.
In contrast to conventional atomic dark matter (ADM) models, the dark photon is part of a richer DR sector, which ensures that it continues to be self-interacting even after recombination.
We show that this model admits a significantly larger value of $H_0$ than $\lcdm$ when fit to CMB and BAO data, while maintaining a comparable goodness of fit. Once the SHOES data set is included, it provides a significantly better fit than $\lcdm$.
}

\maketitle

\section{Introduction}
\label{sec:intro}

There exists a large discrepancy between the value of $H_0$, the current expansion rate of the universe, inferred from the best fit of the parameters in the $\lcdm$ model to cosmological data, and the value obtained from more direct measurements based on the distance ladder technique.
This may be the first serious indication that we need to go beyond the standard $\Lambda$CDM cosmological model.
When comparing two of the most precise measurements, the \lcdm~fit to Planck cosmic microwave background (CMB) data~\cite{Planck:2018vyg} which gives
$H_0 = 67.36 \pm 0.54$ km/s/Mpc, and the supernovae measurements made by the SH0ES collaboration calibrated to Cepheid variable stars~\cite{Riess:2021jrx} which yield
$H_0 = 73.04 \pm 1.04$ km/s/Mpc, one observes a $> 5 \sigma$ tension between them.
While it is possible that this $H_0$ tension could be due to some unknown systematic effects \cite{Kamionkowski:2022pkx,Freedman:2021ahq}, the distance ladder measurements have withstood significant scrutiny and the results have remained stable across different variations in the data and analysis methods employed.%
\footnote{
It is worth noting, however, that the Chicago-Carnegie Hubble Program (CCHP) collaboration reports a direct measurement of $H_0$ that is not in significant tension with the \lcdm fit to Planck data.
The most recent CCHP results can be found in Ref.~\cite{Freedman:2024eph}.}
A wide range of new models beyond \lcdm have been proposed to address this tension (see, \eg, Refs.~\cite{Buen-Abad:2015ova,Lesgourgues:2015wza,Buen-Abad:2017gxg,Zhao:2017cud,DiValentino:2017gzb,Poulin:2018cxd,Smith:2019ihp,Lin:2019qug,Alexander:2019rsc,Agrawal:2019lmo,Escudero:2019gvw,Berghaus:2019cls,Vagnozzi:2019ezj,Ye:2020btb,Das:2020xke,RoyChoudhury:2020dmd,Brinckmann:2020bcn,Krishnan:2020vaf,Seto:2021xua,Ye:2021iwa,Niedermann:2021vgd,Aloni:2021eaq, Dainotti:2021pqg,Odintsov:2022eqm,Berghaus:2022cwf,Schoneberg:2022grr,Joseph:2022jsf,Buen-Abad:2022kgf,Rezazadeh:2022lsf,Brinckmann:2022ajr,Wang:2022bmk,Bansal:2022qbi,Buen-Abad:2023uva,Sandner:2023ptm,Zu:2023rmc,Niedermann:2023ssr,Hughes:2023tcn,Greene:2024qis,Allali:2024anb,Montani:2024pou, Co:2024oek,Cho:2024lhp,Simon:2024jmu,DeSimone:2024lvy} for a partial list of these proposals, and also the models reviewed in~\cite{DiValentino:2021izs, Schoneberg:2021qvd, Abdalla:2022yfr,Poulin:2023lkg,Khalife:2023qbu}).

One of the most promising avenues for addressing the tension is to introduce modifications to the $\lcdm$ paradigm only at early times, during the CMB epoch, in such a way that the sound horizon scale is changed~\cite{Aylor:2018drw}.
In such an approach, even though new physics only directly affects redshifts $z \gtrsim 10^3$, this impacts the extraction of $H_0$ because when the sound horizon scale is changed, one needs to adjust the cosmological parameters to keep the ratio of the sound horizon scale to the angular distance to recombination fixed, thereby changing $H_0$.
In order to address the tension, one needs to decrease the sound horizon at recombination, which implies the need to increase the energy density (and therefore $H$) during the CMB epoch. Recently it has been shown that one can make ``sound-horizon free'' measurements of $H_0$~\cite{Philcox:2022sgj,Zaborowski:2024car} with precisions that are becoming competitive with CMB measurements.
Current analyses have been made using $\Lambda$CDM, but extending this to other scenarios could become an important tool to distinguish and constrain models that address the $H_0$ tension by changing the sound horizon.

There have been a wide range of proposals that introduce new dark sector states to provide the extra energy density needed around the time of recombination.
The simplest scenarios involve new massless degrees of freedom constituting a dark radiation (DR) component (whose abundance is parametrized by $\DNeff$), which may be either free-streaming, similar to neutrinos, or having sizable self-interactions such that the new particles can be described as a perfect fluid.
The biggest challenge for this class of solutions is that the change in the Hubble scale affects the sound horizon scale and the diffusion damping scales differently, which leads to an increase in Silk damping at small scales \cite{Blinov:2020hmc,Hou:2011ec}.
This can be partially compensated for by an increase in the spectral tilt ($n_s$), but since $n_s$ affects all scales whereas Silk damping primarily affects smaller scales, adjusting this parameter only provides a limited improvement in the overall fit.

A natural extension of this framework incorporates a step-like increase in $\DNeff$ occurring during the CMB epoch (``stepped DR") \cite{Aloni:2021eaq}.
It has been shown that such a proposal leads to a significant improvement over the minimal DR scenario \cite{Aloni:2021eaq,Schoneberg:2022grr}.
In this scenario, modes that enter the horizon prior to the step are impacted differently from those that enter afterward.
Such a step can naturally arise from the freeze-out of a massive particle in the DR bath.
Despite the reduction in the Hubble tension, these stepped DR models generally lead to an enhancement in the power spectrum at small scales (primarily due to an increase in $n_s$ compared to $\lcdm$).
This has the effect of exacerbating tensions with small scale measurements from weak lensing and Lyman-$\alpha$ forest data \cite{Goldstein:2023gnw,Rogers:2023upm}.
Further extending this basic scenario to include interactions between dark matter (DM) and DR \cite{Joseph:2022jsf,Buen-Abad:2022kgf} can offer a resolution to this problem \cite{Allali:2023zbi,Buen-Abad:2023uva,Schoneberg:2023rnx,Bagherian:2024obh}.

In this work, we introduce a new class of interacting dark sector models designed to address the Hubble tension.
Our approach extends the standard self-interacting DR (SIDR) scenario by incorporating a subcomponent of DM that interacts strongly with DR.
This interacting dark matter (iDM) subcomponent and the DR form a tightly coupled fluid in the early universe.
The rest of the DM is composed of standard cold dark matter (CDM).
The crucial observation is that, if this subcomponent decouples from the DR during the CMB epoch, the modes entering the horizon before and after this decoupling takes place are affected differently, offering a potential resolution to the problem.
We present a specific realization of this framework based on atomic dark matter (ADM)~\cite{Goldberg:1986nk,Kaplan:2009de} that we refer to as ``{\bf New} {\bf A}tomic {\bf Da}rk {\bf M}atter'', or \nuadam.
In this model, the DM subcomponent interacting with DR consists of dark atoms.
Initially, these dark atoms are ionized, making this iDM subcomponent acoustic.
Eventually, dark recombination takes place during the CMB epoch, and the now neutral dark atoms decouple from the DR.
While the ADM scenario has been studied in detail in Refs.~\cite{Kaplan:2009de,Kaplan:2011yj,Cyr-Racine:2013fsa,Cyr-Racine:2012tfp,Bansal:2021dfh,Bansal:2022qbi,Hughes:2023tcn}, and it also features prominently in ``mirror DM'' models \cite{Blinnikov:1983gh,Hodges:1993yb,Berezhiani:1995am,Mohapatra:2000rk,Berezhiani:2000gw,Foot:2003jt,Berezhiani:2005ek,An:2009vq,Lonsdale:2018xwd}, our version has several unique characteristics.
In contrast to standard ADM models, the dark photon in \nuadam is part of a broader DR sector that continues to exhibit fluid-like behavior even after the dark atoms have decoupled.
We assume that the DR is populated only after Big Bang nucleosynthesis (BBN) concludes, which could naturally occur in scenarios where the dark sector equilibrates with Standard Model (SM) neutrinos at some point after BBN \cite{Aloni:2021eaq}.
Finally, we perform a cosmological data analysis and demonstrate that this model admits a significantly larger value of $H_0$ than $\lcdm$ when fit to CMB and BAO data, while maintaining a comparable goodness of fit. Once the SHOES data set is included, it provides a significantly better fit than $\lcdm$.
The model also provides a better fit to the cosmological data than conventional ADM.

We can obtain a qualitative understanding of the role played by the different ingredients in the model.
The presence of an interacting DR component alters the sound horizon scale, allowing larger $H_0$ values to be realized~\cite{Baumann:2015rya,Brust:2017nmv,Blinov:2020hmc}.
In addition, the interactions between the DR and the iDM subcomponent have two important effects.
Firstly, the pressure the DR imparts on the iDM decreases the power spectrum at small scales which, as we discuss in more detail below, may improve the fit to direct measurements of the matter power spectrum (MPS).
Secondly, the decoupling of the DR from the iDM results in a difference between the evolution of CMB modes that entered the horizon prior to decoupling and those that entered later. 
This is qualitatively similar to the effect of stepped DR~\cite{Aloni:2021eaq,Buen-Abad:2022kgf}, and, in combination with shifts in other cosmological parameters, allows for a significantly larger energy density in DR without degrading the fit to CMB data.
Both these effects only affect smaller scales, since after dark recombination iDM behaves exactly like CDM, and therefore modes that enter the horizon after dark recombination evolve as in \lcdm.
As we shall see, these ingredients allow the model to realize a much larger $H_0$ value, while maintaining a good fit to CMB data and without worsening the fit (or even improving it) to small scale probes of the MPS.

This paper is organized as follows.
In Sec.~\ref{sec:model}, we introduce the model, and discuss the relevant interactions between its different parts, as well as their associated parameters.
In Sec.~\ref{sec:cosmo}, we study the cosmological evolution of the new components of the model and determine the parameter space in which the model reproduces the dynamics that is most relevant for addressing the cosmological tensions.
In Sec.~\ref{sec:num}, we discuss the MCMC fits of our model to different combinations of data and compare results with updated fits to conventional ADM.
We conclude in Sec.~\ref{sec:conclusions}, and include our full numerical results in App.~\ref{appendix}.

\section{The ``New Atomic Dark Matter'' (\nuadam) Model}\label{sec:model}

In this section we describe the model that realizes the framework discussed in the introduction.
In addition to a CDM component that constitutes the majority of DM, we assume that a small fraction, $f_\adm \equiv \rho_\adm / \bl( \rho_\adm + \rho_\cdm \br)$, of the total DM energy density is composed of non-relativistic dark particles analogous to electrons and protons in the SM.
After a period of dark recombination, these particles come together to form dark hydrogen atoms. In this regard, 
our model is similar to the scenario studied in Refs.~\cite{Kaplan:2009de,Kaplan:2011yj,Cyr-Racine:2013fsa,Cyr-Racine:2012tfp,Bansal:2021dfh,Bansal:2022qbi,Hughes:2023tcn}.
However, in contrast to these earlier studies, the DR in our model is self-interacting.
We refer to our setup as the ``{\bf New} {\bf A}tomic {\bf Da}rk {\bf M}atter'' model, or \nuadam, for reasons that will be apparent shortly.
Then the \nuadam sector includes a dark $U(1)$ gauge boson $\Ap^\mu$, which we call the dark photon, as well as a heavy fermion $p'$ and a light fermion $e'$, which we refer to as the dark proton and the dark electron respectively. The dark proton and dark electron transform under the $U(1)$ gauge symmetry with equal and opposite charges.
Accordingly, the \nuadam Lagrangian includes the following terms:
\begin{equation}
    \mathcal{L}_{\text{dark atoms}} = -\frac{1}{4}F'^{\mu \nu}F'_{\mu \nu} + \bar p' \bl( i \slash  \!\!\!\! D - m_{p'} \br) p' +  \bar e' \bl( i \slash  \!\!\!\! D - m_{e'} \br) e' \, ,
\end{equation}
where $F'^{\mu \nu}$ is the field strength for the dark photon $A'^\mu$, $m_{p'}$ ($m_{e'}$) is the dark proton (electron) mass, and $D_\mu p' = \bl( \partial_\mu - i g' A'_\mu \br) p'$ represents the covariant derivative acting on $p'$ (and the analogous one for $e'$, with the opposite dark charge).
As stated above, in \nuadam, in contrast to earlier studies, the DR remains self-interacting even after dark recombination and therefore behaves like a perfect fluid at all the times of interest.
This is realized by including an additional dark $U(1)$ gauge field $X^\mu$, along with $N_f$ flavors of massless Dirac fermions $\nu'_{j}$.
The $\nu'_{j}$ are charged under this new $U(1)$, but not under $A'_{\mu}$, which is why we call them dark ``neutrinos''.
A dark kinetic mixing between both $U(1)$ gauge bosons ensures that the dark photon $A'$ interacts with these new DR components.
The corresponding terms in the Lagrangian are given by
\begin{equation}
    \mathcal{L}_{\rm DR} =  -\frac{1}{4} X^{\mu \nu} X_{\mu \nu} - \frac{\epsilon}{4} F'^{\mu \nu} X_{\mu \nu} + \sum\limits_{j=1}^{N_f} \bar \nu'_{j} \, i \, \slash \!\!\!\! D \nu'_{j} \, ,
\end{equation}
where $X^{\mu \nu} \equiv \partial^\mu X^\nu - \partial^\nu X^\mu$ is the field strength for $X^\nu$, and $D_\mu \nu' = \bl( \partial_\mu - i g_X X_\mu \br) \nu'_{j}$ the covariant derivative of the dark neutrinos, with $g_X$ the dark gauge coupling associated with the $U(1)_X$.

The model contains six new fundamental parameters: the mass of the dark proton $m_{p'}$, the mass of the dark electron $m_{e'}$, the fine structure constant $\alpha'$ of the dark photon $A'_\mu$, the fine structure constant $\alpha_X$ of the gauge boson $X_\mu$, the number $N_f$ of dark neutrino flavors, and the kinetic mixing parameter $\epsilon$.%
\footnote{Note that we are considering a dark sector completely decoupled from the visible sector.
The kinetic mixing parameter $\epsilon$ connects two {\it dark} $U(1)$ gauge fields.}
In addition to these parameters, $T'$ denotes the temperature of the dark sector while $f_\adm$ parametrizes the fraction of DM energy found in dark protons, dark electrons, and dark atoms.

The dark temperature $T'$ determines the energy density in the DR, which is composed of $A'$, $X$, and $N_f$ flavors of $\nu'_{j}$:
\bea
    \rho_\dr & \equiv & \frac{\pi^2}{30} \bl( 2 + 2 + g_{*\nu'} \br) T'^4 \ , \\
    \text{where} \quad g_{*\nu'} & \equiv & \frac{7}{8} \cdot 2 \cdot 2 \cdot N_f = 3.5 \; N_f
\eea
denotes the relativistic degrees of freedom in the $N_f$ flavors of the dark Dirac neutrino fermions $\nu'_{j}$.
We can parametrize this energy density in terms of $\DNeff$, the equivalent number of extra neutrino flavors:
\bea
    \DNeff & \equiv & \frac{\rho_\dr}{\rho_{1 \nu}} = \frac{4}{7} \bl( 4 + g_{*\nu'} \br) \bl( \frac{T'}{T_\nu} \br)^4 = \bl( \frac{11}{4} \br)^{4/3} \frac{4}{7} \bl( 4 + g_{*\nu'} \br) \bl( \frac{T'}{T_\gamma} \br)^4 \\
    & \approx & 2.2016 \times (4 + g_{*\nu'}) \, \xi^4 \ ,
\eea
where $\xi \equiv T' / T_\gamma$.

The fraction $f_\adm$ is related to the non-relativistic \nuadam species of dark electrons, dark protons, and dark atoms, as follows:
\bea
    f_\adm & \equiv & \frac{\rho_\adm}{\rho_\adm + \rho_\cdm} \ , \\
    \text{with} \quad \rho_\adm & \equiv & \rho_{e'} + \rho_{p'} + \rho_{H'} \ , \\
    \text{where} \quad \rho_{i'} & \equiv & m_{i'} n_{i'} \quad \bl( i \in \{ e, \, p, \, H \} \br) \ , \\
    \text{and} \quad n_{e',p'} & 
    \equiv & x' n_\adm \ , \quad n_{H'} \equiv \bl( 1 - x' \br) n_\adm \ .
\eea
Here the index $m_{i'}$ denotes the mass of the $i'$-th species (and $m_{H'} = m_{p'} + m_{e'} - B_{H'}$, where $B_{H'}$ represents the dark hydrogen binding energy), while $n_{e',p'}$ is the number density of {\it free} dark electrons/dark protons, and $x'$ is the dark ionization fraction.
In the expressions above, we have implicitly assumed that the {\it total} abundance of the dark protons and dark electrons is asymmetric at the times of interest and that they have the same number densities in order to ensure dark charge neutrality, \ie~$n_{p'}^{\rm tot} = n_{e'}^{\rm tot} \equiv n_\adm$. The abundance of their anti-particles is taken to be negligible, $n_{\anti{p}'} = n_{\anti{e}'} \approx 0$.
As we discuss later, the cosmological implications of \nuadam depend only on a small number of physical quantities, which leads to degeneracies between the parameters listed above.

We now turn our attention to three important considerations, which lead to certain requirements that the \nuadam model must obey in order to reproduce the dynamics necessary for our purposes.
These are:
\begin{enumerate}[(i)]
    \item {\bf At all times of interest:} The DR must be interacting in order to better fit the CMB data \cite{Blinov:2020hmc}.
    Prior to dark recombination, this is ensured by its interactions with the ionized dark protons and electrons, but after dark recombination other processes are necessary.
    We require the interactions between $\nu'$ and \Ap, induced by the kinetic mixing, to be strong enough to keep the DR self-interacting.
    The same needs to be true of the interactions between $\nu'$ and $X^\mu$, but this is automatically satisfied if $\nu'$ and $A'$ are tightly coupled.
    Furthermore, 2-to-3 processes involving $\nu'$ and $X^\mu$ (\eg, $\nu' \nu' \rightarrow \nu' \nu' X^\mu$) should be efficient, so that the phase-space distribution functions for these particles is given by a thermal distribution function with temperature $T'$ and zero chemical potential.
    \item {\bf Before dark recombination:} The interactions between the dark electrons $e'$ and dark photons $A'$ must be sufficiently strong to keep the dark electron--dark proton plasma tightly coupled to the DR before dark recombination.
    \item {\bf After dark recombination:} In general, the leftover free $e'$ and the newly-minted dark hydrogen $H'$ will remain in kinetic equilibrium for some time after dark recombination, as they do in the SM.
    In order to prevent this from keeping the atomic DM tightly coupled to the DR, we require that the momentum transfer rate between the DR (the $A'$ and $\nu'$) and the dark hydrogen, through scatterings with the free $e'$, is sufficiently small.
\end{enumerate}

\begin{figure}[tb]
	\centering
	\includegraphics[width=.32\linewidth]{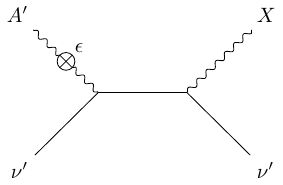}
	\includegraphics[width=.32\linewidth]{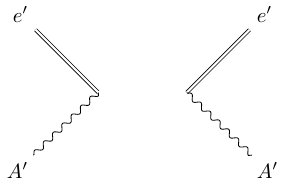}
	\includegraphics[width=.32\linewidth]{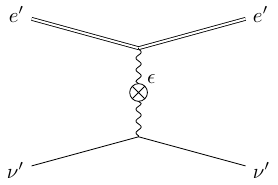}
	\caption{Representative Feynman diagrams for the relevant dark sector interactions.
    {\it Left:} $A' \nu' \to X \nu'$ scattering.
    The associated scattering rate must be large enough to guarantee self-interacting DR (requirement (i)).
    {\it Middle:} $A' e' \to A' e'$ scattering.
    The associated scattering rate must be large enough to keep the two fluids tightly coupled before dark recombination (requirement (ii)).
    {\it Right:} $e' \nu' \to e' \nu'$ scattering between the post-dark recombination leftover free dark electrons $e'$ and the dark neutrinos $\nu'$.
    The corresponding scattering rate must be smaller than the Hubble expansion rate, otherwise $e'$--$H'$ interactions may result in the atomic DM and the DR remaining tightly coupled even after recombination (requirement (iii)).
 }
	\label{fig:diagrams}
\end{figure}

Representative Feynman diagrams for the processes relevant to the requirements listed above are shown in Fig.~\ref{fig:diagrams}.
We devote the next section to the study of these processes and how they realize the requirements listed above, as well as to a detailed description of the evolution equations governing dark recombination, the dark acoustic oscillations (DAO), and the impact that these have on cosmological observables such as the CMB and the MPS.

\section{Cosmology of \nuadam} \label{sec:cosmo}

At high temperatures $T'$, the dark protons and dark electrons constitute an ionized dark plasma, similar to the baryon--photon plasma in $\Lambda$CDM and the dark plasma of standard ADM models.
Once $T'$ drops sufficiently below the binding energy $B_{H'} \equiv \alpha'^2 m_{e'} / 2$, the $p'$ and $e'$ undergo dark recombination, forming neutral atomic dark hydrogen $H'$.
Once the density of free dark electrons becomes sufficiently low, the interacting DM decouples from the DR bath.
We note here that dark recombination in \nuadam proceeds in a manner significantly different from its counterpart in the SM and the more standard atomic DM scenarios, due to the fact that the dark photon is part of an interacting DR bath.
In the present section, we first give a detailed description of dark recombination and the dynamics surrounding the pre- and post-dark recombination eras.
We then turn our attention to the study of the time evolution of cosmological perturbations in the \nuadam model, as well as of their impact on cosmological observables such as the CMB and the MPS.
All the plots shown in this section were created with our implementation of the \nuadam model in the {\tt CLASS} program; see \Sec{sec:num} for more details.

\subsection{Background: Scattering Rates and Thermal History}

We begin by considering the conditions listed in \Sec{sec:model}
that \nuadam must satisfy.

\subsubsection{Self-interacting Dark Radiation}

The main difference between \nuadam and the standard atomic DM scenarios is the fact that the dark photon is part of a richer SIDR component.
If the parameters of the model are such that condition (i) in \Sec{sec:model} is satisfied, the DR will remain self-interacting at all times of interest.
We now focus on this requirement, leaving a discussion of its impact on dark recombination for later.

Let us first consider what is required for the dark photon $A'$ to remain in thermal equilibrium with the rest of the DR, \ie~the $X$ gauge bosons and the $\nu'_j$ dark neutrinos.
The relevant process, shown in the left panel of Fig.~\ref{fig:diagrams}, is $A' \nu'_j \leftrightarrow X \nu'_j$, which is faster than $A' \nu'_j \leftrightarrow A' \nu'_j$ since the latter has an extra $\epsilon^2$ suppression and we are interested in the regime $\epsilon \ll 1$.
The rate for $A' \nu'_j \leftrightarrow X \nu'_j$ scatterings is approximately given by
\begin{equation}\label{eq:gaax}
    \Gamma_{A' \nu' \leftrightarrow X \nu'} \sim N_f \epsilon^2 \alpha_X^2 T' \, ,
\end{equation}
which needs to be larger than the Hubble expansion rate $H \sim T^2/m_\pl$.
Since, as we shall see in the next section, $T'/T \sim \mathcal{O}(0.1)$--$\mathcal{O}(1)$, we conservatively estimate that this condition is satisfied if the DR temperature is
\beq
    T' \lesssim 10~\GeV \, N_f \, \left( \frac{\epsilon}{10^{-6}} \right)^2 \left( \frac{\alpha_X}{10^{-2}} \right)^2 \ .
\eeq
This roughly corresponds to redshifts $z \lesssim 10^{14}$, which includes all times at which the relevant modes are inside the horizon.
Our benchmark parameters are chosen such that the dark photon is a component of the self-interacting DR at all the times of interest.
Since processes such as $X \nu'_{j} \leftrightarrow X \nu'_{j}$ do not have an $\epsilon$ suppression, it is evident that the $X$ boson and the dark neutrinos remain coupled at all relevant times, as long as the dark photon is also coupled to the DR.

Finally, number-changing processes such as $\nu'_j \nu'_k \to \nu'_j \nu'_k X$, if efficient enough, will ensure chemical equilibrium and thus a thermal distribution for the DR with zero chemical potential.
The rate for this process in particular scales roughly like $\Gamma_{\nu'\nu' \to \nu'\nu' X} \sim \alpha_X^3 T'$, which is larger than $H$ for $T' \lesssim 10^9~\TeV \bl( \alpha_X / 10^{-2} \br)^3$.
For the values of $T'/T$ we will consider ($\mO(0.1)$--$\mO(1)$), this is a very weak requirement.

\subsubsection{Before Dark Recombination}

At temperatures significantly above the binding energy $B_{H'}$, the dark electrons and protons are unbound.
This means that the dark electrons can interact efficiently with the dark photons via Compton scattering, keeping them tightly coupled.
For the parameter space of interest, Rutherford scattering between dark electrons and dark protons prior to dark recombination is also very efficient, meaning these two components also remain tightly coupled throughout this era.
Indeed, the dark protons impart momentum to the dark electrons at a rate given roughly by \cite{Cyr-Racine:2012tfp,Dvorkin:2013cea,Xu:2018efh,Boddy:2018wzy,Dvorkin:2020xga}
\bea
    \Gamma_{e'; p'} \sim \frac{\alpha'^2 n_{e'}}{m_{e'}^{1/2} T'^{3/2}} \ln\bl[ \bl( \frac{T' \lambda_D}{\alpha'} \br)^{2} \br] & \approx & 10^{-15}~\eV \ \bl( \frac{f_\adm}{1\%} \br) \bl( \frac{\alpha'}{10^{-2}} \br)^2 \bl( \frac{10^{-3}}{m_{e'} / m_{p'}} \br)^{1/2} \nonumber\\
    & \times & \bl( \frac{1~\GeV}{m_{p'}} \br)^{3/2} \bl( \frac{T'}{100~\eV} \br)^{3/2} \, \ln\bl[ \bl( \frac{T' \lambda_D}{\alpha'} \br)^{2} \br] ,
\eea
where for simplicity we have assumed $m_{e'} \ll m_{p'}$, and we have taken the dark electron ionization fraction to be $x' \approx 1$, since dark recombination has not started yet.
Also, we have regularized the forward-direction log-divergence of the Rutherford scattering with the Debye length $\lambda_D \equiv \sqrt{T' / (4 \pi \alpha' n_{e'})}$ \cite{Dvorkin:2013cea,Cyr-Racine:2012tfp}.
Clearly, this rate is much larger than $H \sim 10^{-23}~\eV \times \bl( T / 100~\eV \br)^2$ at the time when the shortest CMB modes entered the horizon.
One can also see that the heat-exchange rate between the dark electrons and dark protons is highly efficient, resulting in both fluids having the same temperature.

As a result, the dark electrons and dark protons are tightly coupled, responding as a single plasma fluid to the pressure imparted on each of them by the dark photons.
The dominant coupling between this plasma and the dark photon is through $e' A' \leftrightarrow e' A'$ Compton scattering.
The momentum-exchange rate from the dark photons on the dark plasma is therefore:
\beq
    \Gamma_{e'+p';A'} \sim \sigma_{\rm Td} n_{A'} \frac{T'}{m_{p'}} \sim \frac{\alpha'^2}{m_{e'}^2}\frac{T'^4}{m_{p'}} \approx 10^{-25}~\eV \ \bl( \frac{\alpha'}{10^{-2}} \br)^2 \bl( \frac{10^{-3}}{m_{e'} / m_{p'}} \br)^2 \bl( \frac{1~\GeV}{m_{p'}} \br)^3 \bl( \frac{T'}{1~\eV} \br)^4 \, ,
\eeq
which is also larger than the Hubble expansion rate for these temperatures.
In these expressions, $\sigma_{\rm Td} = 8 \pi \alpha'^2 / 3 m_{e'}^2$ is the dark Thomson cross section.
In light of this, we see that condition (ii) in \Sec{sec:model} is satisfied.

\subsubsection{Dark Recombination}

Since the dark photon is part of a self-interacting DR fluid, the recombination in the dark sector is significantly altered compared to the visible sector.
Specifically, unlike in the SM case, direct recombination to the ground state $p' e' \leftrightarrow H'(1s) \, A'$ is no longer negligible because the mean free path for $A' \nu' \leftrightarrow X \nu'$ is shorter than that for dark photon absorption, which ensures the dark photons will not immediately reionize the dark atoms.

Indeed, the cross section for absorbing photons near the threshold energy is \cite{Gottfried:2003aaa,Karzas:1961aaa,Spitzer:1978aaa}
\beq
    \sigma_\text{abs} \sim 6 \times 10^{-18} \left( \frac{10^{-2}}{\alpha'} \right) \left( \frac{\MeV}{m_{e'}} \right)^2 \cm^2 \ ,
\eeq
whereas the rate for $A' \nu' \leftrightarrow X \nu'$ scattering is given by \Eq{eq:gaax}.
This means that the ratio of the dark photon's mean free path due to absorption by a ground state (1s) dark hydrogen to that due to DR interactions is
\bea
    \frac{\ell_\text{abs}}{\ell_\text{SI}} & \simeq & \frac{\Gamma_{A'\nu' \leftrightarrow X \nu'}}{\sigma_{\rm abs} n_{H'(1s)}} \nonumber\\
    & \sim & 600 \ \frac{N_f}{x_{H' (1s)}} \bl( \frac{1\%}{f_\adm} \br) \bl( \frac{\alpha'}{10^{-2}} \br) \bl( \frac{\epsilon}{10^{-6}} \br)^2 \bl( \frac{\alpha_X}{10^{-2}} \br)^2 \bl( \frac{m_{e'} / m_{p'}}{10^{-3}} \br)^2 \bl( \frac{m_{p'}}{1~\GeV} \br)^3 \bl( \frac{1~\eV}{T'} \br)^2 \nonumber\\
    & \sim & 24 \ \frac{N_f}{x_{H' (1s)}} \bl( \frac{1\%}{f_\adm} \br) \bl( \frac{10^{-2}}{\alpha'} \br)^3 \bl( \frac{\epsilon}{10^{-6}} \br)^2 \bl( \frac{\alpha_X}{10^{-2}} \br)^2 \bl( \frac{m_{p'}}{1~\GeV} \br) \ ,
    \label{eq:rec-ratio-lengths}
\eea
where for the last equality we have set the temperature to be near the onset of recombination,  $T' = B_{H'}/10$.
When $\epsilon$ and $\alpha_X$ are sufficiently large, we can see from Eq.~\ref{eq:rec-ratio-lengths} that $\ell_\text{abs} > \ell_\text{SI}$.
This means that any dark photons emitted during the recombination process are quickly re-assimilated into the DR through its interactions, before they have the chance to reionize a 1s dark hydrogen atom and thus thwart recombination.
The two length scales involved become comparable only when the dark recombination is nearly complete.

\begin{figure}[h!]
	\centering
	\includegraphics[width=.75\linewidth]{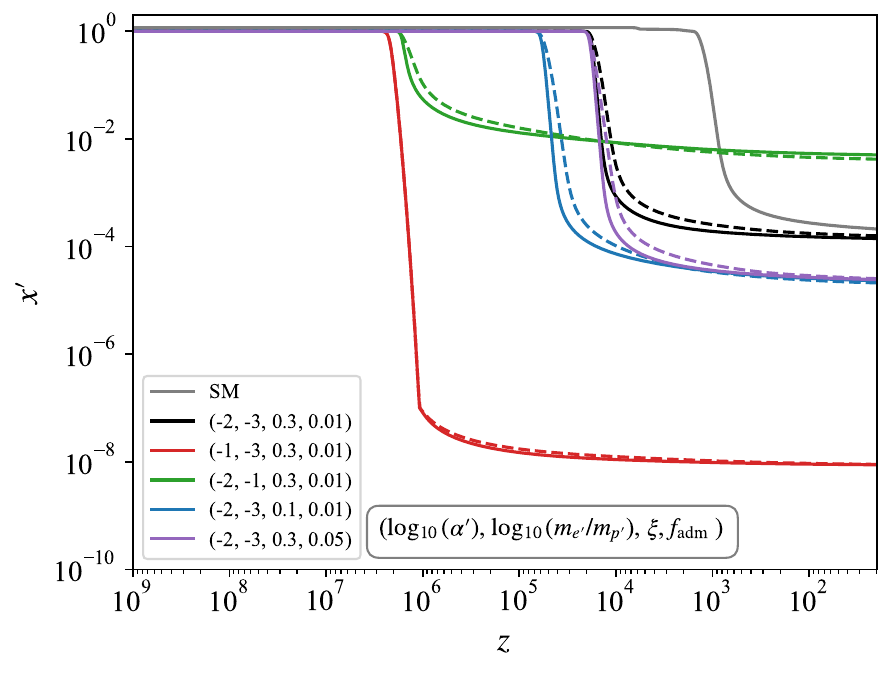}
	\caption{The ionization fraction as a function of redshift for \nuadam (solid) and ADM (dashed) with respect to different parameters, labeled as $(\log_{10}(\alpha'), \allowbreak \logmm, \allowbreak \xi, \allowbreak f_\adm)$.
 }
	\label{fig:drecom}
\end{figure}

As stated above, this means that for large $\epsilon$ and $\alpha_X$ recombination to the dark hydrogen ground state is preferred.
Therefore, when solving the equations that describe dark recombination, one must sum over all dark hydrogen final states when solving the recombination equations, including the ground state $H'(1s)$.
In fact, recombination to the ground state is the dominant process.
This is commonly known in the literature as ``Case A recombination''.
On the other hand, if $\epsilon$ or $\alpha_X$ are not large enough and $\ell_{\rm abs} / \ell_{\rm SI} < 1$, then recombination to the ground state is not possible and one needs to consider recombination into the excited states~\cite{Peebles:1968ja}, in what is known as ``Case B recombination''.
Since the ratio $\ell_{\rm abs} / \ell_{\rm SI}$ depends on the temperature, for some parts of the parameter space the dark recombination process may transition from Case A to Case B.
It is challenging to create a code that numerically solves the equations in this most general case.
Since dark recombination is purely Case A for a significant portion of the parameter space of interest, and since the difference between cases A and B has only a temporary impact on the cosmological evolution of $x'$ (compare the solid and dashed lines in Fig.~\ref{fig:drecom}), we consider only Case A recombination throughout this paper.
This can be easily realized if $10^{-6} \lesssim \epsilon \lesssim 10^{-8}$, and $1 \gtrsim \alpha_X \gtrsim 10^{-2}$; between these intervals one has $\ell_\text{abs} > \ell_\text{SI}$.

Assuming that all the dark hydrogen states are in thermal equilibrium, the equation governing the evolution of the free dark electron fraction $x'$ \cite{Peebles:1968ja,Spitzer:1978aaa} takes the simplified form:
\beq
    \frac{\dd x'}{\dd t} = - \left[ \mathcal{A}_A(T'_\adm) n_\adm x'^2 - \mathcal{A}_A(T') x_{H'(1s)} \left( \frac{m_{e'} T'}{2\pi} \right)^{3/2} e^{-B_{H'}/T'}\right] \ ,
\label{eq:xdcA}
\eeq
where, as introduced in \Sec{sec:model}, $n_\adm$ denotes the total dark baryon number density (\ie, dark protons, both free and bound in dark hydrogen); $x' = n_{e'} / n_\adm$ and $x_{H'(1s)} = n_{H'(1s)}/n_\adm \approx x_{H'} = 1 - x'$.
The coefficient $\mathcal{A}_A$ is the Case A recombination rate, which can be approximated as \cite{Pequignot:1991}
\beq
\mathcal{A}_A(T') = \frac{(\alpha'/\alpha_e)^2}{(\mu'/m_e)^2}
\frac{5.596\times10^{-13}~T'^{-0.6038}}{1+0.3436~T'^{0.4479}} \cm^3\seg \ ,
\eeq
where $\mu'$ represents the reduced mass of dark proton and electron.
$T'_\adm$ and $T'$ denote the temperatures of the atomic DM and the DR, respectively.
While in the early universe the ionized dark plasma maintained $T'_\adm = T'$, the two temperatures no longer track each other after dark recombination.
The temperature of atomic DM is governed by the following equation:
\beq
    (1+z)\frac{\dd T'_\adm}{\dd z} = 2 T'_\adm + \frac23 \frac{\Pi_\text{p-r} - \Pi_\text{p-i} - \Pi_\text{ff} + \Pi_\text{R}}{n_\adm (1 + x')H(z)} + \Gamma_T \frac{T'_\adm - T'}{H(z)} \ ,
\eeq
where $\Gamma_T$ is the Compton heating rate, and $\Pi_\text{p-r}$, $\Pi_\text{p-i}$, $\Pi_\text{ff}$ and $\Pi_\text{R}$ are the volumetric rates for photo-recombination heating, photo-ionization cooling, Bremsstrahlung, and Rayleigh scattering, respectively.
In this study, we exclude the photo-recombination and photo-ionization contributions to the matter temperature.
This omission is justified by the complexity of the detailed atomic physics required to accurately model their interactions with the DR, and by the relatively minor impact these processes have on the overall recombination history~\cite{Bansal:2022qbi}.
The rates of the processes we are considering are then given by \cite{Cyr-Racine:2012tfp, Seager:1999km}
\begin{equation}
\begin{aligned}
    \Pi_\text{ff} & \simeq \frac{16\alpha'^3 \bar g_{\rm ff} \sqrt{2\pi T'} x'^2 n_\adm^2}{(3 \mu')^{3/2}} \left( \frac{\pi^2 \delta(1+2\delta) - 6\zeta(3)\delta^2}{6}\right) \ , & \\
    \Pi_\text{R} & \simeq \frac{430080\zeta(9)\alpha'^2 n_\adm(1-x')T'^9 \delta}{\pi^2 B_{H'}^4 m_{H'}m_{e'}} \ , & \\
    \Gamma_T & = \frac{64\pi^3\alpha'^2 T'^4}{135m_{e'}^3} \frac{x'}{1+x'} \left[ 1 + \left(\frac{m_{e'}}{m_{p'}} \right)^3\right] \ , &
\end{aligned}
\label{eq:tdrates}
\end{equation}
where $g_{\rm ff} = 1.3$ is the Gaunt factor, and $\delta \equiv (T' - T'_\adm)/T'$.
As we discuss in more detail in the next section, we solve \Eq{eq:xdcA} with the aid of a modified version of the \texttt{HyRec-2} code \cite{Switzer:2007sq,Hirata:2008ny,Ali-Haimoud:2010tlj,Ali-Haimoud:2010hou,Lee:2020obi}.

In Figure~\ref{fig:drecom}, we show a few examples of the evolution of $x'$ as a function of redshift for several values of $\alpha'$, $m_{e'}/m_{p'}$, $\xi$, and $f_\adm$ in the \nuadam model, having fixed all other parameters to their best-fit values, according to our likelihood analysis of the baseline dataset $\mD$, as described in \Sec{sec:num}.
The solid lines in Fig.~\ref{fig:drecom} represent the evolution of $x'$ in the \nuadam model (Case A recombination), while the dashed lines correspond to the standard ADM model (\ie, purely Case B recombination).
Note that the difference between the two models is relatively small and, in the end, the asymptotic values of $x'$ are very similar; the two recombination scenarios impact the dynamics at the time of dark recombination while leaving the end result mostly identical.
The main difference is  that for Case A recombination $x'$ drops earlier and more quickly.
This is to be expected, since in Case A recombination the dark hydrogen ground state is available as a final state of the recombination process and, it being the most tightly bound state, allows for a more efficient recombination soon after the dark temperature drops below $B_{H'}$.

Increasing $\alpha'$ (red curves) raises both the binding energy and the Thomson scattering rate, leading to earlier recombination and lower residual $x'$.
A larger $m_{e'}/m_{p'}$ ratio (green lines) also increases the binding energy, but reduces the Thomson scattering rate, resulting in an earlier, shallower recombination.
Decreasing $\xi$ (blue curves) makes the dark sector colder, and it similarly shifts recombination to an earlier time.
While increasing $f_\adm$ (purple lines) does not directly impact the redshift of dark recombination, it does alter the final ionization fraction by changing $n_\adm$ in Eq.~\ref{eq:xdcA}.
Compared to the ordinary ADM results (dashed curves), dark recombination occurs more quickly in the \nuadam scenario due to the inclusion of the direct recombination to the dark hydrogen ground state, but the final value of $x'$ is nearly identical in both cases.

\subsubsection{After Dark Recombination}

After dark recombination, the dark hydrogen can still scatter with the leftover fraction $x'$ of free dark electrons.
The momentum-exchange rate from the dark electrons on the dark hydrogen is \cite{Cyr-Racine:2012tfp}
\bea
    \Gamma_{H';e'} \sim \frac{m_{e'}}{m_{p'}} \sigma_{e' H'} n_{e'} v_{e'} & \approx & 10^{-27}~\eV \bl( \frac{f_\adm}{1\%} \br) \bl( \frac{x'}{10^{-4}} \br) \bl( \frac{10^{-2}}{\alpha'} \br)^2 \nonumber\\
    && \times \bl( \frac{10^{-3}}{m_{e'}/m_{p'}} \br)^{3/2} \bl( \frac{1~\GeV}{m_{p'}} \br)^{7/2} \bl( \frac{T'}{1~\eV} \br)^{7/2} \ ,
\eea
where $\sigma_{e' H'} = {25\pi \alpha'^2}/{4B_{H'}^2}$
\footnote{The formula here differs from the one given in Ref.~\cite{Cyr-Racine:2012tfp} by a factor of about $16$, but agrees numerically with the results of Ref.~\cite{STEPANEK200399}.}
and we have assumed $m_{e'} \ll m_{p'} \approx m_{H'}$ and $T' = T'_\adm \ll B_{H'}$.
Comparing this rate to the Hubble expansion rate during the era of radiation domination, $H \sim 10^{-27}~\eV \bl( T/\eV \br)^2$, we can see that the dark hydrogen remains tightly coupled to even the small fraction of dark electrons that remain free after dark recombination.

Given that the dark hydrogen remains tightly coupled to the free dark electrons, there is a danger that, if the leftover free dark electrons remain tightly coupled to the DR, the entire atomic DM would remain coupled to DR.
This would mean that a fraction $f_\adm$ of the DM would still have dark acoustic oscillations well after dark recombination, as in the partially acoustic dark matter scenario~\cite{Chacko:2016kgg}, which does not provide as good a fit to data~\cite{Buen-Abad:2017gxg}.
To ensure this does not occur, the \nuadam parameters must satisfy requirement (iii) of \Sec{sec:model}.
The momentum-exchange rate from the $A'$ to the tightly-coupled $e'$--$H'$ system is:
\beq
    \Gamma_{e'+H'; A'} \sim \frac{x' \alpha'^2}{m_{e'}^2}\frac{T'^4}{m_{p'}} \approx 10^{-29}~\eV \ \bl( \frac{x'}{10^{-4}} \br) \bl( \frac{\alpha'}{10^{-2}} \br)^2 \bl( \frac{10^{-3}}{m_{e'} / m_{p'}} \br)^2 \bl( \frac{1~\GeV}{m_{p'}} \br)^3 \bl( \frac{T'}{1~\eV} \br)^4 \, ,
\eeq
which for the benchmark values in this equation is indeed below the Hubble rate.
Through the kinetic mixing between $A'$ and $X$, the $e'$ has a coupling to the $X$ and $\nu'_j$ particles.
The leading contribution comes from the $t$-channel process shown in the right panel of Fig.~\ref{fig:diagrams}, which has a momentum-transfer rate $\Gamma_{e'; \nu'_j} \sim N_f \epsilon^2 \alpha' \alpha_X T'^2 / m_{e'}$.
This in turn results in a momentum-exchange rate of the $\nu'_j$ on the entire dark plasma given by
\bea
    \Gamma_{e'+H'; \nu'_j} \sim \frac{N_f x' \epsilon^2 \alpha' \alpha_X T'^2}{m_{p'}} & \approx & 10^{-29}~\eV \ \bl( \frac{N_f}{3} \br) \bl( \frac{x'}{10^{-4}} \br) \bl( \frac{\epsilon}{10^{-6}} \br)^2 \nonumber\\
    && \times \bl( \frac{\alpha'}{10^{-2}} \br) \bl( \frac{\alpha_X}{10^{-2}} \br) \bl( \frac{1~\GeV}{m_{p'}} \br) \bl( \frac{T'}{1~\eV} \br)^2 \ ,
\eea
having assumed $T'_\adm \approx T'$.
For the benchmark values shown in the expression above, we see that the rate is safely below Hubble, meaning that after dark recombination the atomic DM is no longer tightly coupled to the DR.

\subsection{Perturbations: Impact on Cosmological Observables}
\label{subsec:perts}

Having described the background history of the \nuadam model, we now turn our attention to its cosmological perturbations and their impact on the anisotropies of the CMB and on the MPS.
In the conformal Newtonian gauge, the equations governing the evolution of the energy density and velocity gradient perturbations in the atomic DM and the self-interacting DR ($\delta_\adm$ and $\theta_\adm$, and $\delta_\dr$ and $\theta_\dr$, respectively) are \cite{Ma:1995ey,Cyr-Racine:2012tfp,Cyr-Racine:2013fsa,Bansal:2021dfh,Bansal:2022qbi,Hughes:2023tcn}

\bea
    \dot{\delta}_\adm & = & -\theta_\adm + 3 \dot\phi \ , \\
    \dot{\theta}_\adm & = & - \mathcal{H} \theta + c_{d,s}^2 k^2 \delta_\adm + k^2 \psi + S \gamma \Gamma_d \bl( \theta_\dr - \theta_\adm \br) \ , \\
    \dot{\delta}_\dr & = & - \frac{4}{3} \theta_\dr + 4 \dot\phi \ ,\\
    \dot{\theta}_\dr & = & \frac{1}{4}\delta_\dr + k^2 \psi + \gamma \Gamma_d \bl( \theta_\adm - \theta_\dr \br) \ ,
\eea
where the dots denote derivatives with respect to conformal time $\tau$, $\mathcal{H} \equiv \dot{a}/a$ is the conformal Hubble rate, $\phi$ and $\psi$ are the gravitational potentials in the spacetime metric, $k$ is the wavenumber of the perturbation in question, and $S \equiv 4 \rho_\dr / 3 \rho_\adm$. The parameter $\gamma \equiv 2 / \bl(4 + g_{*\nu'} \br)$ accounts for the fact that the atomic DM only scatters off the dark photon component of the DR, and $\Gamma_d$ is the momentum-exchange rate from the non-relativistic atomic DM components to $A'$, given by \cite{Bansal:2022qbi}:
\bea
    \Gamma_d & \equiv & \Gamma_{\rm Compton} + \Gamma_{\rm Rayleigh} + \Gamma_{\rm p-i} \ , \\
    \text{with} \quad \Gamma_{\rm Compton} & \equiv & a\, x' n_\adm \sigma_{\rm Td} \bl( 1 + \bl( \frac{m_{e'}}{m_{p'}} \br)^2 \br) \ , \\
    \Gamma_{\rm Rayleigh} & \equiv & 32 \pi^4 a\, n_\adm \bl( 1 - x' \br) \sigma_{\rm Td} \bl( \frac{T'}{B_{H'}} \br)^4 \quad \bl( \text{for }T' \ll B_{H'} \br) \ , \\
    \text{and} \quad \Gamma_{\rm p-i} & \equiv & a\, n_\adm \bl( 1 - x' \br) e^{-B_{H'}/T'} \frac{\sqrt{\pi} m_{e'}^{3/2}}{4 \sqrt{2} \zeta(3) T'^{3/2}} \bl( \frac{\alpha'}{\alpha_{\rm em}} \br)^3 \mathcal{A}_A(T') \ ;
\eea
where we have included contributions from Compton, Rayleigh, and photo-ionization scattering.
Note that since the DR is self-interacting, the higher moments $F_{\dr,\,\ell\geq 2}$ of the Boltzmann hierarchy vanish.

Soon after dark recombination occurs, the neutralized dark baryons cease to feel the pressure from the dark photons, in what is known as the (dark) baryon drag or decoupling epoch.
The two times of interest, dark recombination and dark baryon decoupling, can be defined with the help of $\Gamma_d$.
Indeed, dark recombination is said to have taken place when the dark visibility\footnote{A delightful oxymoron.} function $g_d$, defined as the probability that dark photons last scattered within $\dd \tau$ of $\tau$, reaches its maximum.
In terms of the conformal time $\tau$, the visibility function is
\beq
    g_d(\tau) \equiv \Gamma_d(\tau) \, \exp\bl[ - \int\limits_{\tau}^{\tau_0} \dd\tau' \, \Gamma_d(\tau') \br] \ ,
\eeq
where $\tau_0$ is the conformal age of the universe.
The conformal time at recombination is therefore $\tau_{\rm d,\, rec} \equiv \arg\max[g_d(\tau)]$; we will frequently quote instead the corresponding redshift $z_{\rm d,\, rec}$.
On the other hand, dark baryon decoupling is defined in terms of the dimensionless dark baryon drag $\overline{\tau}_d$:
\beq
    \overline{\tau}_d(\tau) \equiv \int\limits_{\tau}^{\tau_0} \dd\tau' \, S(\tau') \, \Gamma_d(\tau') \ .
\eeq
The conformal time at dark baryon decoupling is when $\overline{\tau}_d (\tau_{\rm d,\, dec}) = 1$; we will instead quote the corresponding redshift $z_{\rm d, \, dec}$.

\subsubsection{Matter Power Spectrum}

Let us now devote our attention to the impact of the \nuadam model on the MPS.
The pressure arising from the dark photons prior to dark recombination prevents the overdensities in the dark proton and dark electron fluids from growing, and instead leads them to undergo acoustic oscillations.
The fact that not all matter perturbations are growing decreases the gravitational potential perturbations (compared to \lcdm) and slows the growth of the CDM component.
This leads to a suppression in the MPS for modes that enter the horizon before the decoupling of the ionized dark plasma fluid from the dark photons.

\begin{figure}[h!]
    \centering
    \includegraphics[width=.49\linewidth]{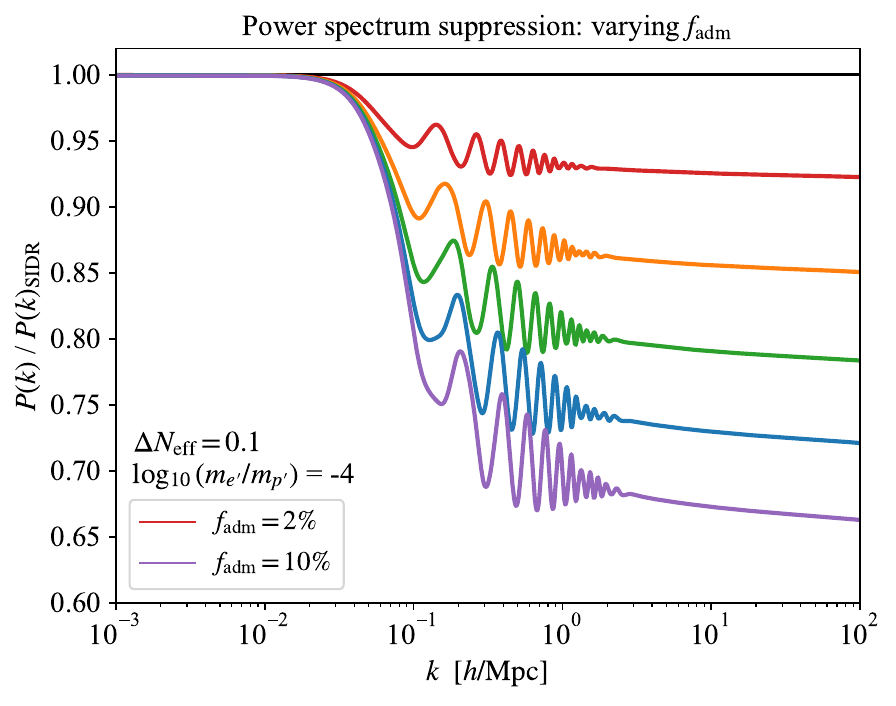}
    \includegraphics[width=.49\linewidth]{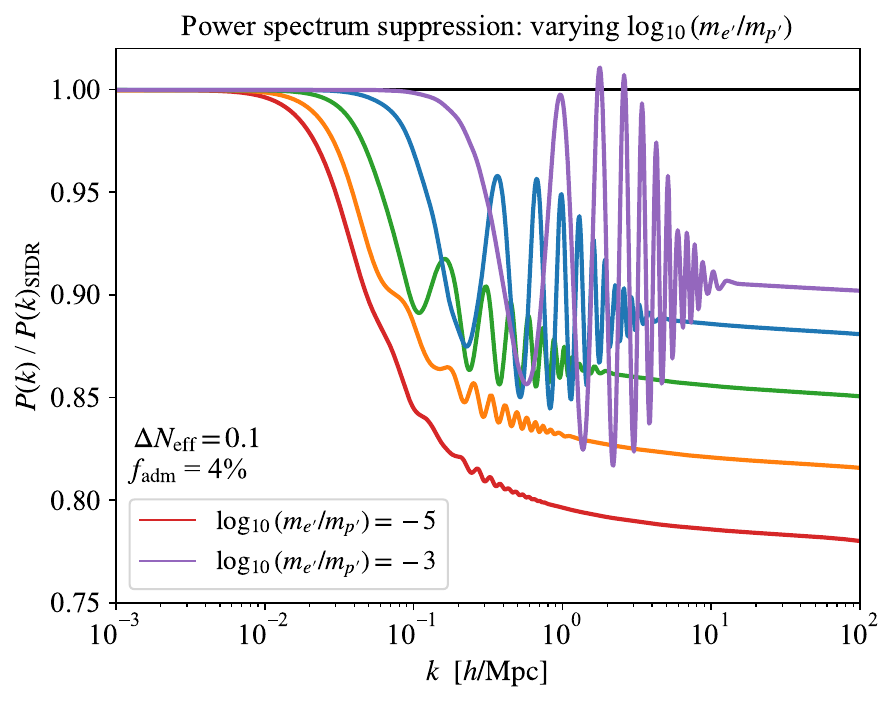}
	\caption{Ratio of the MPS of the \nuadam model to that of SIDR, for different choices of $f_\adm$ (left panel) and $\logmm$ (right panel).
    We have fixed $\DNeff = 0.1$ throughout.
    The $\theta_s$, $\omega_b$, $A_s$, $n_s$, and $\tau_{\rm reio}$ parameters have been fixed to their mean values in Planck's \lcdm fit to TT+TE+EE+lowE+lensing+BAO data \cite{Planck:2018vyg}.
    In order to keep the time of matter-radiation equality the same as in Ref.~\cite{Planck:2018vyg} in the presence of additional DR ($z_\eq = 3387$), we choose $\omega_\dm = 0.121$.
    {\it Left}: varying $f_\adm$ from $2\%$ to $10\%$ in $2\%$ increments, while fixing $\logmm = -4$.
    This corresponds to $z_{\rm d,\,rec} \approx 1900$ and $z_{\rm d,\,dec} \approx 2000$.
    {\it Right}: varying $\logmm$ from $-5$ to $-3$ in $0.5$ increments, while fixing $f_\adm = 4\%$.
    In order of increasing values of $\logmm$, this corresponds $(z_{\rm d,\,rec}, z_{\rm d,\,dec}) \approx (180, 210), \,(580, 640), \,(1900, 2000), \,(6200, 6200), \,\text{and } (21000, 19000)$, respectively.}
	\label{fig:rainbow_pk}
\end{figure}

In Fig.~\ref{fig:rainbow_pk} we show the ratio of the MPS of the \nuadam model to that of a SIDR model (\ie, \lcdm+$\DNeff$, where the extra DR is self-interacting).
The cosmological parameters $\theta_s$, $\omega_b$, $A_s$, $n_s$, and $\tau_{\rm reio}$ have been fixed to their mean values in the Planck Collaboration's \lcdm fit to TT+TE+EE+lowE+lensing+BAO data (last column of Table 2 of Ref.~\cite{Planck:2018vyg}).
We have taken $\DNeff = 0.1$ and consequently $\omega_\dm = 0.121$, in order to keep the redshift of matter-radiation equality fixed to $z_\eq = 3387$, which is the mean value given in Ref.~\cite{Planck:2018vyg}.
As we can see, the MPS in \nuadam is suppressed relative to that in SIDR; this suppression depends on the fraction of DM energy density that is in the dark electron/proton fluid (left panel), as well as on the time of recombination by changing the dark electron mass (right panel).
As expected, the suppression is more severe as we increase the amount of iDM, since having a larger fraction of DM not being able to gravitationally clump leads to a bigger slowdown of the growth of the cold component perturbations.
The suppression also moves to smaller scales (larger $k$) if dark recombination occurs earlier (larger $m_{e'}$), since only smaller scale modes would have entered the horizon prior to dark recombination.
Furthermore, one can also see that the acoustic oscillations in the dark fluid get imprinted in the MPS, which is completely analogous to the baryon acoustic oscillations (BAO) imprinted by the interactions of baryons and photons prior to recombination.
A similar phenomenon has been previously studied in Refs.~\cite{Cyr-Racine:2012tfp,Cyr-Racine:2013fsa,Bansal:2022qbi,Buen-Abad:2022kgf,Buen-Abad:2023uva,Hughes:2023tcn}.
As in the case of BAO, the DAO disappear for sufficiently small length scales (large $k$s) due to diffusion damping.
As can be seen in the right panel of Fig.~\ref{fig:rainbow_pk}, larger $\logmm$ leads to both a larger amplitude in the DAO and a less severe overall suppression.
Both effects are due to the fact that as \logmm increases, so does $z_{\rm d,\, dec}$.
Indeed, a larger $z_{\rm d,\, dec}$ means that the CDM perturbations have less time to grow larger than the iDM perturbations, before the latter decouple from the DR and begin to grow as well.
Therefore, the DAO in the iDM represent a larger share of the total DM perturbations at that time, and their impact on the evolution of the CDM component is also more significant.
As a result, the amplitude of the imprinted DAO is larger.
An earlier decoupling time also means that the gravitational potentials remain shallow (due to the non-growth of iDM because of DR pressure) for a shorter span of time.
Because of this, the growth of CDM perturbations is stunted for a shorter period of time as well, and therefore the overall suppression of the MPS is less severe.

\subsubsection{Cosmic Microwave Background Anisotropies}

\begin{figure}[h!]
    \centering
    \includegraphics[width=.49\linewidth]{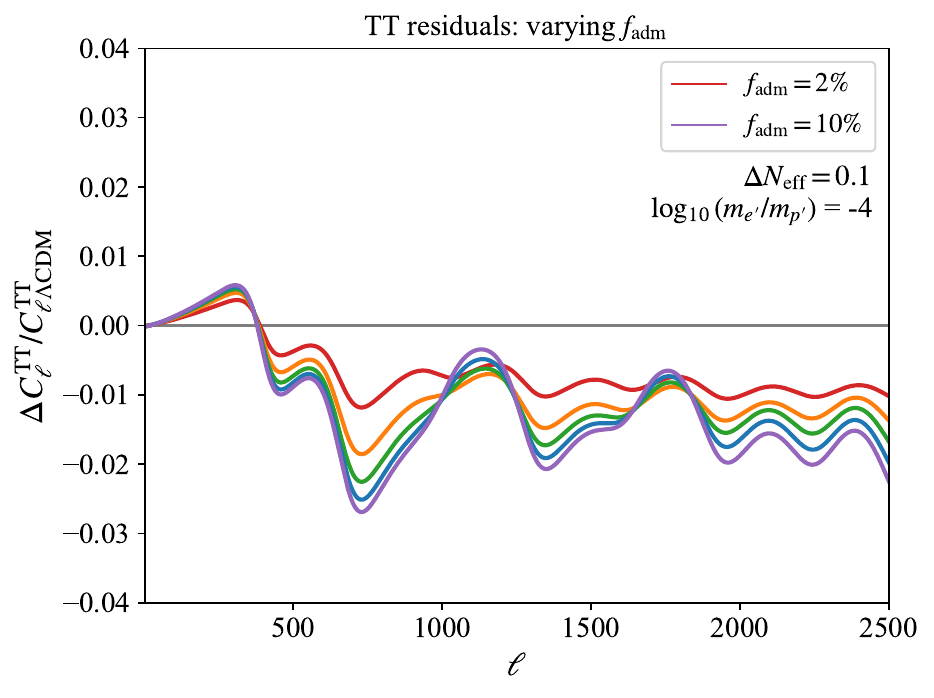}
    \includegraphics[width=.49\linewidth]{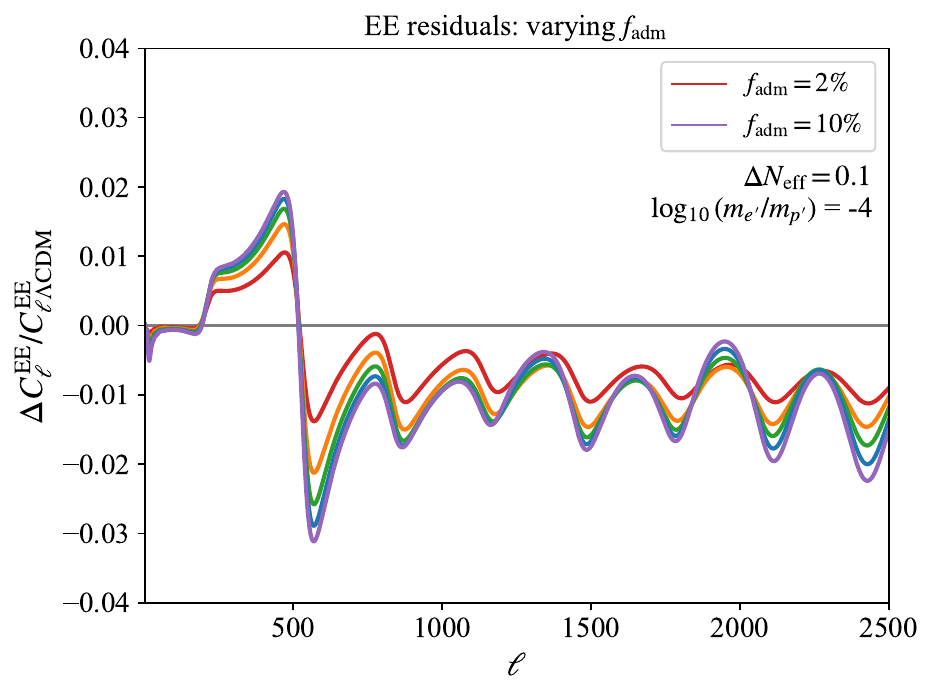}
    \includegraphics[width=.49\linewidth]{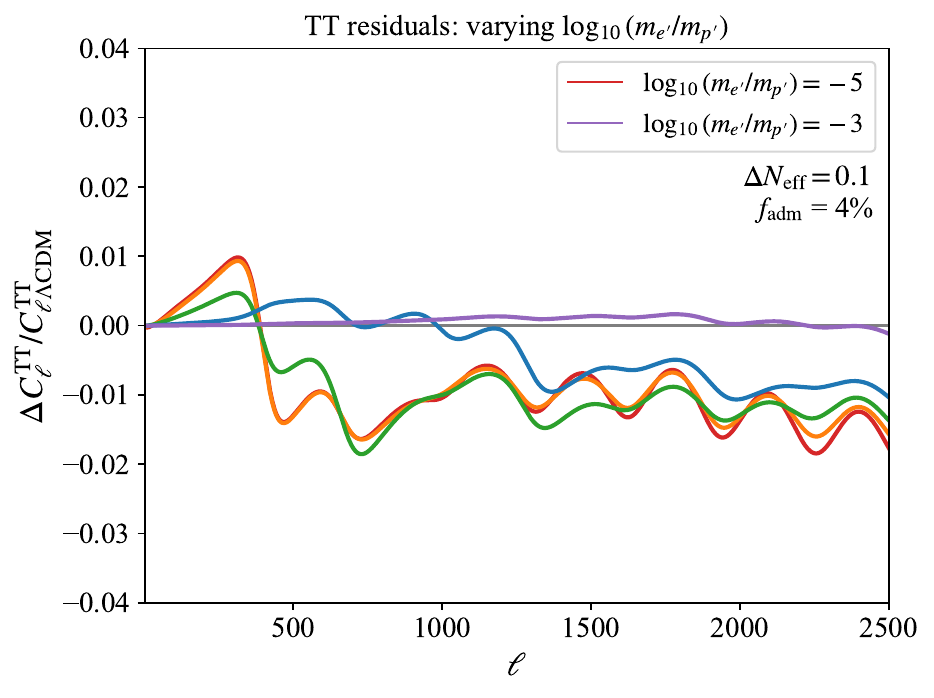}
    \includegraphics[width=.49\linewidth]{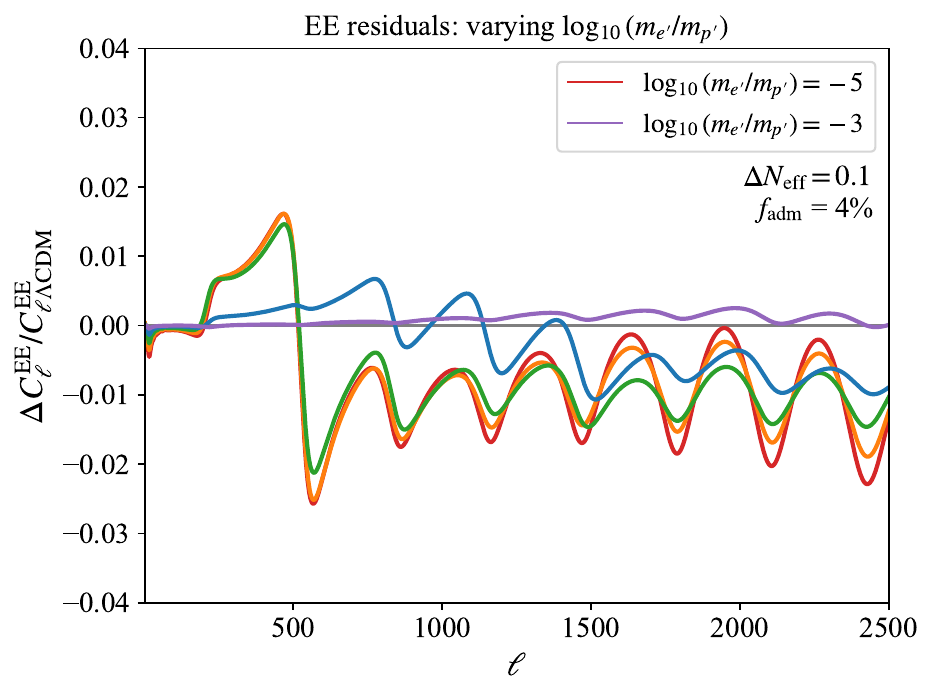}
	\caption{Residuals of the CMB's $C_\ell^{TT}$ (left) and $C_\ell^{EE}$ (right) for \nuadam compared to those of SIDR, for different choices of $f_\adm$ (top) and $\logmm$ (bottom).
    All other parameters are fixed as in Fig.~\ref{fig:rainbow_pk}.}
	\label{fig:rainbow_cmb}
\end{figure}

The impact of a SIDR fluid on the CMB is well understood (see, \eg, Refs.~\cite{Chacko:2003dt,Friedland:2007vv,Baumann:2015rya,Brust:2017nmv,Blinov:2020hmc}).
In general, the origin of the effects that a dark sector has on the CMB can partly be traced to the changes in the gravitational potential perturbations induced by the dark fluid perturbations.
In Fig.~\ref{fig:rainbow_cmb}, we show the residuals of the CMB power spectrum of the \nuadam model compared to those of the SIDR model, with the same cosmological parameters as in Fig.~\ref{fig:rainbow_pk}.
The left (right) column shows the residuals in $C^{TT}_\ell$ ($C^{EE}_\ell$), while the top (bottom) row depicts their change as we vary $f_\adm$ ($\logmm$).

Modes that enter the horizon early (high $\ell$) have a gravitational potential that decays faster than in the standard SIDR scenario, resulting in a change to the radiation driving force and gravitational potential contribution to the Sachs-Wolf effect (see appendix of Ref.~\cite{Schoneberg:2023rnx} for a detailed discussion).
This in turn leads to a power suppression at high $\ell$, as seen in Fig.~\ref{fig:rainbow_cmb}.
The interactions between DR and the iDM also decrease the effective sound speed of the dark fluid, which leads to a change in the location of the high $\ell$ peaks~\cite{Ghosh:2024wva}.
This manifests itself in the form of an oscillatory pattern, as can be seen in the residual plots.

The step-like suppression in the CMB power spectra is qualitatively similar to what was observed in stepped DR models~\cite{Aloni:2021eaq}.
As in the case of stepped DR, this allows \nuadam to have a much larger $\DNeff$ contribution, and thus increase $\HO$, without worsening the fit to CMB data.
This can be understood, at least partially, by recalling that increasing $\DNeff$ (and exploring approximate degeneracies with other parameters) results in a more pronounced Silk damping \cite{Bashinsky:2003tk}.
Because Silk damping only affects $\ell$ above the damping scale, it creates an $\ell$ dependent change in the CMB power spectra.
While $n_s$ can be used to partially offset the extra suppression due to enhanced Silk damping from increasing $\Neff$, it affects all $\ell$.
Having additional dynamics that only affects high $\ell$ allows for extra approximate degeneracies with the cosmological parameters to offset the increased Silk damping.

\section{Numerical Analysis}
\label{sec:num}

We modify the publicly available code {\tt CLASS} v3.1 \cite{Lesgourgues:2011re,Blas:2011rf,Lesgourgues:2011rg,Lesgourgues:2011rh}\footnote{\href{https://github.com/lesgourg/class_public}{\tt github.com/lesgourg/class\_public}.} in order to implement the \nuadam model and solve for the cosmic evolution of the universe.
We build upon the previous work of the authors of Refs.~\cite{Bansal:2021dfh,Bansal:2022qbi}, who have graciously made publicly available their own modifications of {\tt CLASS} implementing standard atomic DM.\footnote{\href{https://github.com/srbPhy/class_twin}{\tt github.com/srbPhy/class\_twin}, \href{https://github.com/jp-barron/class_adm-3.1}{\tt github.com/jp-barron/class\_adm-3.1}.}
Our implementation of \nuadam involved modifying these previous codes in order to include Case A dark recombination (\ie, dark recombination to the dark hydrogen ground state) in the {\tt HyRec-2} \cite{Switzer:2007sq,Hirata:2008ny,Ali-Haimoud:2010tlj,Ali-Haimoud:2010hou,Lee:2020obi} and ``{\tt thermodynamics}'' modules, the additional DR degrees of freedom represented by the $N_f$ dark neutrinos $\nu'_j$ and the additional gauge boson $X$, the DR self-interactions, as well as the proper treatment of the tight-coupling and radiation-streaming approximations in the dark sector \cite{Blas:2011rf}.

We employ our modified version of {\tt CLASS} in conjunction with the MCMC sampler {\tt MontePython}~\cite{Audren:2012wb,Brinckmann:2018cvx} in order to study how well the \nuadam model fits a variety of cosmological datasets, and how this fit compares to \lcdm.
We use the {\it Halofit} model \cite{Smith:2002dz,Takahashi:2012em} to treat the non-linear regime of the MPS.%
\footnote{The Halofit model has been extensively tested for \lcdm.
While it is possible that this may not be true for models with a non-\lcdm-like linear MPS, Ref.~\cite{Schoneberg:2023rnx} showed that Halofit makes only a minute difference in models where DR and iDM interact strongly (\eg, SPartAcous \cite{Buen-Abad:2022kgf,Buen-Abad:2023uva}).
Since these models and \nuadam have qualitatively the same step-like suppression in the linear MPS, we expect this conclusion to hold also for our model.
}
In {\tt MontePython} we use the Metropolis-Hastings algorithm, and take a Gelman-Rubin (GR) criterion of $R < 1.01$, where $R$ is the GR statistic~\cite{Gelman:1992zz}, to be an indication that the MCMC chains have converged.

In addition to the standard \lcdm parameters $\{ \omega_b, \, \omega_\dm, \, \theta_s, \, \ln \bl( 10^{10} A_s \br), \, n_s, \, \tau_\text{reio} \}$, we scan over three other \nuadam parameters: the amount of self-interacting DR \DNeff,%
\footnote{\label{foot5}We assume that the DR is populated only after BBN ($T_\gamma \sim 1~\MeV$) via some unspecified mechanism.
This can easily be done with the inclusion of another dark sector component which deposits its energy density in the DR once BBN has finished.
This late DR thermalization allows \DNeff to take values relevant to the $H_0$ problem without at the same time running afoul of independent measurements of the primordial abundance of helium $Y_{\rm He}$.
We implement this in our code simply by setting $\DNeff = 0$ in the {\tt thermodynamics\_helium\_from\_bbn} routine of the {\tt thermodynamics.c} module.}
the fraction $f_\adm$ of the total DM energy density that can form dark atoms, and the ratio of the dark electron to dark proton masses in log-space, $\logmm$.%
We fix $\alpha_2 = 10^{-2}$, and $m_{p'} = 1~\GeV$, since these are parameters highly degenerate with the ones we vary, along with $N_f = 3$.
The precise values of $\alpha_X$ and $\epsilon$ are irrelevant as long as requirements (i)--(iii) in \Sec{sec:model} are satisfied.
The \nuadam priors we use are $\tau_{\rm reio} \geq 0.04$, $-3.5 \geq \logmm \geq -4.5$, $\DNeff \geq 0$, and $1 \geq f_\adm \geq 0$; all other parameters are allowed to vary freely.
Our choice of the $\logmm$ prior is informed by the fact that, for parameter values above this interval, \nuadam reduces to the SIDR limit.
Indeed, since a larger $m_{e'}/m_{p'}$ ratio corresponds to an earlier onset of dark recombination (see Fig.~\ref{fig:drecom}), for sufficiently large values of this parameter the dark recombination takes place well before any of the length-scales probed by CMB or MPS experiments re-enter the horizon.
Therefore, for all intents and purposes, the DM consists entirely of neutral dark hydrogen identical to CDM all throughout the evolution of the modes in question.
Thus, the only departure from \lcdm is in the presence of the additional DR $\DNeff$, which due to the kinetic mixing described in the previous section, is self-interacting; in other words, the SIDR model.
In this region there is a second minimum, corresponding to the well-known preference of various cosmological datasets that include the SH0ES measurements of $H_0$ (the $\mH$ dataset described in the next subsection) for a non-vanishing presence of SIDR \cite{Blinov:2020hmc}.
Since we are interested in the novel \nuadam model and not in SIDR, we exclude this region of parameter space from our analysis.

\subsection{Experiments and Methodology}
\label{subsec:exps}

We submit \nuadam to a full likelihood analysis using several combinations of the most up-to-date cosmological observations.
We divide our suite of datasets into three groups:

\begin{itemize}
    \item $\mD$: the baseline dataset.
    This includes the following experiments:
    \begin{itemize}
        \item {\bf Planck: } measurements of TT, TE, and EE CMB anisotropies and lensing from Planck 2018 PR3~\cite{Planck:2018vyg} (the likelihoods dubbed {\tt `Planck\_highl\_TTTEEE'}, {\tt `Planck\_lowl\_EE'}, {\tt `Planck\_lowl\_TT'}, and {\tt `Planck\_lensing'} in {\tt MontePython}).
        \item {\bf BAO: } measurements of the baryon acoustic oscillations.
        These include measurements of $D_V/r_\mathrm{drag}$ from the Six-degree Field Galaxy Survey (6dFGS) at $z = 0.106$ \cite{Beutler:2011hx} and the Sloan Digital Sky Survey (SDSS) from the MGS galaxy sample at $z = 0.15$ \cite{Ross:2014qpa} ({\tt `bao\_smallz\_2014'}), as well as measurements from the Data Release 16 (DR16) of the SDSS-IV extended Baryon Oscillation Spectroscopic Survey (eBOSS) \cite{eBOSS:2020yzd} of emission line galaxies (ELG) at $0.6 < z < 1.1$ \cite{eBOSS:2020qek,eBOSS:2020fvk} ({\tt 'bao\_eBOSS\_DR16\_ELG'}), quasars (QSOs) at $0.8 < z < 2.2$ \cite{eBOSS:2020gbb,eBOSS:2020uxp} ({\tt 'bao\_eBOSS\_DR16\_gal\_QSO'}), and Lyman-$\alpha$ forests \cite{eBOSS:2020tmo} ({\tt 'bao\_eBOSS\_DR16\_Lya\_auto'}, {\tt 'bao\_eBOSS\_DR16\_Lya\_cross\_QSO'}).
        \item {\bf Pantheon+: } measurements of the apparent magnitude of 1550 Type Ia supernovae (SNIa), at redshifts $0.001 < z < 2.26$ \cite{Scolnic:2021amr,Brout:2022vxf} ({\tt `Pantheon+'}).
    \end{itemize}
    \item $\mF$: ``Full-shape'' MPS likelihood based on BOSS DR12 LRG and eBOSS DR16 QSO data, analyzed with Effective Field Theory of Large Scale Structure (EFTofLSS) methods, and implemented in the {\tt PyBird} publicly available code ({\tt 'eftboss'} and {\tt 'efteboss'}).\footnote{\href{https://github.com/pierrexyz/pybird}{\tt github.com/pierrexyz/pybird}.}
    \item $\mH$: Late-universe measurements of the Hubble parameter $H_0$ today.
    The SH0ES collaboration obtains their value of $H_0$ from their measurements of the absolute magnitude $M_B$ of SNIa.
    Thus, we use a Gaussian likelihood based on their $M_B = -19.253 \pm 0.027$ result \cite{Riess:2021jrx}.
\end{itemize}

\subsection{Other datasets}

In this section we briefly review some recent datasets that have recently suggested the existence of further discrepancies between observations and \lcdm, and discuss our reasons for not having included them in our likelihood analysis.

\subsubsection{Lyman-$\alpha$ forest}
\label{subsubsec:lya}

The Lyman-$\alpha$ forest is sensitive to smaller non-linear scales, and is thus uniquely poised to constrain new physics that impacts structure formation at non-linear scales.
Interestingly, recent studies have claimed that Lyman-$\alpha$ observations are in significant tension with \lcdm fits to CMB, BAO, and supernovae results \cite{Goldstein:2023gnw,Rogers:2023upm}.
Although tempting, in particular due to the impact that \nuadam has on small scales, we choose not to include likelihoods quantifying the Lyman-$\alpha$ observables at these small scales, such as the so-called $\{ \alpha, \beta, \gamma, \delta \}$--parametrization \cite{Hooper:2022byl}, or the ``{\it Compressed Lyman-$\alpha$ likelihood}'' \cite{Pedersen:2022anu}.
This is because the shape of the power spectrum in \nuadam differs from the templates used to build these likelihoods.
In particular, lacking dedicated hydrodynamical simulations, it is unclear what the impact is, if any, of the DAO caused by the DR--dark plasma interactions and imprinted in the MPS on these length scales.

Even though we are not using Lyman-$\alpha$ data in our fits, we performed a simplified study of the implications of $\nuadam$ to these observables.
Our approach is to find an ``average'' function $\overline{\Delta}^2(k)$ that approximately represents the non-oscillatory part of the MPS in the scales of relevance to Lyman-$\alpha$ data.
We then use this function to estimate the amplitude $\Delta_L^2 \equiv k_*^3 P_{\rm lin}(k_*, z_*) / 2 \pi^2$ and tilt $n_L = \bl. \dd \ln P_{\rm lin}(k, z) / \dd \ln k \br \vert_{k_*,z_*}$ of the (non-oscillatory) linear MPS at pivot wavenumber $k_* \equiv 0.009~\seg/\km$ and redshift $z_* \equiv 3$, which have been found to represent the Lyman-$\alpha$ data fairly well \cite{eBOSS:2018qyj,Pedersen:2022anu}.
To do this we propose the function
\beq\label{eq:template_mpk}
    \overline{\Delta}^2(k; \boldsymbol{\theta}) = \theta_1 \bl( \frac{k}{k_*} \br)^{\theta_2+3} \bl( \frac{\ln(k \theta_3)}{\ln(k_* \theta_3)} \br)^2 \bl[ \bl( 1 - \theta_7 \br)\bl( 1 + \bl( \theta_4 k \br)^{\theta_5} \br)^{\theta_6} + \theta_7 \br] \ ,
\eeq
which ``cuts'' through the DAO.
This function is made of three distinct factors, each of them independently motivated.
The first is a power law at the point of interest, the second characterizes the logarithmic growth of small scales during the radiation domination era, and the third the plateau-like matter-power spectrum suppression caused by the pressure of the DR imparted upon the dark atoms, in the spirit of Ref.~\cite{Hooper:2022byl}.
We then use the {\tt optimize.curve\_fit} function of the {\tt Python} \cite{van1995python,10.5555/1593511} {\tt SciPy} package \cite{Virtanen:2019joe} to fit its parameters $\boldsymbol{\theta} \equiv \bl\{ \theta_1, \theta_2, \theta_3, \theta_4, \theta_5, \theta_6, \theta_7 \br\}$ to the dimensionless linear MPS $k^3 P_{\rm lin}(k, z)/2 \pi^2$ computed by {\tt CLASS} and evaluated at $z_* = 3$, for the wavenumber window $0.001~\seg/\km \leq k \leq 0.08~\seg/\km$ corresponding to the scales to which XQ-100/HIRES+MIKE \cite{Irsic:2017sop,Viel:2013fqw} and eBOSS \cite{eBOSS:2018qyj,eBOSS:2020jck,eBOSS:2020tmo} are sensitive.
From this fit we then numerically extract $\Delta_L^2$ and $n_L$ for each distinct point of our scanned \nuadam parameter space.
This allows us to make a qualitative comparison to the Lyman-$\alpha$ forest results from Refs.~\cite{eBOSS:2018qyj, Pedersen:2022anu,Goldstein:2023gnw,Rogers:2023upm} without relying on dubious assumptions about the non-linear evolution of the DAO and the applicability of any one likelihood derived from them to the \nuadam model.
For more details about our procedure, please see \App{appendix_lyman}.

\subsubsection{$S_8$}

It is well known that ``direct'' measurements of $\sigma_8$ (the variance of matter fluctuations smoothed out on scales of $8~\Mpc/h$), most of which rely on late universe weak lensing and galaxy clustering measurements, are in conflict with the value inferred from \lcdm~fits to Planck data.
In terms of the more common clustering amplitude parameter, $S_8 \equiv \sigma_8 \sqrt{\Omega_m / 0.3}$, the tension between some of these measurements and \lcdm with Planck (which gives $S_8 = 0.825 \pm 0.011$ when including lensing and BAO data \cite{Planck:2018vyg}) reaches the level of $2$--$3\sigma$ (see Refs.~\cite{Abdalla:2022yfr} for a recent review).
For example, the Kilo-Degree Survey (KiDS-1000) finds $S_8 = 0.766^{+0.020}_{-0.014}$ \cite{Heymans:2020gsg}.
For its part, the Year 3 results of the Dark Energy Survey (DES) found $S_8 = 0.775^{+0.026}_{-0.024}$ \cite{DES:2021wwk}; a recent tomographic analysis using harmonic space finds $0.704 \pm 0.029$ and $0.753 \pm 0.024$ for two different galaxy catalogues.
A DES Y3+KiDS-1000 joint analysis gives a slightly larger value, $S_8 = 0.790^{+0.018}_{-0.014}$ \cite{Kilo-DegreeSurvey:2023gfr}.
Recent work based on full-shape analysis with simulation-based priors on EFTofLSS even finds $S_8 \approx 0.68 \pm 0.11$ \cite{Ivanov:2024xgb}, although the full-shape analysis with more conservative priors on the EFTofLSS parameters shows almost no tension.

The prospect of new physics beyond \lcdm~being behind these discrepancies is an exciting one, and it has garnered much attention in recent years \cite{Battye:2014qga,Buen-Abad:2015ova,Lesgourgues:2015wza,Enqvist:2015ara,Murgia:2016ccp,Kumar:2016zpg,Chacko:2016kgg,Poulin:2016nat,Buen-Abad:2017gxg,Buen-Abad:2018mas,Dessert:2018khu,Archidiacono:2019wdp,Heimersheim:2020aoc,Clark:2020miy,Bansal:2021dfh,FrancoAbellan:2021sxk,Ye:2021iwa,Schoneberg:2022grr,Joseph:2022jsf,Buen-Abad:2022kgf,Wang:2022bmk,Bansal:2022qbi,Zu:2023rmc,Cruz:2023lmn,Buen-Abad:2023uva,Hughes:2023tcn}.
However, one cannot directly compare predictions from models that have non-trivial (\ie, non \lcdm like) transfer functions to the $S_8$ reported by these analyses.
While the calculation of $\sigma_8$ (and hence $S_8$) for any given model is well defined in terms of a simple integral of the power spectrum (see, \eg, \cite{Weinberg:2008zzc}), the weight of each $k$-mode in the power spectrum to these experiments varies.
In practice, this means that the range of $k$ that gives the most relevant contributions to the experimental results, in general, is not the one that dominates the integral defining $\sigma_8$.
Thus, implicit in all of these results, an assumption must be made about the way in which the MPS and the lensing power (or galaxy clustering) are related: the values of $S_8$ quoted by these lensing experiments are model-dependent \cite{DES:2022qpf,Zhou:2024igb}.
Being the standard model of cosmology, \lcdm is used in these analyses.
The impact that this assumption has in mapping from observations to reported $S_8$ is far from evident.
In particular, it is possible that models involving DAO or a pronounced suppression of the MPS (such as SPartAcous \cite{Buen-Abad:2022kgf,Buen-Abad:2023uva}, NADMDR/SIDR+ and WZDR+ \cite{Buen-Abad:2015ova,Lesgourgues:2015wza,Buen-Abad:2017gxg,Joseph:2022jsf}, or \nuadam) may result in very different values of $S_8$ being inferred.

Because of this important subtlety outlined above, we limit ourselves to merely reporting $S_8$ as extracted directly from the \nuadam MPS, and comparing it to the more moderate DES Y3+KiDS-1000 joint result, {\it without} making this measurement part of our numerical analysis.
Finally, we want to point out that if there really is a discrepancy between the MPS data and the \lcdm prediction, this would be encoded in the full-shape likelihood $\mF$.
However, due to the large number of extra parameters required to account for non-linear structure formation, small-scale astrophysics, and galaxy bias, the significance of this discrepancy, even if present, would be reduced.\footnote{In this regard, it is worth noticing that  Ref.~\cite{Ivanov:2024xgb} has performed a new full-shape analysis of the BOSS and eBOSS data using a simulation informed prior on the EFT parameters, and found a significant change in the extracted value for $\sigma_8$, with significant tension with the expected value from the value inferred from the best fit to Planck data.}

\subsection{Results}
\label{subsec:results}

We now turn our attention to the results of our numerical analysis.
Table~\ref{tab:lcdm_bf95} shows the total $\chi^2$ of the \lcdm model's best fit to the $\mD$, $\mDH$, $\mDF$, and $\mDHF$ datasets.
Also shown are the best fit values of the $H_0$ and $S_8$ parameters (in parenthesis), as well as their mean and $95\%$ C. R. intervals (\ie, the upper and lower $95\%$ C. R. limits, minus the mean).
It is evident that \lcdm is unable to address the Hubble tension, since at most it can give an upper 95\% C. R. value of $H_0 \approx 69.3~\km/\seg/\Mpc$ when fitting directly to $\mDH$.

Compare this to the analogous results for \nuadam, shown in Table~\ref{tab:nuadam_bf95}.
Here we include the best fit, mean, and $95\%$ C. R. intervals of the additional parameters $\DNeff$, $f_\adm$, and $\logmm$, as well as the $\chi^2$ difference between \nuadam's best fit point and \lcdm's, which we denote by $\Delta \chi^2$.
To obtain the Akaike Information Criterion estimator (AIC) \cite{Akaike:1974vps}, one only needs to add $+6$ to the $\Delta \chi^2$ values listed here, to account for the three additional model parameters being varied.
While \nuadam's fit to $\mD$ is only marginally better than that of \lcdm (with a {\it positive} AIC), it leads to a $ 95\% $ C.R. that allows much larger values of $H_0$ than \lcdm, corresponding to a significant reduction in the tension.
The fit to the SH0ES dataset $\mH$ is significantly better, reaching $H_0 \approx 74.1~\km/\seg/\Mpc$ as the upper end of its 95\% C. R., with $\Delta \chi^2 \approx -34$ ($\mathrm{AIC} \approx -28$).
The \nuadam model is therefore capable of greatly alleviating the Hubble tension, and in order to do so it requires nonvanishing values of the DR and interacting DM ($\DNeff$ and $f_\adm$, respectively) at more than $2\sigma$.
The inclusion of the MPS full-shape $\mF$ dataset leads to some reduction in the maximum allowed value for $H_0$, seen in both the $\mDF$ and $\mDHF$ fits.
For $\mDHF$, it degrades the fit somewhat (making $f_\adm$ compatible with 0\% to $2\sigma$), although still improving upon \lcdm by $\Delta\chi^2 \approx -24$ ($\mathrm{AIC} \approx -18$), largely due to the increase in $H_0$.
On the other hand, it has a more significant impact on $\mDF$, significantly reducing the maximum allowed value for $H_0$. The two parameters that are most affected by the inclusion of the full-shape data are $\omega_{cdm}$ and $\DNeff$, which are significantly correlated due to the sensitivity of data to the redshift at equality.
Given that the matter power spectrum is particularly sensitive to $\omega_{cdm}$, it is likely that the largest impact of including full-shape data is a tightening of the allowed range for $\omega_{cdm}$.
Although the inclusion of the MPS full-shape dataset appears to disfavor larger values of $H_0$, given that the results of these analyses are potentially sensitive to the choice of priors, further study is warranted before a firm conclusion can be reached (c.f. \cite{Simon:2022lde,Holm:2023laa,Ivanov:2024xgb,Chudaykin:2024wlw}).

\begin{table}[h!]
\centering
\begin{adjustbox}{max width=\columnwidth}
\begin{tabular}{|c|c|c|c|}
\toprule
\lcdm & $\chi^2$ & $H_0~[\km/\seg/\Mpc]$ & $S_8$ \\
\midrule
$\mD$ & 4198.13 & $(67.7) \, 67.9^{+1.0}_{-1.0}$ & $(0.824) \, 0.821^{+0.024}_{-0.025}$ \\
$\mDH$ & 4236.49 & $(68.5) \, 68.5^{+0.7}_{-0.7}$ & $(0.809) \, 0.810^{+0.019}_{-0.019}$ \\
$\mDF$ & 4468.58 & $(67.7) \, 67.7^{+0.7}_{-0.7}$ & $(0.825) \, 0.824^{+0.018}_{-0.018}$ \\
$\mDHF$ & 4503.91 & $(68.4) \, 68.4^{+0.6}_{-0.6}$ & $(0.811) \, 0.810^{+0.017}_{-0.017}$ \\
\bottomrule
\end{tabular}
\end{adjustbox}
\caption{Results of the \lcdm fits to $\mD$, $\mDH$, $\mDF$, and $\mDHF$ datasets.
Shown are the best fit $\chi^2$, as well as the (best fit value) mean $\pm 95\%$ confidence regions for the $H_0$ and $S_8$ parameters.}
\label{tab:lcdm_bf95}
\end{table}

\begin{table}[h!]
\centering
\begin{adjustbox}{max width=\columnwidth}
\begin{tabular}{|c|c|c|c|c|c|c|}
\toprule
\nuadam        & $\Delta\chi^2$ &  $\DNeff$ & $f_\adm [\%]$ & $\logmm$ & $H_0~[\km/\seg/\Mpc]$ & $S_8$ \\ \midrule
$\mD$   & -1.93 & $(0.21) \, 0.27^{+0.33}_{-0.27}$ & $(1.9) \, 1.8^{+1.9}_{-1.8}$  & $(-3.8) \, -3.8^{+0.3}_{-0.4}$ & $(68.6) \, 69.0^{+2.1}_{-1.9}$ & $(0.825) \, 0.825^{+0.020}_{-0.020}$ \\
$\mDH$   & -33.94 & $(0.90) \, 0.87^{+0.30}_{-0.29}$ & $(3.9) \, 3.7^{+2.1}_{-2.1}$  & $(-3.7) \, -3.7^{+0.1}_{-0.1}$ & $(72.6) \, 72.5^{+1.6}_{-1.5}$ & $(0.821) \, 0.821^{+0.019}_{-0.019}$ \\
$\mDF$   & -3.85 & $(0.02) \, 0.08^{+0.16}_{-0.08}$ & $(1.8) \, 1.4^{+2.1}_{-1.4}$  & $(-4.0) \, -3.9^{-0.4^*}_{+0.3^*}$ & $(67.8) \, 68.2^{+1.3}_{-1.1}$ & $(0.816) \, 0.819^{+0.021}_{-0.022}$ \\
$\mDHF$   & -25.02 & $(0.68) \, 0.66^{+0.26}_{-0.26}$ & $(0.9) \, 0.9^{+0.8}_{-0.9}$  & $(-4.0) \, -4.0^{+0.4^*}_{-0.4^*}$ & $(72.0) \, 71.9^{+1.4}_{-1.4}$ & $(0.817) \, 0.818^{+0.018}_{-0.017}$ \\
\bottomrule
\end{tabular}
\end{adjustbox}
\caption{Results of the \nuadam fits to $\mD$, $\mDH$, $\mDF$, and $\mDHF$ datasets.
Shown are the best fit $\Delta\chi^2$ values compared to \lcdm, as well as the (best fit value) mean $\pm 95\%$ confidence regions for the $\DNeff$, $f_\adm$, $\logmm$, $H_0$, and $S_8$ parameters.
The posteriors for the fits including the $\mF$ dataset are highly non-Gaussian.
In particular, the $\logmm$,  marked with an asterisk, is very broad and multimodal, which leads to a poor determination of the $95\%$ C. R.; thus we derive directly from the posterior distributions rather than using Gaussian.
To obtain the best fit AIC number, simply add $+6$ to the $\Delta\chi^2$ values shown here.
}
\label{tab:nuadam_bf95}
\end{table}

\begin{figure}[h!]
    \centering
    \includegraphics[width=.6\linewidth]{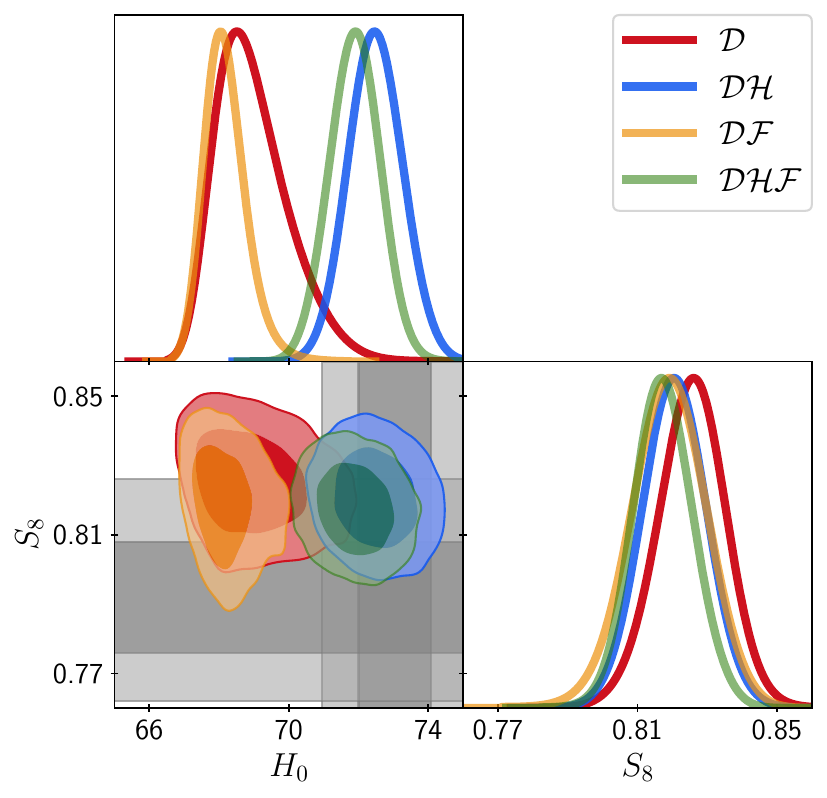}
	\caption{1D and 2D posteriors ($1\sigma$ and $2\sigma$ contours) for $H_0$ and $S_8$, from the fits of the \nuadam model to the $\mD$ (red), $\mDH$ (blue), $\mDF$ (yellow), and $\mDHF$ (green) datasets.
    The gray bands show the $1\sigma$ and $2\sigma$ contours of the $H_0$ measurement by the SH0ES collaboration used in our $\mH$ dataset \cite{Riess:2021jrx}, as well as those of the $S_8$ parameter, as determined by the recent DES Y3+KiDS-1000 joint analysis \cite{Kilo-DegreeSurvey:2023gfr}.}
	\label{fig:triangle_h0s8_1}
\end{figure}

\begin{figure}[h!]
    \centering
    \includegraphics[width=.49\linewidth]{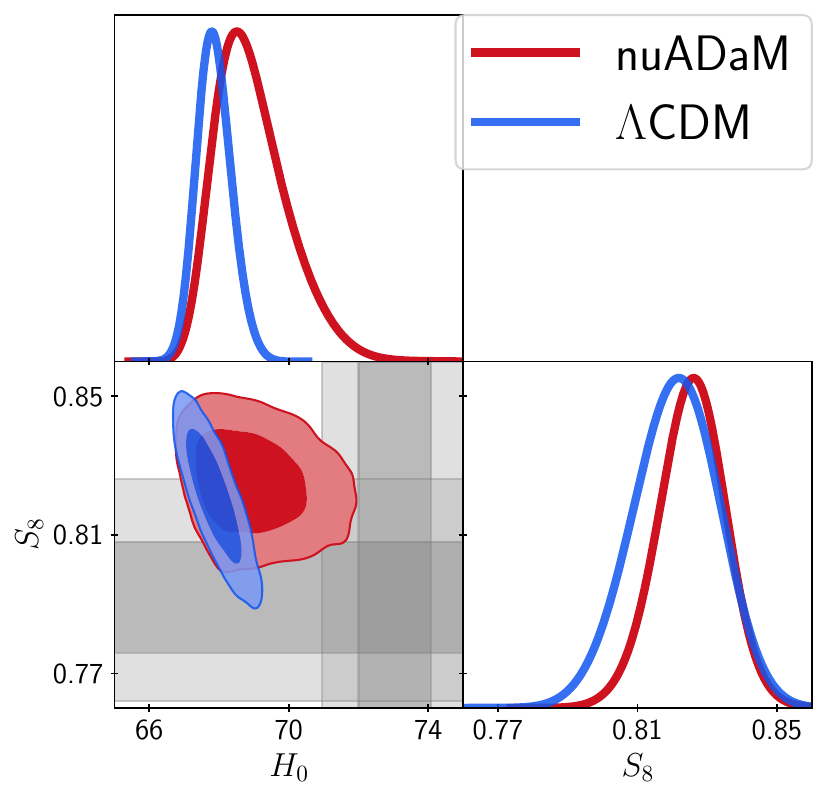}
    \includegraphics[width=.49\linewidth]{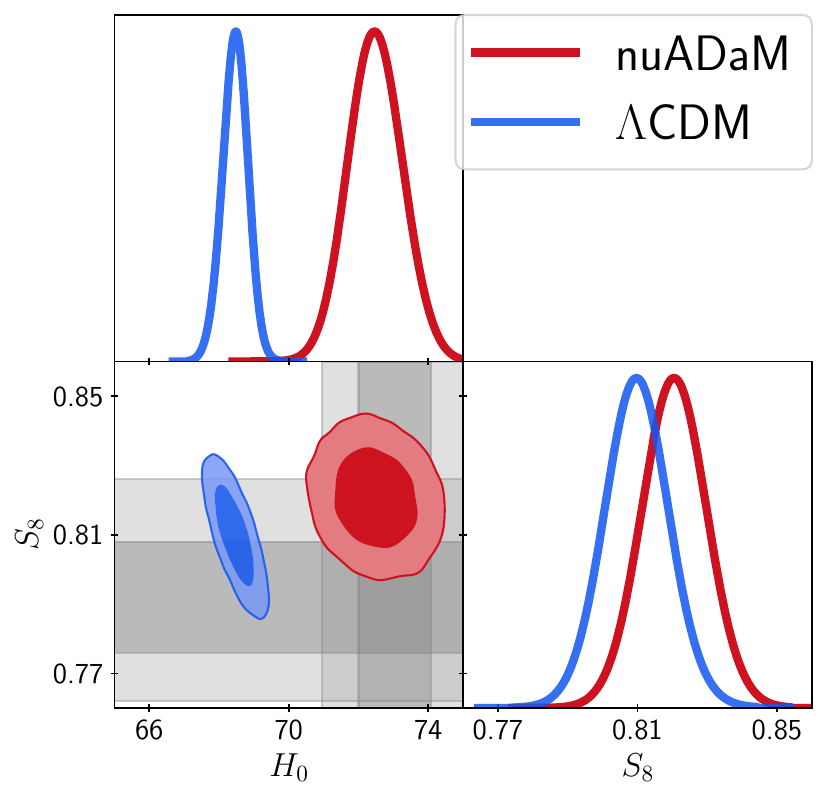}
    \includegraphics[width=.49\linewidth]{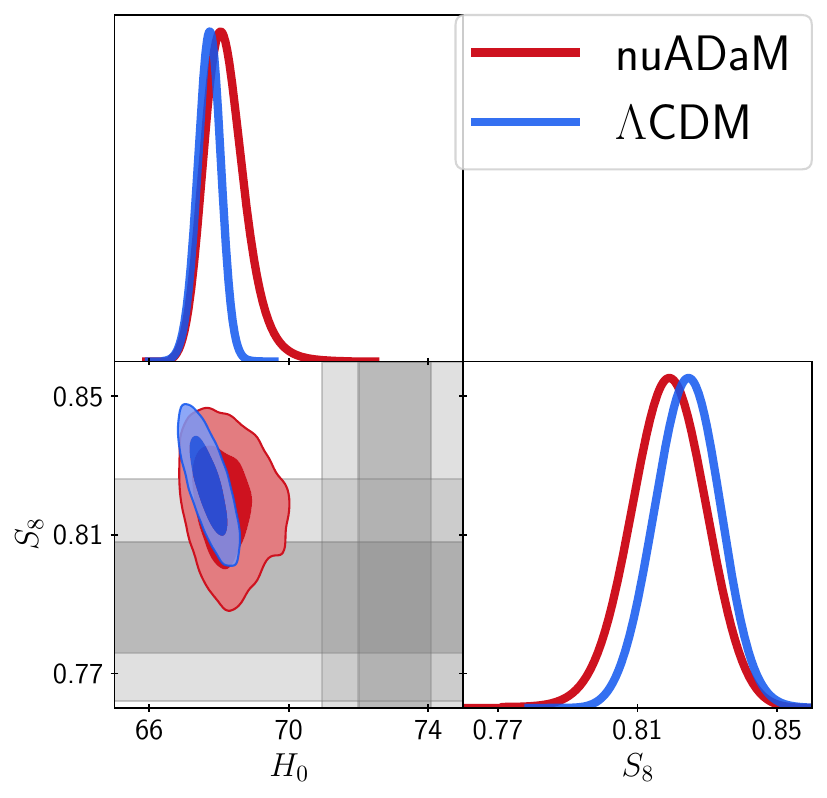}
    \includegraphics[width=.49\linewidth]{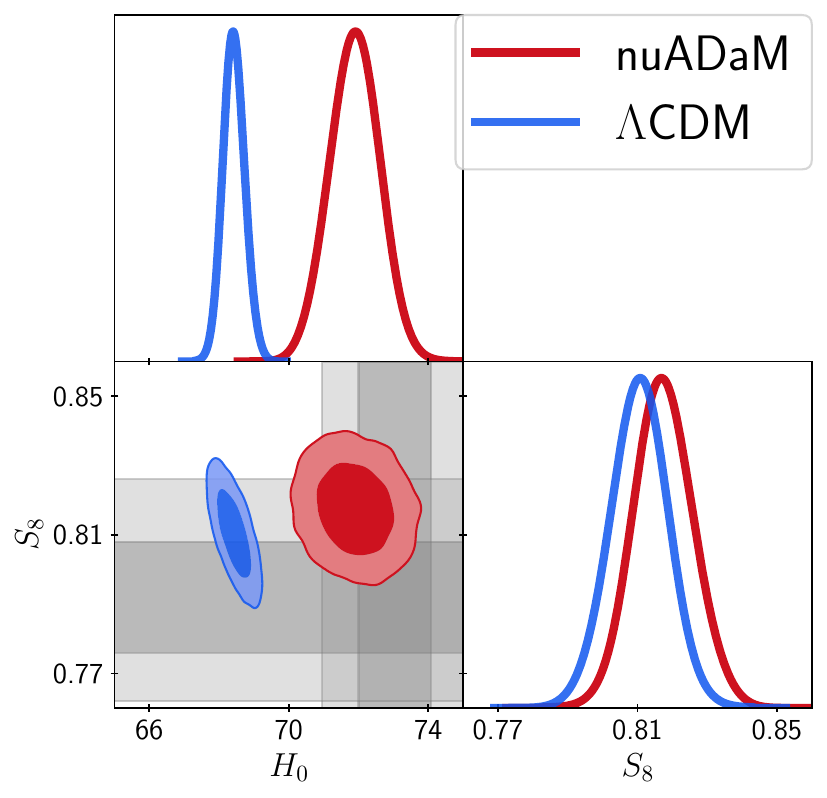}
 \caption{1D and 2D posteriors ($1\sigma$ and $2\sigma$ contours) for $H_0$ and $S_8$, from the fits of the \nuadam (red), and \lcdm (blue) models to the $\mD$ (top left), $\mDH$ (top right), $\mDF$ (bottom left), and $\mDHF$ (bottom right) datasets.
    The gray bands show the $1\sigma$ and $2\sigma$ contours of the $H_0$ measurement by the SH0ES collaboration used in our $\mH$ dataset \cite{Riess:2021jrx}, as well as those of the $S_8$ parameter, as determined by the recent DES Y3+KiDS-1000 joint analysis \cite{Kilo-DegreeSurvey:2023gfr}.}
	\label{fig:triangle_h0s8_2}
\end{figure}

In \Fig{fig:triangle_h0s8_1} we show the 2D contours of the $H_0$ and $S_8$ parameters in the \nuadam's fit to the $\mD$, $\mDH$, $\mDF$, and $\mDHF$ datasets, while in \Fig{fig:triangle_h0s8_2} we display the contours for both \nuadam and \lcdm, for each one of these datasets separately.
We include the $1\sigma$ and $2\sigma$ bands of the $H_0$ measurement by the SH0ES collaboration used in our $\mH$ dataset \cite{Riess:2021jrx}, as well as those of the $S_8$ measurement by the recent DES Y3+KiDS-1000 joint analysis \cite{Kilo-DegreeSurvey:2023gfr}.
These plots highlight that even when the SH0ES dataset is included, \lcdm is sufficiently constrained by the CMB data that it cannot accommodate significantly larger values of $H_0$. They also showcase
how \nuadam improves the fit without simultaneously increasing $S_8$ (and thus avoids worsening a possible tension in this data as well).
This is true, even though the datasets used are not directly sensitive to $S_8$, highlighting the fact that the DR-iDM interactions can improve both tensions simultaneously.
Indeed, the $2\sigma$ contour of \nuadam's fit to $\mD$ overlaps with the $1\sigma$ intervals of the SH0ES and DES Y3+KiDS-1000 measurements, and the fit to $\mDH$ is in excellent agreement with these experiments.
The inclusion of $\mF$ decreases both the best-fit and maximum allowed value for $H_0$, and also somewhat degrades the fit.
Nonetheless, the model still provides an excellent fit to the combined data, which is a substantial improvement over other well-known solutions to the $H_0$ tension (\eg, Ref.~\cite{McDonough:2023qcu}).

\begin{figure}[h!t]
    \centering
    \includegraphics[width=.61\linewidth]{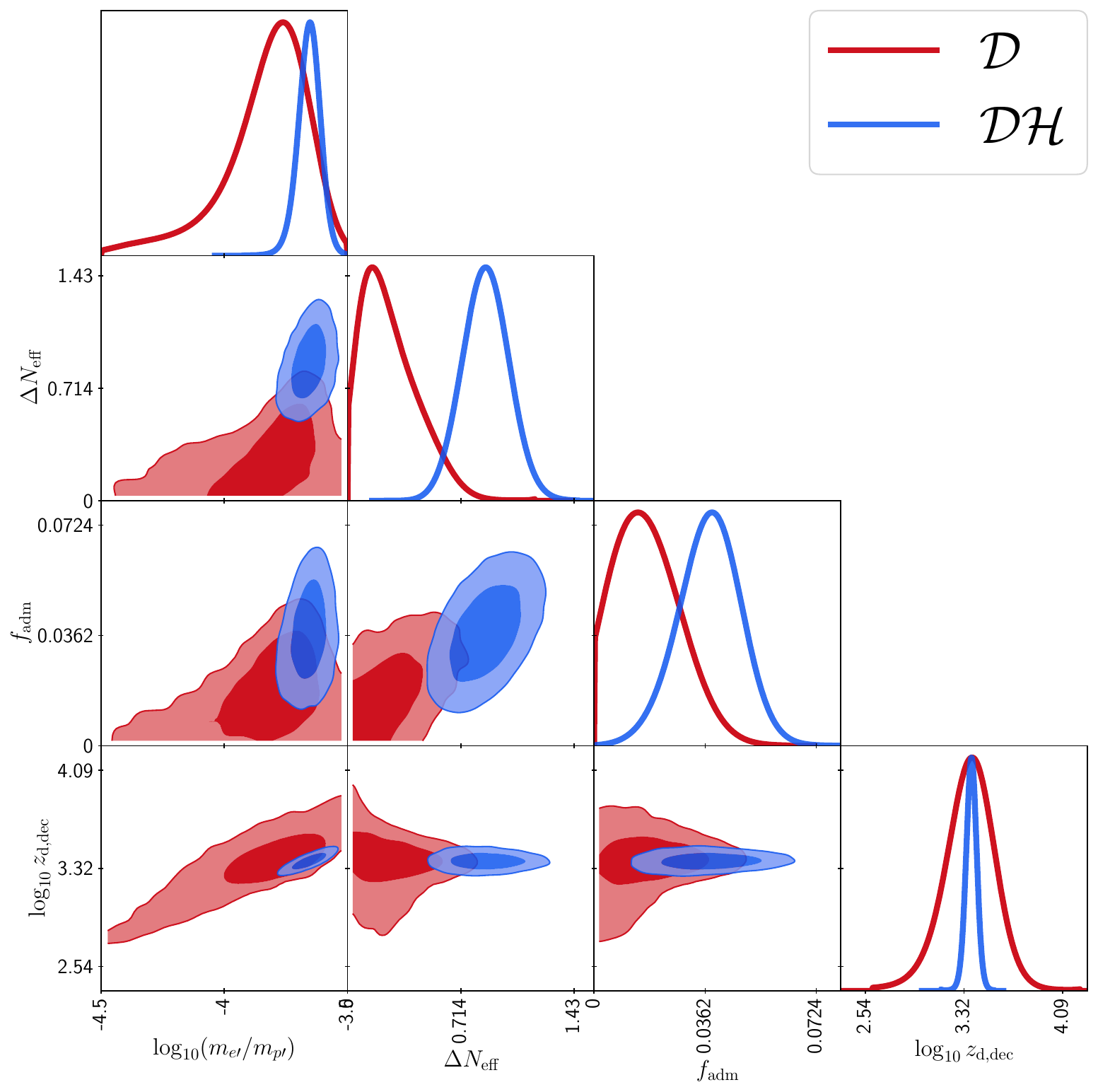}
    \includegraphics[width=.61\linewidth]{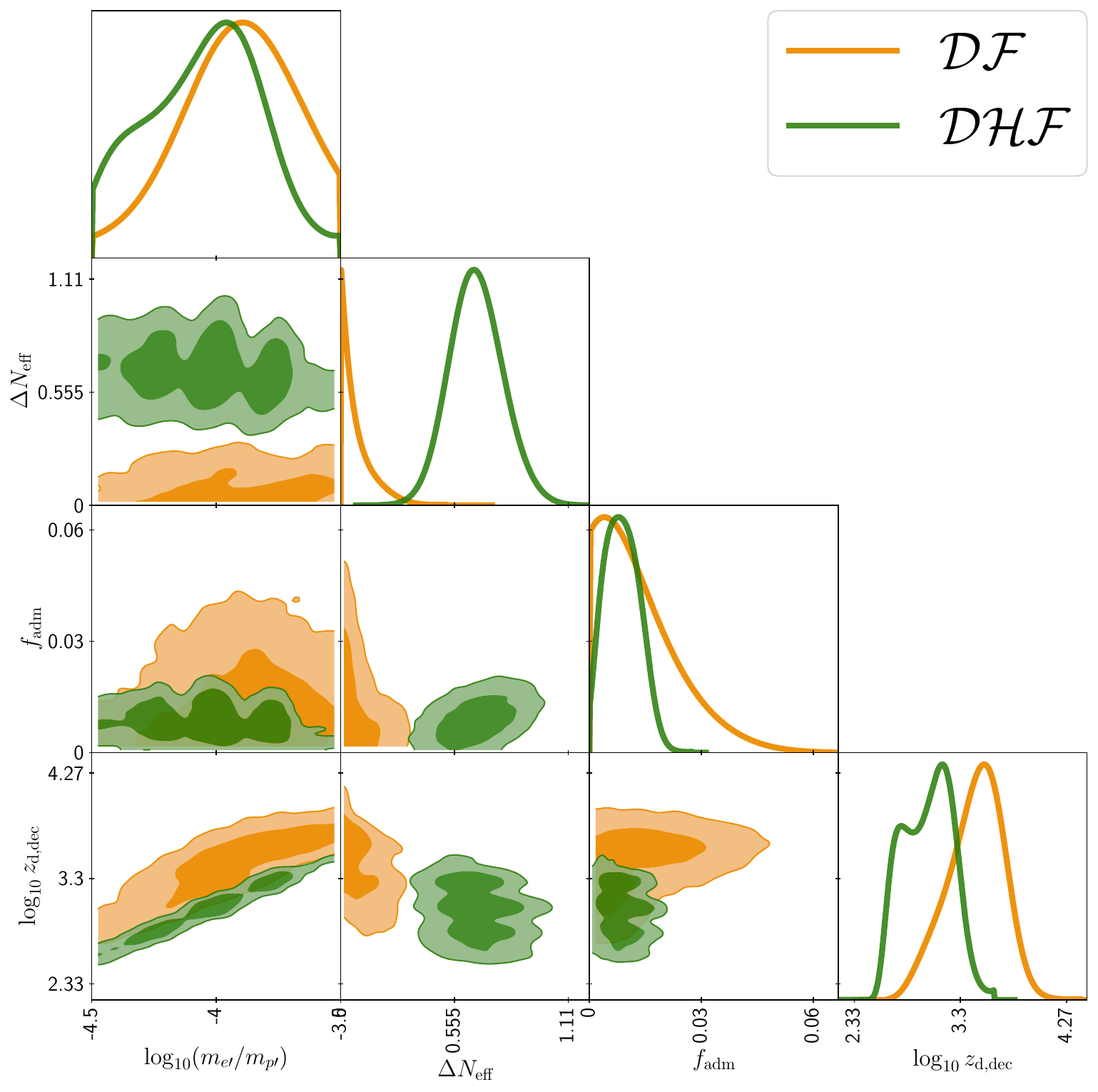}
	\caption{1D and 2D posteriors ($1\sigma$ and $2\sigma$ contours) of the \nuadam model parameters $\logmm$, $\DNeff$, $f_\adm$, and $\log_{10} z_{\rm d, \, dec}$.
    {\it Top:} results from fitting \nuadam the $\mD$ and $\mDH$ datasets.
    {\it Bottom:}: results from fitting \nuadam to the $\mDF$ and $\mDHF$ datasets.}
	\label{fig:triangle_nuadam2}
\end{figure}

The posteriors of the $\DNeff$, $f_\adm$, and $\logmm$ \nuadam parameters, whose values are listed in Table~\ref{tab:nuadam_bf95}, are shown in \Fig{fig:triangle_nuadam2}.
Clearly, \nuadam fits to $\mD$ yield results compatible at $1\sigma$ or $2\sigma$ with \lcdm (\ie, vanishing $\DNeff$ or $f_\adm$, and very large $\logmm$).
However, for all three parameters the maxima of their posteriors are displaced from their \lcdm limits.
This preference for non-\lcdm values is only enhanced when including the $\mH$ dataset; in fact, values $\DNeff \sim \text{few} \times 0.1$ and $f_\adm \sim \text{few } \%$ are preferred.
The posteriors of the $\logmm$ parameter, which is directly correlated with the redshifts $z_{\rm d,\, rec}$ and $z_{\rm d,\,\dec}$ discussed in \Subsec{subsec:perts}, also peak at values corresponding to a dark baryon drag epoch taking place shortly after matter-radiation equality, a time to which the CMB data is sensitive.
Note that the posteriors for the the fits including $\mF$ are highly non-Gaussian, specially for $\log_{10}(m_{e'}/m_{p'})$ (see algo Fig.~\ref{fig:triangle_nuadam2}), and thus the $95\%$ C. R. was not well determined by MontePython.
This might be due to the large number of extra nuisance parameters required by the full-shape analysis, or may be arising from the model being unable to simultaneously fit both data sets well, with $\mH$ and $\mF$ pulling the fit towards different regions of parameter space.
It will be interesting to study whether upcoming DESI data and improvements in the treatment of the priors of these extra parameters will lead to significant changes in the results.

\begin{figure}[h!]
    \centering
    \includegraphics[width=.75\linewidth]{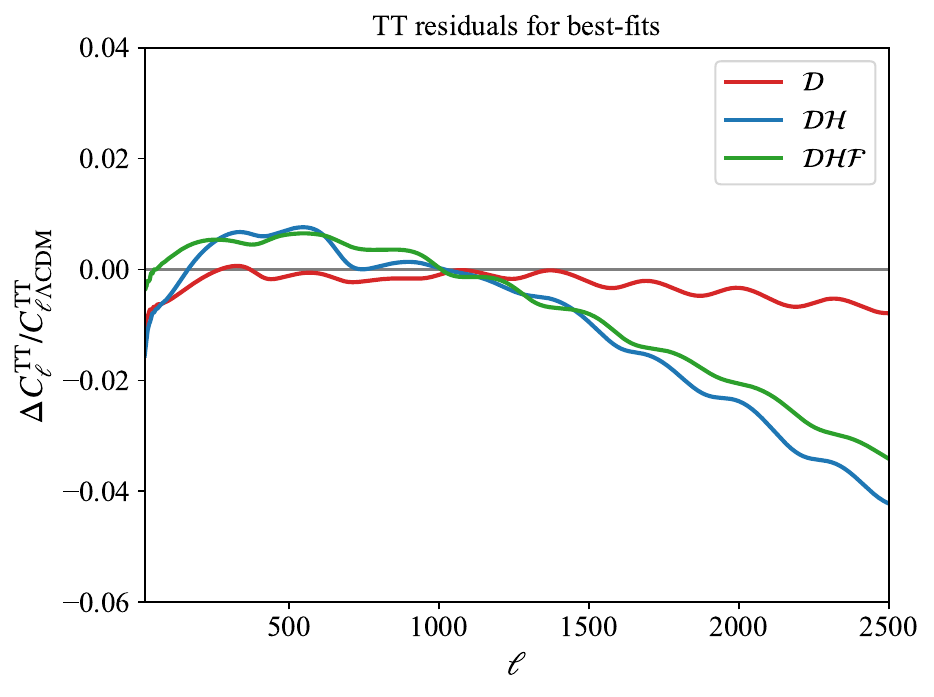}
    \includegraphics[width=.75\linewidth]{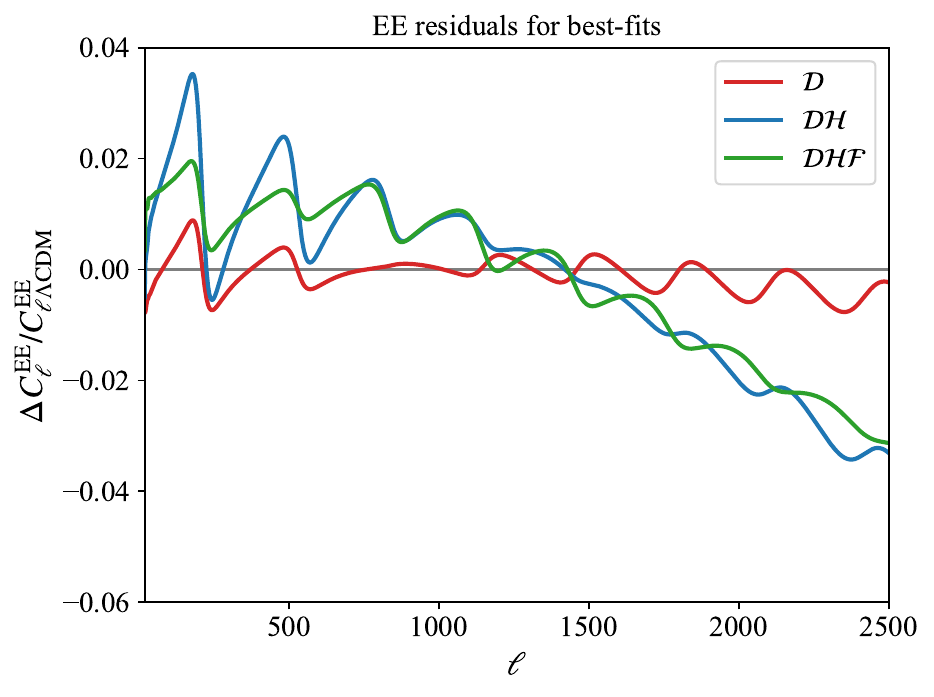}
	\caption{$C_\ell^{TT}$ (top) and $C_\ell^{EE}$ (bottom) residuals of \nuadam compared to \lcdm, for the best fit point to the $\mD$ (red), $\mDH$ (blue), and $\mDHF$ (green) datasets.}
	\label{fig:bf_cmb}
\end{figure}

\begin{figure}[h!]
    \centering
    \includegraphics[width=.75\linewidth]{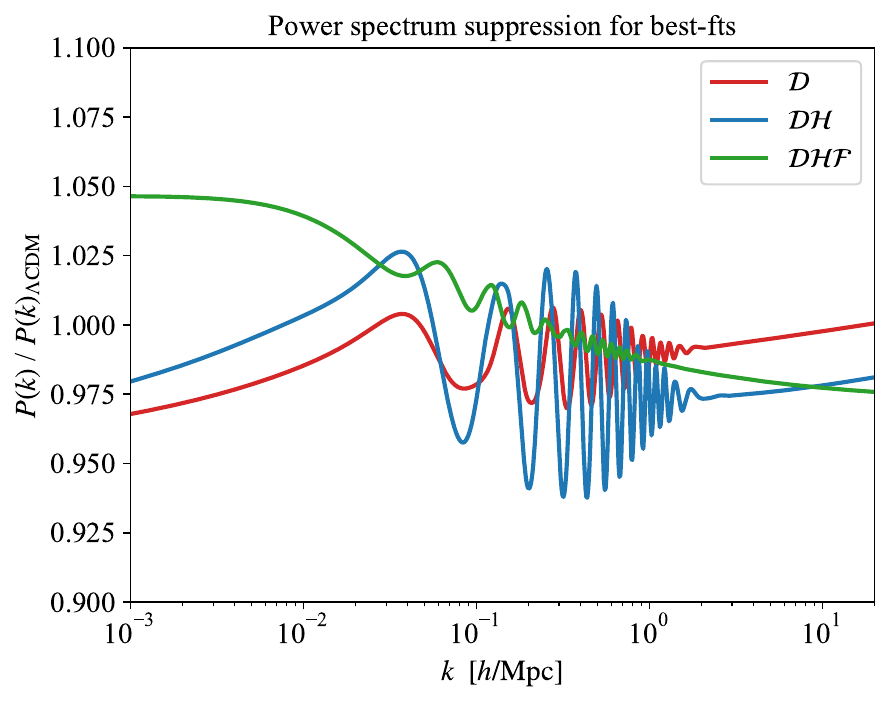}
	\caption{Ratios of the linear MPS in the \nuadam model to that in the \lcdm model, for the best fit point to the $\mD$ (red), $\mDH$ (blue), and $\mDHF$ (green) datasets.}
	\label{fig:bf_pk}
\end{figure}

The impact of the \nuadam best fit points on the CMB and matter power spectra can be seen in Figs.~\ref{fig:bf_cmb} (as residuals) and \ref{fig:bf_pk} (as ratios).
The curves shown correspond to fits to the $\mD$, $\mDH$, and $\mDHF$ datasets.
It is clear that the \nuadam best fit to $\mD$ is such that both its CMB and matter power spectra approximate that of \lcdm.
When including $\mH$ in the fit, the spectra can be very different from each other; \eg, at large $\ell$/$k$ (small scales) the \nuadam spectra are suppressed at the level of $\text{few }\%$ when compared to those of \lcdm.
However, this does {\it not} result in a degraded $\chi^2$ fit to the $\mD$ datasets.
Indeed, as can be seen in Table~\ref{tab:val_chi2} in the next subsection, the best fits of the \nuadam model to the CMB, lensing, BAO, and SNIa datasets always yield a similar $\chi^2$ to that of the \lcdm best fits; the difference is typically $\vert\Delta\chi^2\vert \sim \mO(1)$.
This is a major advantage of \nuadam compared to other models: it is able to ease the tension in the Hubble measurements without degrading the model's fit to other datasets, such as Planck's.
Given the increased difference between the residuals at high $\ell$, and the percent level changes in the MPS in the $10^{-2} \, - \, 10^{-1}$h/Mpc range, upcoming CMB and LSS data will play an important role in determining whether \nuadam is preferred over \lcdm.

\begin{figure}[h!]
    \centering
    \includegraphics[width=.49\linewidth]{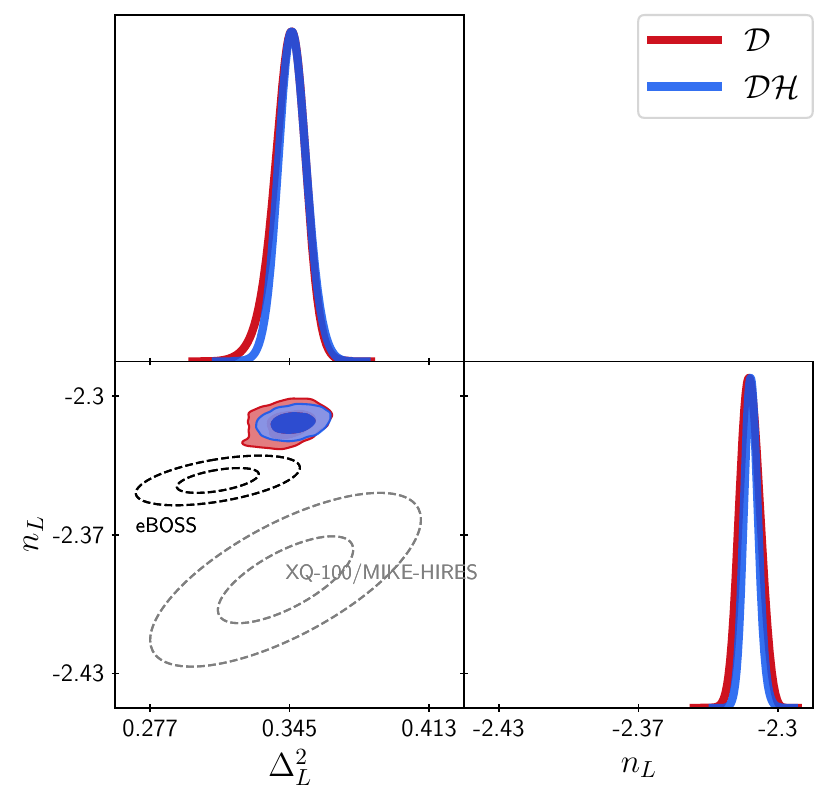}
    \includegraphics[width=.49\linewidth]{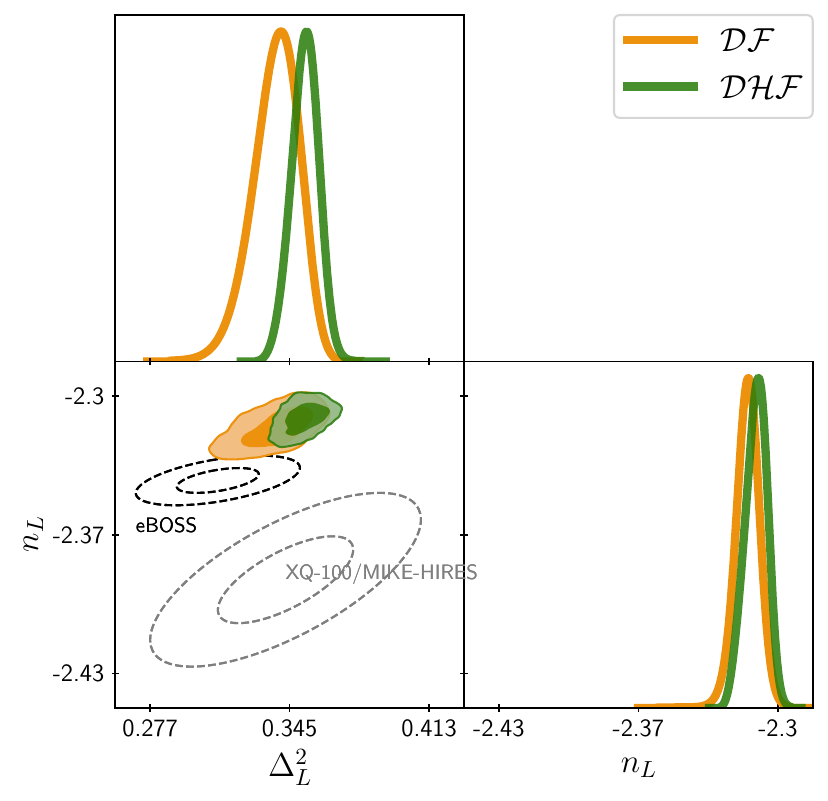}
	\caption{1D and 2D posteriors ($1\sigma$ and $2\sigma$ contours) for the $\Delta^2_L$ and $n_L$ parameters (estimated in accordance to \Subsec{subsubsec:lya}), from the fits of the \nuadam model to the $\mD$ (red), $\mDH$ (blue), $\mDF$ (yellow), and $\mDHF$ (green) datasets.
    In black are the 2D contours from the eBOSS experiment \cite{eBOSS:2018qyj,eBOSS:2020jck,eBOSS:2020tmo}; those of XQ-100/MIKE-HIRES \cite{Irsic:2017sop,Viel:2013fqw} are shown in gray.}
	\label{fig:triangle_lya_1}
\end{figure}

\begin{figure}[h!]
    \centering
    \includegraphics[width=.49\linewidth]{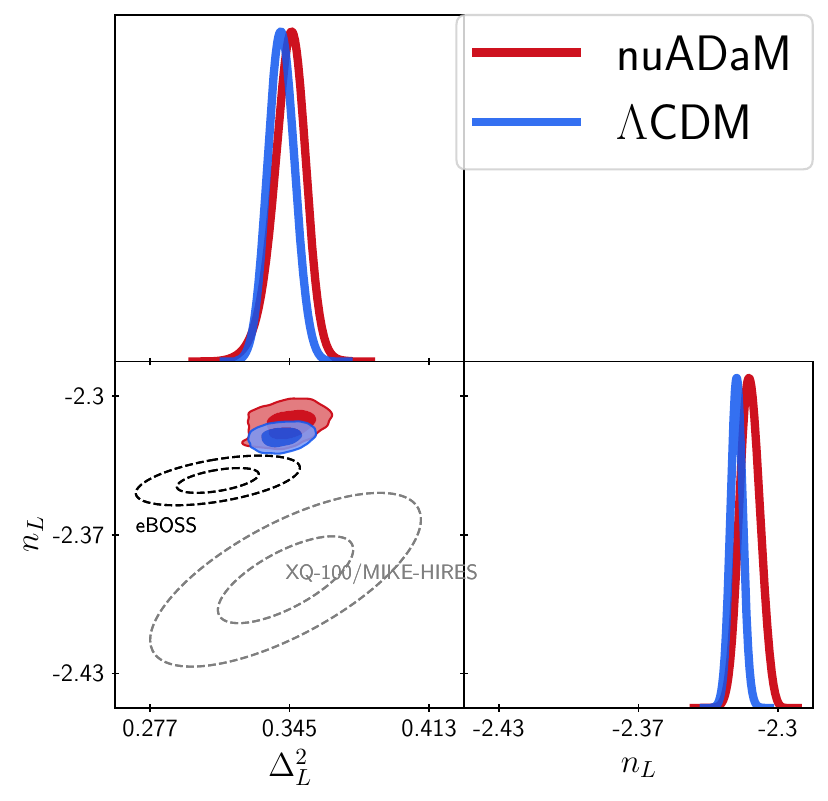}
    \includegraphics[width=.49\linewidth]{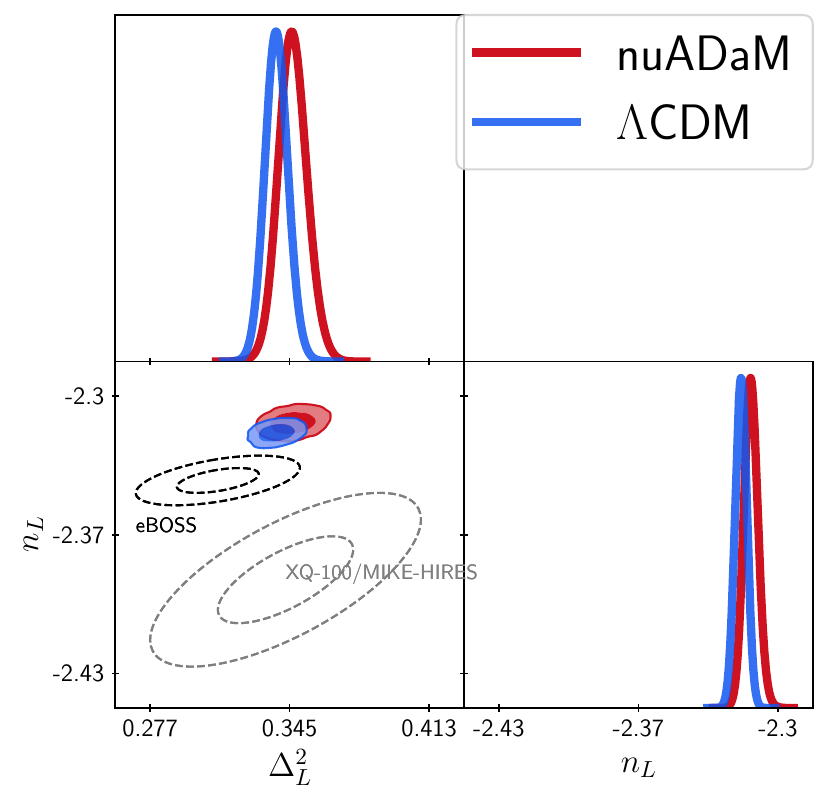}
        \includegraphics[width=.49\linewidth]{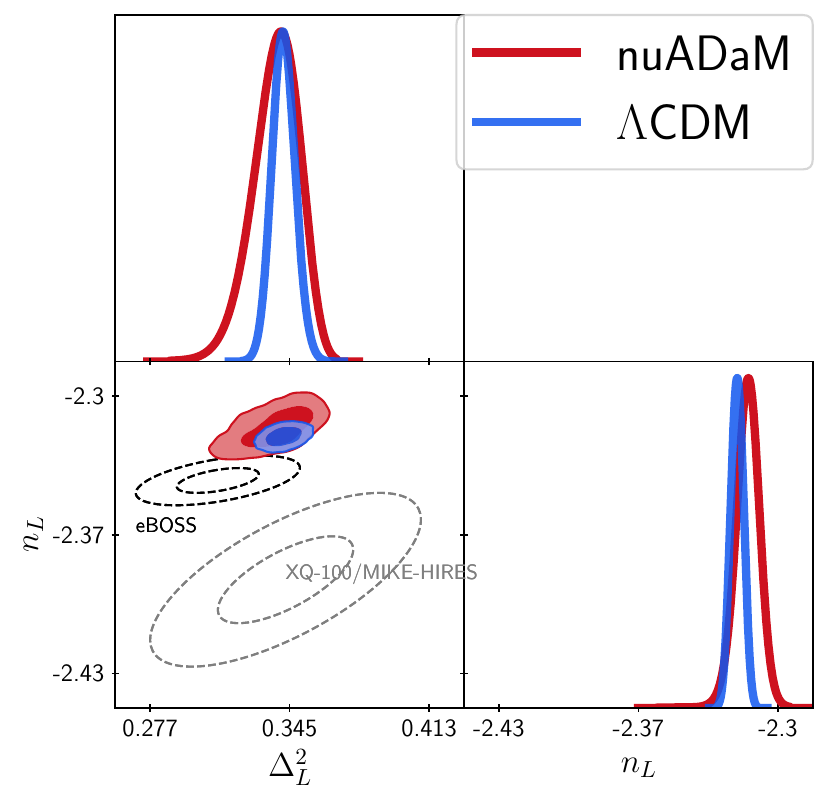}
    \includegraphics[width=.49\linewidth]{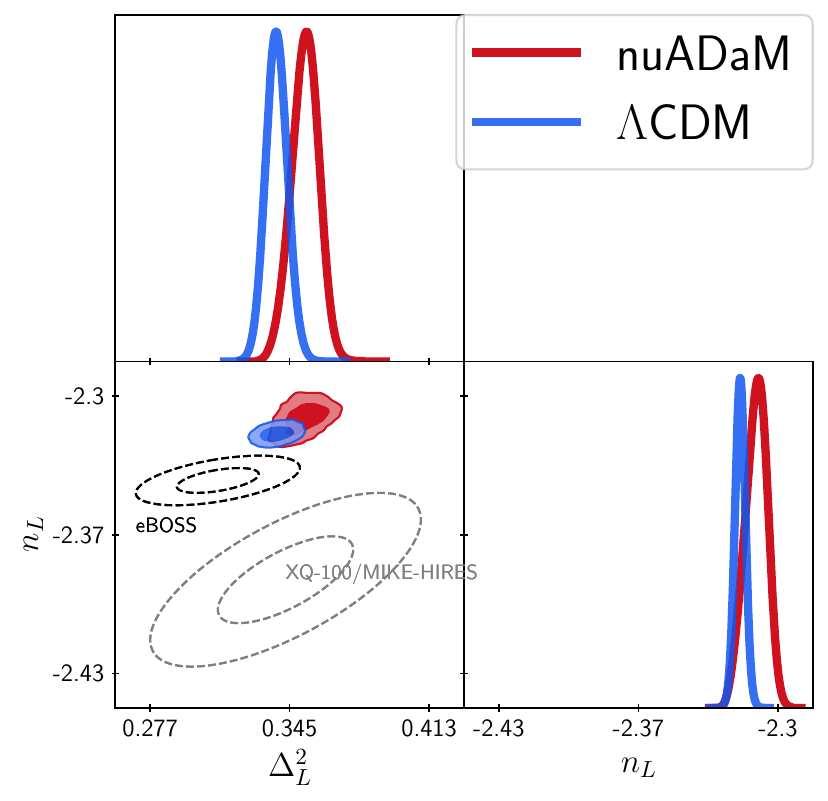}
	\caption{1D and 2D posteriors ($1\sigma$ and $2\sigma$ contours) for the $\Delta^2_L$ and $n_L$ parameters (estimated in accordance to \Subsec{subsubsec:lya}), from the fits of the \nuadam (red) and \lcdm (blue) models to the $\mD$ (top left), $\mDH$ (top right), $\mDF$ (bottom left), and $\mDHF$ (bottom right) datasets.
    In black are the 2D contours from the eBOSS experiment \cite{eBOSS:2018qyj,eBOSS:2020jck,eBOSS:2020tmo}; those of XQ-100/MIKE-HIRES \cite{Irsic:2017sop,Viel:2013fqw} are shown in gray.}
	\label{fig:triangle_lya_2}
\end{figure}

Following the discussion of \Subsec{subsubsec:lya}, we have estimated the $\Delta^2_L$ and $n_L$ parameters, which describe the amplitude and slope of the linear MPS at the wavenumbers and redshifts associated with Lyman-$\alpha$ observations.
In Figs.~\ref{fig:triangle_lya_1} and \ref{fig:triangle_lya_2}, we show the associated 1D and 3D posteriors of $\Delta^2_L$ and $n_L$, for the \lcdm and \nuadam fits to $\mD$, $\mDH$, $\mDF$, and $\mDHF$.
As stated in \Subsec{subsubsec:lya}, the values we provide for these parameters must not be taken as having been rigorously derived.
However, they provide useful qualitative information about our model's MPS at the scales relevant for Lyman-$\alpha$.
In fact, note how the \lcdm model's fit to $\mD$ yields a contour (blue, top left panel of \Fig{fig:triangle_lya_2}) close to the one in Ref.~\cite{Goldstein:2023gnw} (gray contour, Fig. 2).
Since this is the case despite both sets of data not being identical, our approach to the Lyman-$\alpha$ parameters seems to be validated at a qualitative level.
As can be deduced from these contours, the \nuadam model is unable to suppress the MPS at the requisite levels in order to address the Lyman-$\alpha$ tension.

\subsubsection{Comparison with the standard Atomic Dark Matter model}

We complemented our \lcdm and \nuadam analyses of the $\mD$, $\mH$, and $\mF$ datasets with a study of the standard ADM model.
As stated in the introduction and in \Sec{sec:model}, \nuadam differs from ADM in that its DR has additional degrees of freedom that make it self-interacting.
Due to the DR self-interactions, in \nuadam dark recombination is Case A, while in ADM it is Case B.
Given the small differences in the evolution of the free dark electron fraction shown in Fig.~\ref{fig:drecom}, we expect that this only results in minor differences in the CMB prediction for both scenarios.
The main difference should be due to the fact that DR becomes free-streaming after dark recombination in ADM, while it remains a perfect fluid in \nuadam.
Since this effect becomes less important if dark recombination occurs after matter-radiation equality, it is worth examining whether \nuadam in fact provides a better fit compared to ADM.
In order to have both models on equal footing, we assume that the DR in ADM (parametrized by $\DNeff$) is also populated only after BBN (see Footnote \ref{foot5}), which has been shown to allow larger $\Neff$ in DR models.
Aside from how DR is treated, ADM has the same content as \nuadam, and therefore the same parameters.

\begin{figure}[h!]
    \centering
    \includegraphics[width=0.95\linewidth]{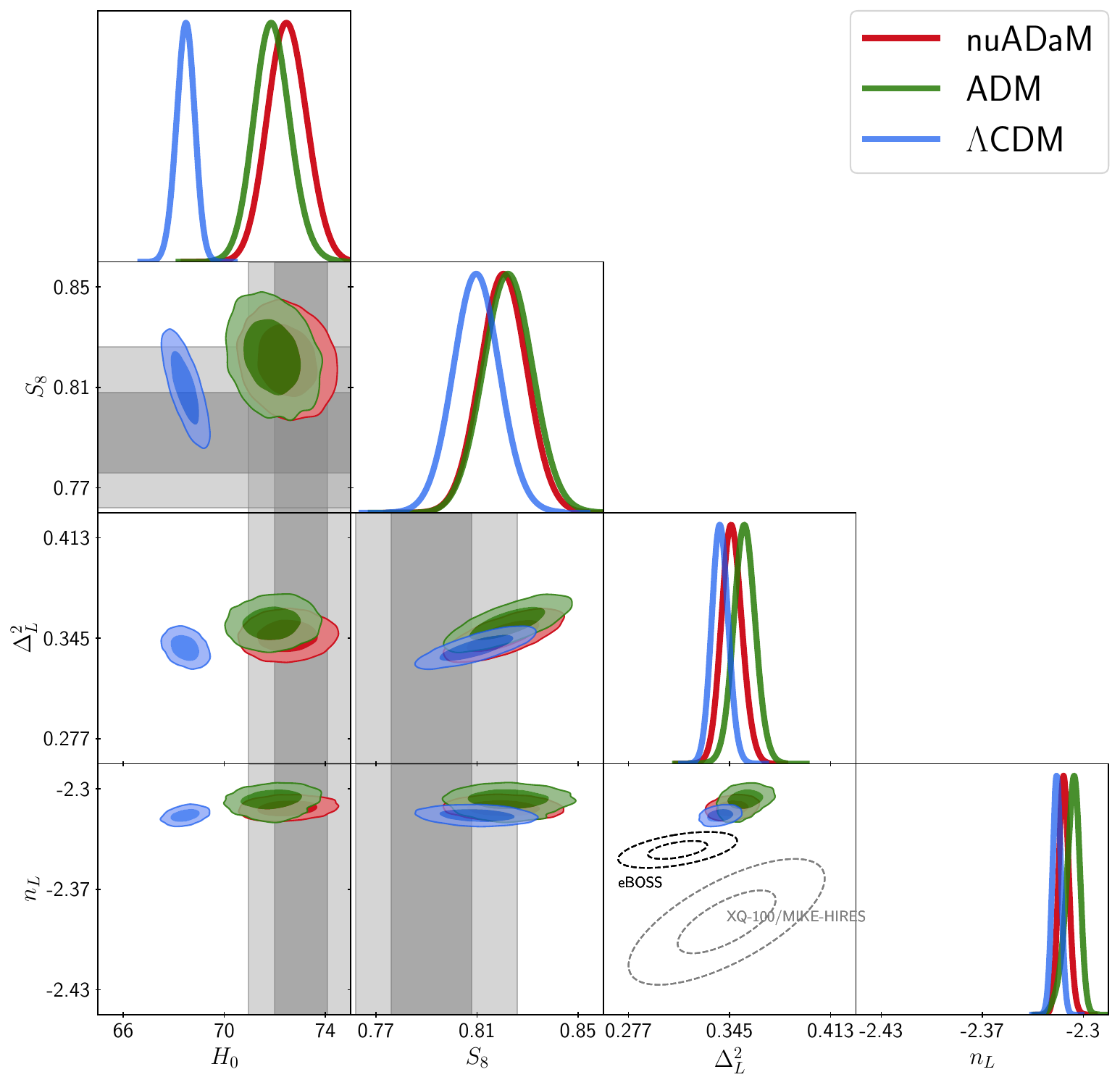}
    \caption{1D and 2D posteriors ($1\sigma$ and $2\sigma$ contours) for $H_0$, $S_8$, $\Delta^2_L$ and $n_L$, from the fits of the \nuadam (red), ADM (green), and \lcdm (blue) models to the $\mDH$ dataset.
    The gray bands show the $1\sigma$ and $2\sigma$ contours of the $H_0$ measurement by the SH0ES collaboration used in our $\mH$ dataset \cite{Riess:2021jrx}, as well as those of the $S_8$ parameter, as determined by the recent DES Y3+KiDS-1000 joint analysis \cite{Kilo-DegreeSurvey:2023gfr}.
    On the $n_L$ and $\Delta^2_L$ plane, the 2D contours from the eBOSS experiment \cite{eBOSS:2018qyj,eBOSS:2020jck,eBOSS:2020tmo} and XQ-100/MIKE-HIRES \cite{Irsic:2017sop,Viel:2013fqw} are shown in black and gray respectively.}
	\label{fig:triangle_h0s8_3}
\end{figure}

The results of our numerical analyses are summarized in Table~\ref{tab:val_chi2}.
In addition, \Fig{fig:triangle_h0s8_3} shows the 1D and 2D posteriors of the $H_0$ and $S_8$ parameters of the \lcdm, \nuadam, and ADM models, when fit to $\mDH$.
It is noteworthy that our results show ADM is a much better solution to the $H_0$ tension than had been previously reported \cite{Bansal:2022qbi,Hughes:2023tcn}.
These improvements stem from both allowing $\Neff$ at BBN to be uncorrelated to the one probed by CMB, and by decreasing the number of relevant parameters to only those that affect the cosmological implications of the model (thus significantly improving the MCMC scan).
The best fit $H_0$ values we obtained for the ADM model were $H_0 = 71.6~\km/\seg/\Mpc$ for the $\mDH$ datasets, and $H_0 = 71.9~\km/\seg/\Mpc$ for the $\mDHF$ datasets (see Table~\ref{tab:adm_val_bf}).
These can be compared with the \nuadam best fit values in Table~\ref{tab:nuadam_bf95} ($72.6~\km/\seg/\Mpc$ and $72.0~\km/\seg/\Mpc$, respectively).

As can be seen from our results, and as already discussed previously in this section, \nuadam marginally outperforms \lcdm when fit to $\mD$ and $\mDF$, but is much better when $\mH$ is included.
Although it provides better fits to these datasets than \lcdm, ADM's fits are nevertheless consistently worse than \nuadam's, particularly with regard to its fit to CMB and SH0ES data.
This indicates that the self-interacting nature of the DR plays an important role in improving the fit of ADM models to cosmological data; this is perhaps unsurprising in light of the work done in Ref.~\cite{Blinov:2020hmc}.
To corroborate this, we consider an ADM model where the DR is also self-interacting, but there are no additional fermions $\nu'$ that can realize such self-interactions (\ie, $N_f = 0$).
This ensures that for a fixed $\DNeff$, the DR temperature is the same compared to ADM.
We have dubbed this unphysical model nu$^0$ADaM; its fit to $\mDH$ can also be seen in Table~\ref{tab:val_chi2}.
It is clear that the fit of this model very closely resembles \nuadam, and is also better than the standard ADM scenario.

\begin{table}[]
\centering
\begin{adjustbox}{max width=\columnwidth}
\begin{tabular}{@{}|c|c|c|ccccccc|@{}}
\toprule
\toprule
\multicolumn{1}{|c|}{Dataset} & Model & $\chi^2_\text{tot}$ & $\chi^2_\text{CMB}$ & $\chi^2_\text{PL.lens}$ & $\chi^2_\text{BAO}$ & $\chi^2_\text{Pantheon}$ & $\chi^2_\text{SH0ES}$ & $\chi^2_{\text{EFT}_\text{BOSS}}$ & $\chi^2_{\text{EFT}_\text{eBOSS}}$ \\ \midrule
\multirow{2}{*}{$\mD$}            & $\lcdm$  & $4198.13$ & $2765.49$ & $8.92$ & $12.55$ & $1411.17$ & $-$ & $-$ &  $-$ \\
                              & ADM   & $4197.55$ & $2765.02$ & $8.87$ & $12.72$ & $1410.94$ & $-$ & $-$ & $-$ \\
                              & \nuadam   & $4196.20$ & $2763.57$ & $8.90$ & $12.85$ & $1410.87$ & $-$ & $-$ & $-$ \\ \midrule
\multirow{2}{*}{$\mDH$}           & $\lcdm$  & $4236.49$ & $2769.00$ & $8.99$ & $12.44$ & $1414.58$ & $31.48$ & $-$ & $-$ \\
                              & ADM   & $4210.39$ & $2772.85$ & $9.27$ & $12.25$ & $1412.65$ & $3.37$ & $-$ &  $-$ \\
                              & \nuadam   & $4202.55$ & $2768.09$ & $9.20$ & $12.03$ & $1412.09$ & $1.14$ & $-$ &  $-$ \\
                              & nu$^0$ADaM   & $4202.97$ & $2768.60$ & $9.19$ & $12.03$ & $1411.73$ & $1.42$ & $-$ &  $-$ \\
                              \midrule
\multirow{2}{*}{$\mDF$}           & $\lcdm$  & $ 4468.58$ & $2765.96$ & $8.85$ & $12.22$ & $1411.82$ & $-$ & $177.16$ & $92.56$ \\
                              & ADM   & $4465.13$ & $2766.53$ & $8.83$ & $12.28$ & $1412.30$ & $-$ & $173.75$ & $91.43$ \\
                              & \nuadam   & $4464.73$ & $2766.44$ & $8.84$ & $12.36$ & $1412.24$ & $-$ & $173.17$ & $91.66$ \\ \midrule
\multirow{2}{*}{$\mDHF$}         & $\lcdm$  & $4503.91$ & $2768.93$ & $8.92$ & $12.46$ & $1413.13$ & $33.19$ & $175.73$ & $91.54$ \\
                              & ADM   & $4480.35$ & $2773.07$ & $9.23$ & $12.63$ & $1413.95$ & $3.43$ & $176.98$ & $91.84$ \\
                              & \nuadam   & $4478.89$ & $2772.64$ & $9.35$ & $12.92$ & $1413.46$ & $3.02$ & $176.06$ & $91.44$ \\ \bottomrule \bottomrule
\end{tabular}
\end{adjustbox}
\caption{Table of the best-fit $\chi^2$ values of the \nuadam, ADM and $\lcdm$ models for datasets studied in this work.
For the purposes of validating \nuadam, we have included fits from an {\it unphysical} model nu$^0$ADaM, where $N_f = 0$ and yet the DR is still taken to be self-interacting.}
\label{tab:val_chi2}
\end{table}

\section{Conclusions}
\label{sec:conclusions}

In this paper, we have proposed a new framework for addressing the Hubble tension based on an interacting dark sector. We have considered a scenario in which a subcomponent of DM is tightly coupled to fluid-like DR at early times. The DR decouples from the DM during the CMB epoch but continues to be fluid-like. Thus, CMB modes that entered the horizon before this decoupling occurred evolve differently from those that entered subsequently, allowing a solution to the tension.
We have presented a model based on this framework that provides a significantly better fit to the SH0ES data than $\lcdm$, without spoiling the fit to other cosmological datasets such as the CMB, lensing, the BAO, and supernovae.

The model we present combines some of the characteristic features of the well-known SIDR and ADM scenarios. In our construction, the fluid-like DR is composed of massless fermions and gauge bosons interacting with each other, while the acoustic subcomponent of DM is composed of dark analogues of the SM electron and proton that interact via a dark photon. The dark photon is tightly coupled to the rest of the DR. As in the standard ADM framework, the dark electrons and dark protons constitute an ionized plasma that exhibits acoustic behavior.
As the universe expands and the dark sector cools, dark recombination eventually takes place. This proceeds differently than in standard ADM due to the large self-interactions of the DR, which causes the mean-free-path of the dark photon to be much smaller than in standard ADM.
This alters the dynamics of dark recombination: dark electrons and protons can recombine directly into the ground state of dark hydrogen, instead of having to go through the excited states (the so-called ``Peebles correction''). The DR decouples from dark hydrogen following dark recombination, and the dark acoustic oscillations cease. The CMB peaks corresponding to modes that entered the horizon prior to dark recombination evolve differently from those that entered subsequently, allowing a good fit to CMB data even in the presence of a sizable DR component. Indeed, this generalization of the SIDR and ADM scenarios, which we dub the ``New Atomic Dark Matter'' (\nuadam) model, provides a global fit to the combined cosmological data from the CMB, BAO, large scale structure and supernovae that 
represents a significant improvement over \lcdm. 
It also provides a better fit to the data than conventional ADM.

Our analysis assumes that the DR is populated after BBN through a dark sector process, the nature of which we leave unspecified.
Given that the cosmological length scales currently observable are much larger than those that could be sensitive to a post-BBN thermalization of the DR, the specifics of this process ultimately have little impact on our study.

The good fit of \nuadam to the data shows that the Hubble tension is significantly alleviated without degrading the fit to CMB temperature, polarization, and Planck lensing data, and without worsening the tension between direct and indirect measurements of the $S_8$ parameter.
While fits that do not include the SH0ES measurement of $H_0$ in their priors do not show a significant improvement in the goodness of the fit to the CMB and BAO data, they allow for a much larger maximum value for $H_0$, reducing the tension to  $\lesssim 3 \sigma$.
When including the SH0ES result as a prior, \nuadam shows a very large improvement in the fit compared to $\lcdm$, with its best fit $\chi^2$ improving by almost $34$ (and the associated AIC by $28$). For the best-fit point, we find $H_0 = 72.6$ km/s/Mpc, a sizable amount of DR corresponding to $\DNeff = 0.87$, and a clear preference for an $\mathcal{O}(1\%)$ of DM being interacting.

When information about the full-shape MPS from BOSS and eBOSS data is included, the improvement over $\lcdm$ is still very significant ($\Delta \chi^2 \approx -25$), but noticeably reduced compared to the fits without that input.
Despite this reduction in the goodness of the fit, the fact that the model is still very clearly preferred over $\lcdm$ when
the SH0ES dataset is included must be considered promising. This is particularly the case given that, for other well-known proposals to explain the $H_0$ tension, the inclusion of full-shape power spectrum analysis has limited the extent to which they are able to address the problem~\cite{McDonough:2023qcu}.

In order to properly compare the results of $\nuadam$ with the previously proposed ADM scenario, we have used a similar pipeline to fit ADM to the same datasets.
We found that ADM provides a much better solution to the $H_0$ tension than had previously been reported, if one also assumes that the dark photon is populated only after BBN.
In particular, when fitting to the $\mDH$ dataset, the best fit point for ADM has a $\chi^2$ lower by $26$ compared to the $\lcdm$ fit to the same data, and a corresponding $H_0 = 71.6~\km/\seg/\Mpc$. Nevertheless, the fit is not as good as $\nuadam$, showing that, as expected, keeping the radiation interacting at all times allows for a better solution to the $H_0$ tension.
When the full-shape analysis is included, both models perform comparably, and future data will be important to determine which one offers the better solution to the problem when more precise information about the MPS is included.

While our model does not appear well-suited to address the budding Lyman-$\alpha$ tension or the more established $S_8$ tension, it is worth noting that these two problems are on less firm ground than the Hubble tension.
In the case of the Lyman-$\alpha$ tension, the distinct features of the \nuadam model that are imprinted on the MPS (\eg, power suppression, and dark acoustic oscillations) preclude a straightforward analysis of the available Lyman-$\alpha$ likelihoods.
Therefore, dedicated hydrodynamical simulations may be required to elucidate whether our model can alleviate the tension in this data.
In the case of $S_8$, it is unclear the level to which the \lcdm assumptions embedded in the analysis of weak lensing data impact the computation of $S_8$, and how this would be altered if a MPS originating within a \nuadam scenario were to be assumed instead.
More work is needed before an ADM scenario (whether in its original incarnation or in the generalized version proposed here) can be deemed to be ruled out or favored by cosmological data.

\acknowledgments{

The authors thank Jared Barron, Thejs Brinckmann, David Curtin, Kylar Greene, 
Colin Hill, Melissa Joseph, Evan McDonough, Vivian Poulin, Adrien La Posta, Michael Toomey, and Yuhsin Tsai for helpful discussions and feedback.
MBA thanks Stephanie Buen Abad for proofreading this manuscript, as well as the New Adam, for everything.
MBA, IF and ZC are supported in part by the National Science Foundation under Grant Number PHY-2210361, and the Maryland Center for Fundamental Physics.
The research of MBA was supported in part by grant no. NSF PHY-2309135 to the Kavli Institute for Theoretical Physics (KITP).
GMT is supported in part by the National Science Foundation under Grant Number PHY-2412828.
ZC and GMT are also supported in part by the US-Israeli BSF Grant 2018236.
The research of CK is supported by the National Science Foundation Grant Number PHY-2210562.
TY is funded by the NSF grant PHY-2309456 and the Samsung Science and Technology Foundation under Project Number SSTF-BA2201-06.
}

\appendix

\section{Numerical Results}
\label{appendix}

In this appendix we present more details of our numerical fits. The best fit values of the  \lcdm, ADM, and \nuadam model parameters to all the dataset combinations discussed in the paper are shown in Tables~\ref{tab:lcdm_val_bf}, \ref{tab:adm_val_bf}, \ref{tab:nuadam_val_bf} respectively. Tables~\ref{tab:lcdm_val_mn}, \ref{tab:adm_val_mn}, and \ref{tab:nuadam_val_mn} show the mean $\pm1\sigma$ parameter values for the same three models. 

\begin{table}[h!]
\centering
\begin{adjustbox}{max width=\columnwidth}
\begin{tabular}{@{}|c|cccc|@{}}
\toprule
\toprule
Parameter & $\mD$ & $\mDH$ & $\mDF$ & $\mDHF$ \\ \midrule
$100~\theta_s$  & $1.0419$ &  $1.0421$  & $1.0419$ & $1.0421$ \\
$100~\omega_{b}$     & $2.241$ & $2.261$ & $2.243$ & $2.261$ \\
$\omega_\dm$       & $0.1197$ & $0.1179$ & $0.1194$ & $0.1178$ \\
$\ln 10^{10} A_s$        & $3.047$ & $3.054$ & $3.046$ & $3.056$ \\
$n_s$        & $0.9674$ & $0.9720$ & $0.9668$ & $0.9715$ \\
$\tau_\mathrm{reio}$        & $0.05552$ & $0.06010$ & $0.06140$ & $0.05906$ \\ \midrule
$M_B$        & $-19.434$ & $-19.404$ & $-19.428$ & $-19.409$ \\
$H_0~[\km/\seg/\Mpc]$        & $67.54$ & $68.46$ & $67.68$ & $68.45$ \\
$\sigma_8$    & $0.8119$ & $0.8090$ & $0.8098$ & $0.8095$  \\
$S_8$        & $0.8293$ & $0.8106$ & $0.8244$ & $0.8110$ \\ 
\bottomrule
\bottomrule
\end{tabular}
\end{adjustbox}
\caption{Best-fit values of the \lcdm model to all the datasets studied in this work.}
\label{tab:lcdm_val_bf}
\end{table}

\begin{table}[h!]
\centering
\begin{adjustbox}{max width=\columnwidth}
\begin{tabular}{@{}|c|cccc|@{}}
\toprule
\toprule
Parameter & $\mD$ & $\mDH$ & $\mDF$ & $\mDHF$ \\ \midrule
$100~\theta_s$  & $1.0421$ &  $1.0431$  & $1.0420$ & $1.0430$ \\
$100~\omega_{b}$     & $2.246$ & $2.290$ & $2.250$ & $2.288$ \\
$\omega_\dm$       & $0.1198$ & $0.1312$ & $0.1180$ & $0.1304$ \\
$\ln 10^{10} A_s$        & $3.049$ & $3.045$ & $3.051$ & $3.046$ \\
$n_s$        & $0.9689$ & $0.9747$ & $0.9689$ & $0.9732$ \\
$\tau_\mathrm{reio}$        & $0.05559$ & $0.05707$ & $0.05735$ & $0.05906$ \\
$\Delta\Neff$       & $0.0326$ & $0.7030$ & $0.0153$ & $0.66972$ \\
$f_\adm [\%]$        & $0.7226$ & $1.4284$ & $1.5628$ & $1.0789$ \\
$\logmm$     & $-3.789$ & $-3.616$ & $-3.729$ & $-3.689$ \\
\midrule
$\log_{10}z_{\rm d,\, rec}$     & $3.367$ & $3.210$ & $3.506$ & $3.143$ \\
$\log_{10}z_{\rm d,\, dec}$     & $3.380$ & $3.165$ & $3.553$ & $3.095$ \\
$M_B$        & $-19.431$ & $-19.303$ & $-19.424$ & $-19.303$ \\
$H_0~[\km/\seg/\Mpc]$        & $67.68$ & $71.62$ & $67.77$ & $71.88$ \\
$\sigma_8$    & $0.8079$ & $0.8184$ & $0.8010$ & $0.8189$  \\
$S_8$        & $0.8264$ & $0.8226$ & $0.8160$ & $0.8198$ \\ 
\bottomrule
\bottomrule
\end{tabular}
\end{adjustbox}
\caption{Best-fit values of the ADM model to all the datasets studied in this work.}
\label{tab:adm_val_bf}
\end{table}

\begin{table}[h!]
\centering
\begin{adjustbox}{max width=\columnwidth}
\begin{tabular}{@{}|c|cccc|@{}}
\toprule
\toprule
Parameter & $\mD$ & $\mDH$ & $\mDF$ & $\mDHF$ \\ \midrule
$100~\theta_s$  & $1.0425$ &  $1.0437$  & $1.0420$ & $1.0429$ \\
$100~\omega_{b}$     & $2.261$ & $2.316$ & $2.251$ & $2.289$ \\
$\omega_\dm$       & $0.1224$ & $0.1334$ & $0.1179$ & $0.1304$ \\
$\ln 10^{10} A_s$        & $3.050$ & $3.048$ & $3.053$ & $3.043$ \\
$n_s$        & $0.9721$ & $0.9792$ & $0.9697$ & $0.9720$ \\
$\tau_\mathrm{reio}$        & $0.05560$ & $0.05718$ & $0.05733$ & $0.05927$ \\
$\Delta\Neff$       & $0.2045$ & $0.8956$ & $0.0205$ & $0.6775$ \\
$f_\adm [\%]$        & $1.8582$ & $3.8720$ & $1.8044$ & $0.9292$ \\
$\logmm$     & $-3.830$ & $-3.656$ & $-3.947$ & $-3.988$ \\ 
\midrule
$\log_{10}z_{\rm d,\, rec}$     & $3.382$ & $3.393$ & $3.514$ & $3.096$ \\
$\log_{10}z_{\rm d,\, dec}$     & $3.372$ & $3.369$ & $3.541$ & $3.062$ \\
$M_B$        & $-19.402$ & $-19.282$ & $-19.424$ & $-19.300$ \\
$H_0~[\km/\seg/\Mpc]$        & $68.57$ & $72.60$ & $67.77$ & $72.00$ \\
$\sigma_8$    & $0.8059$ & $0.8094$ & $0.8007$ & $0.8183$  \\
$S_8$        & $0.8253$ & $0.8208$ & $0.8164$ & $0.8174$ \\ 
\bottomrule
\bottomrule
\end{tabular}
\end{adjustbox}
\caption{Best-fit values of the \nuadam model to all the datasets studied in this work.}
\label{tab:nuadam_val_bf}
\end{table}

\begin{table}[h!]
\centering
\begin{adjustbox}{max width=\columnwidth}
\begin{tabular}{@{}|c|cccc|@{}}
\toprule
\toprule
Parameter & $\mD$ & $\mDH$ & $\mDF$ & $\mDHF$ \\ \midrule
$100~\theta_s$  & $1.0420_{-0.0003}^{+0.0003}$  &  $1.0421_{-0.0003}^{+0.0003}$  & $1.0419_{-0.0003}^{+0.0003}$ & $1.0421_{-0.0003}^{+0.0003}$ \\
$100~\omega_{b}$     & $2.247_{-0.016}^{+0.015}$ & $2.261_{-0.013}^{+0.013}$ & $2.243_{-0.013}^{+0.013}$ & $2.260_{-0.012}^{+0.012}$ \\
$\omega_\dm$       & $0.1190_{-0.0010}^{+0.0012}$ & $0.1179_{-0.0008}^{+0.0008}$ & $0.1193_{-0.0008}^{+0.0008}$ & $0.1179_{-0.0007}^{+0.0007}$ \\
$\ln 10^{10} A_s$        & $3.048_{-0.016}^{+0.014}$ & $3.053_{-0.015}^{+0.014}$ & $3.046_{-0.014}^{+0.013}$ & $3.054_{-0.014}^{+0.014}$ \\
$n_s$        & $0.9671_{-0.0038}^{+0.0040}$ & $0.9704_{-0.0033}^{+0.0033}$ & $0.9668_{-0.0035}^{+0.0035}$ & $0.9700_{-0.0029}^{+0.0029}$ \\
$\tau_\mathrm{reio}$        & $0.05700_{-0.00820}^{+0.00713}$ & $0.06008_{-0.00779}^{+0.00682}$ & $0.05583_{-0.00708}^{+0.00634}$ & $0.06074_{-0.00740}^{+0.00666}$ \\ \midrule
$M_B$        & $-19.425_{-0.016}^{+0.014}$ & $-19.406_{-0.010}^{+0.011}$ & $-19.429_{-0.010}^{+0.010}$ & $-19.408_{-0.009}^{+0.009}$ \\
$H_0~[\km/\seg/\Mpc]$        & $67.87_{-0.56}^{+0.47}$ & $68.45_{-0.38}^{+0.37}$ & $67.71_{-0.35}^{+0.36}$ & $68.43_{-0.32}^{+0.32}$ \\
$\sigma_8$    & $0.8095_{-0.0066}^{+0.0064}$ & $0.8082_{-0.0062}^{+0.0057}$ & $0.8100_{-0.0059}^{+0.0054}$ & $0.8086_{-0.0058}^{+0.0055}$ \\
$S_8$        & $0.8210_{-0.0118}^{+0.0136}$ & $0.8099_{-0.0096}^{+0.0094}$ & $0.8242_{-0.0096}^{+0.0095}$ & $0.8104_{-0.0086}^{+0.0082}$ \\ 
\bottomrule
\bottomrule
\end{tabular}
\end{adjustbox}
\caption{Mean $\pm 1\sigma$ values of the \lcdm model to all the datasets studied in this work.}
\label{tab:lcdm_val_mn}
\end{table}

\begin{table}[h!]
\centering
\begin{adjustbox}{max width=\columnwidth}
\begin{tabular}{@{}|c|cccc|@{}}
\toprule
\toprule
Parameter & $\mD$ & $\mDH$ & $\mDF$ & $\mDHF$ \\ \midrule
$100~\theta_s$  & $1.0422_{-0.0003}^{+0.0003}$  &  $1.0431_{-0.0005}^{+0.0004}$  & $1.0421_{-0.0003}^{+0.0003}$ & $1.0428_{-0.0004}^{+0.0004}$ \\
$100~\omega_{b}$     & $2.243_{-0.014}^{+0.013}$ & $2.289_{-0.015}^{+0.014}$ & $2.250_{-0.018}^{+0.012}$ & $2.287_{-0.013}^{+0.013}$ \\
$\omega_\dm$       & $0.1215_{-0.0027}^{+0.0018}$ & $0.1314_{-0.0027}^{+0.0026}$ & $0.1187_{-0.0019}^{+0.0025}$ & $0.1305_{-0.0026}^{+0.0025}$ \\
$\ln 10^{10} A_s$        & $3.048_{-0.013}^{+0.012}$ & $3.045_{-0.015}^{+0.013}$ & $3.052_{-0.013}^{+0.013}$ & $3.045_{-0.014}^{+0.014}$ \\
$n_s$        & $0.9686_{-0.0035}^{+0.0033}$ & $0.9745_{-0.0046}^{+0.0043}$ & $0.9694_{-0.0035}^{+0.0034}$ & $0.9725_{-0.0040}^{+0.0038}$ \\
$\tau_\mathrm{reio}$        & $0.05537_{-0.00671}^{+0.00580}$ & $0.05764_{-0.00702}^{+0.00637}$ & $0.05743_{-0.00680}^{+0.00628}$ & $0.05921_{-0.00709}^{+0.00643}$ \\
$\Delta\Neff$       & $0.1322_{-0.1321}^{+0.0332}$ & $0.7122_{-0.1349}^{+0.1285}$ & $0.0519_{-0.0519}^{+0.0072}$ & $0.6708_{-0.1306}^{+0.1257}$ \\
$f_\adm [\%]$        & $0.9497_{-0.9494}^{+0.2006}$ & $1.4529_{-0.9837}^{+0.5846}$ & $1.5162_{-1.5155}^{+0.3026}$ & $1.0117_{-0.5512}^{+0.4191}$ \\
$\logmm$     & $-3.824_{-0.075}$ & $-3.717_{-0.034}$ & $-3.803_{-0.075}$ & $-3.842_{-0.232^*}$ \\ 
\midrule
$\log_{10}z_{\rm d,\, rec}$     & $3.235_{-0.172}^{+0.296}$ & $3.110_{-0.058}^{+0.216}$ & $3.366_{-0.177}^{+0.287}$ & $2.993_{-0.236^*}^{+0.204^*}$ \\
$\log_{10}z_{\rm d,\, dec}$     & $3.224_{-0.198}^{+0.296}$ & $3.066_{-0.056}^{+0.216}$ & $3.387_{-0.203}^{+0.302}$ & $2.949_{-0.225^*}^{+0.189^*}$ \\
$M_B$        & $-19.414_{-0.027}^{+0.015}$ & $-19.302_{-0.024}^{+0.021}$ & $-19.419_{-0.015}^{+0.011}$ & $-19.303_{-0.021}^{+0.020}$ \\
$H_0~[\km/\seg/\Mpc]$        & $68.21_{-0.86}^{+0.52}$ & $71.90_{-0.75}^{+0.73}$ & $68.00_{-0.53}^{+0.38}$ & $71.86_{-0.73}^{+0.70}$ \\
$\sigma_8$    & $0.8083_{-0.0067}^{+0.0086}$ & $0.8177_{-0.0074}^{+0.0072}$ & $0.8030_{-0.0072}^{+0.0111}$ & $0.8180_{-0.0067}^{+0.0065}$ \\
$S_8$        & $0.8262_{-0.0098}^{+0.0104}$ & $0.8223_{-0.0099}^{+0.0097}$ & $0.8173_{-0.0101}^{+0.0128}$ & $0.8190_{-0.0089}^{+0.0086}$ \\ 
\bottomrule
\bottomrule
\end{tabular}
\end{adjustbox}
\caption{Mean $\pm 1\sigma$ values of the ADM model to all the datasets studied in this work.
Since the right end of the distribution for $\logmm$ is truncated by the prior in all datasets, we only report the left end of the $\sigma$ interval.
The $\pm 1 \sigma$ values marked with an asterisk in $\mDHF$ are derived directly from the posterior distributions rather than using Gaussian fits, due to the strongly non-Gaussian nature of the posterior.}
\label{tab:adm_val_mn}
\end{table}

\begin{table}[h!]
\centering
\begin{adjustbox}{max width=\columnwidth}
\begin{tabular}{@{}|c|cccc|@{}}
\toprule
\toprule
Parameter & $\mD$ & $\mDH$ & $\mDF$ & $\mDHF$ \\ \midrule
$100~\theta_s$  & $1.0426_{-0.0005}^{+0.0004}$  &  $1.0436_{-0.0004}^{+0.0004}$  & $1.0421_{-0.0003}^{+0.0003}$ & $1.0428_{-0.0004}^{+0.0004}$ \\
$100~\omega_{b}$     & $2.263_{-0.021}^{+0.017}$ & $2.314_{-0.016}^{+0.016}$ & $2.254_{-0.014}^{+0.014}$ & $2.288_{-0.014}^{+0.014}$ \\
$\omega_\dm$       & $0.1237_{-0.0037}^{+0.0026}$ & $0.1330_{-0.0029}^{+0.0028}$ & $0.1195_{-0.0025}^{+0.0021}$ & $0.1303_{-0.0026}^{+0.0026}$ \\
$\ln 10^{10} A_s$        & $3.048_{-0.014}^{+0.013}$ & $3.048_{-0.014}^{+0.013}$ & $3.051_{-0.014}^{+0.013}$ & $3.042_{-0.014}^{+0.013}$ \\
$n_s$        & $0.9707_{-0.0043}^{+0.0039}$ & $0.9786_{-0.0045}^{+0.0044}$ & $0.9696_{-0.0036}^{+0.0039}$ & $0.9718_{-0.0039}^{+0.0038}$ \\
$\tau_\mathrm{reio}$        & $0.05558_{-0.00673}^{+0.00629}$ & $0.05736_{-0.00686}^{+0.00631}$ & $0.05719_{-0.00698}^{+0.00629}$ & $0.05899_{-0.00689}^{+0.00660}$ \\
$\Delta\Neff$       & $0.2743_{-0.2361}^{+0.1009}$ & $0.8717_{-0.1529}^{+0.1468}$ & $0.0810_{-0.0810}^{+0.0127}$ & $0.6649_{-0.1355}^{+0.1298}$ \\
$f_\adm [\%]$        & $1.7908_{-1.3795}^{+0.7661}$ & $3.7147_{-1.0459}^{+1.0737}$ & $1.3909_{-1.3905}^{+0.3432}$ & $0.8834_{-0.5307}^{+0.4050}$ \\
$\logmm$     & $-3.830_{-0.088}^{+0.194}$ & $-3.657_{-0.043}^{+0.050}$ & $-3.917_{-0.191}^{+0.275}$ & $-4.039_{-0.262^*}^{+0.218^*}$ \\
\midrule
$\log_{10}z_{\rm d,\, rec}$     & $3.385_{-0.114}^{+0.149}$ & $3.397_{-0.045}^{+0.043}$ & $3.455_{-0.185}^{+0.280}$ & $3.050_{-0.254^*}^{+0.244^*}$ \\
$\log_{10}z_{\rm d,\, dec}$     & $3.376_{-0.112}^{+0.154}$ & $3.372_{-0.042}^{+0.042}$ & $3.453_{-0.283^*}^{+0.232^*}$ & $3.009_{-0.264^*}^{+0.222^*}$ \\
$M_B$        & $-19.389_{-0.041}^{+0.023}$ & $-19.283_{-0.024}^{+0.022}$ & $-19.414_{-0.021}^{+0.012}$ & $-19.303_{-0.021}^{+0.021}$ \\
$H_0~[\km/\seg/\Mpc]$        & $68.98_{-1.32}^{+0.76}$ & $72.49_{-0.81}^{+0.76}$ & $68.16_{-0.67}^{+0.42}$ & $71.89_{-0.73}^{+0.72}$ \\
$\sigma_8$    & $0.8072_{-0.0071}^{+0.0081}$ & $0.8096_{-0.0074}^{+0.0072}$ & $0.8046_{-0.0074}^{+0.0100}$ & $0.8178_{-0.0064}^{+0.0064}$ \\
$S_8$        & $0.8252_{-0.0096}^{+0.0093}$ & $0.8205_{-0.0096}^{+0.0093}$ & $0.8189_{-0.0100}^{+0.0118}$ & $0.8175_{-0.0085}^{+0.0089}$ \\ 
\bottomrule
\bottomrule
\end{tabular}
\end{adjustbox}
\caption{Mean $\pm 1\sigma$ values of the \nuadam model to all the datasets studied in this work.
The $\pm 1 \sigma$ values marked with an asterisk in $\mDF$ and $\mDHF$ are derived directly from the posterior distributions rather than using Gaussian fits, due to the strongly non-Gaussian nature of the posterior.
}
\label{tab:nuadam_val_mn}
\end{table}

\begin{figure}[h!]
    \centering
    \includegraphics[width=.95\linewidth]{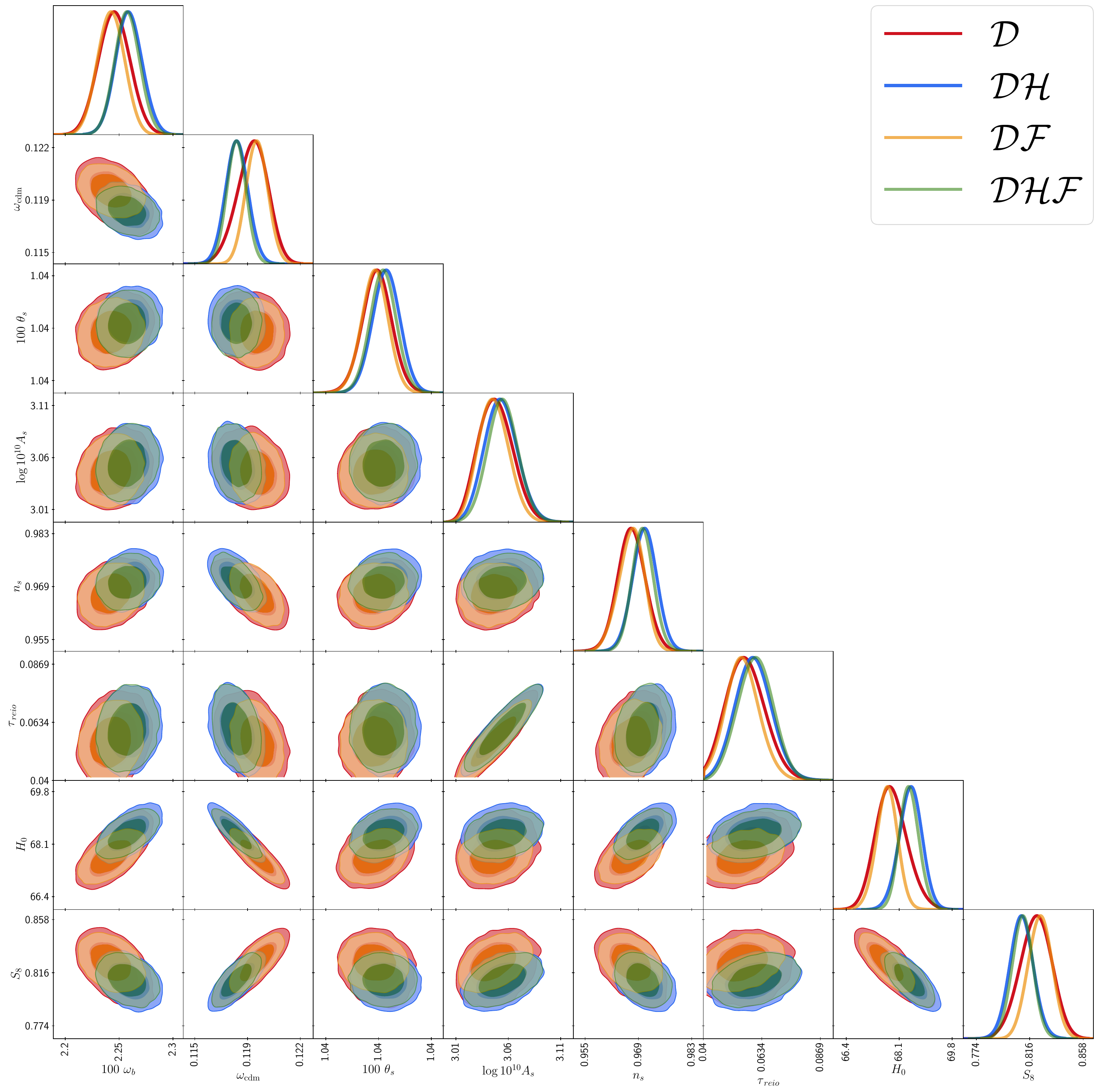}
	\caption{1D and 2D posterior distributions ($1\sigma$ and $2\sigma$ contours) of the parameters of the \lcdm model, fitted to all the datasets considered in this paper.}
	\label{fig:triangle_full_lcdm}
\end{figure}

\begin{figure}[h!]
    \centering
    \includegraphics[width=.95\linewidth]{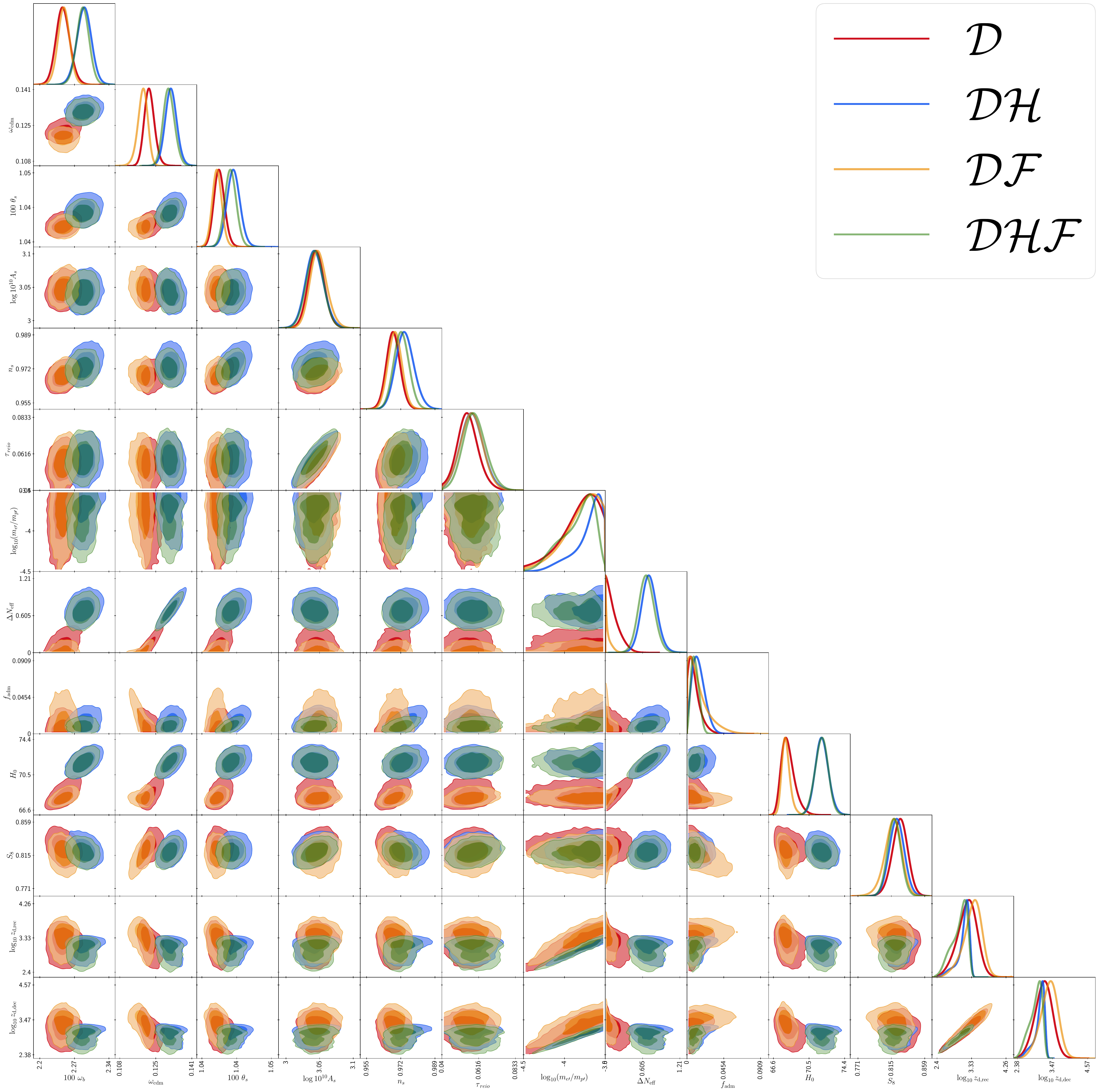}
	\caption{1D and 2D posterior distributions ($1\sigma$ and $2\sigma$ contours) of the parameters of the ADM model, fitted to all the datasets considered in this paper.}
	\label{fig:triangle_full_adm}
\end{figure}

\begin{figure}[h!]
    \centering
    \includegraphics[width=.95\linewidth]{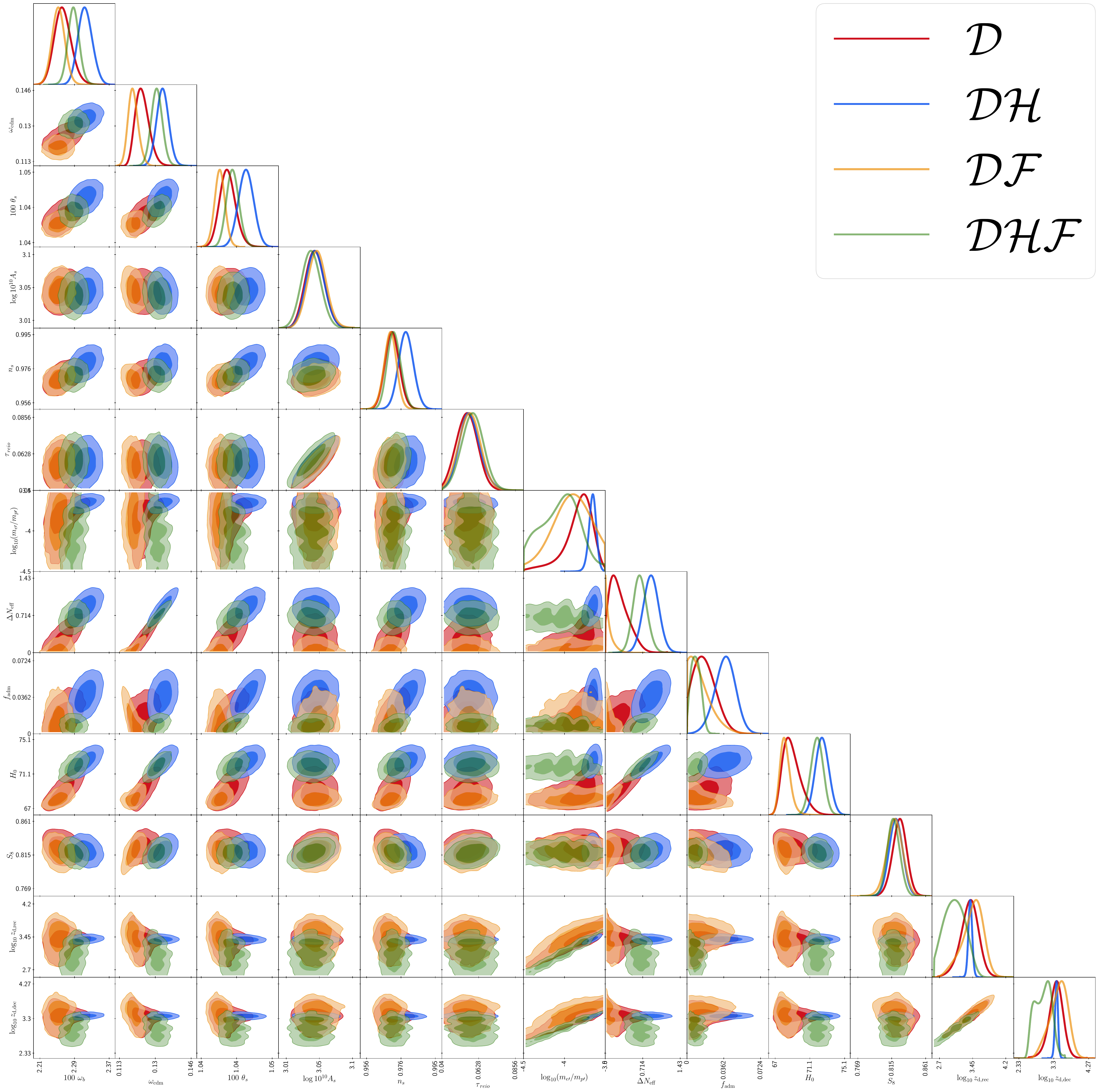}
	\caption{1D and 2D posterior distributions ($1\sigma$ and $2\sigma$ contours) of the parameters of the \nuadam model, fitted to all the datasets considered in this paper.}
	\label{fig:triangle_full_nuadam}
\end{figure}

In figures \ref{fig:triangle_full_lcdm}, \ref{fig:triangle_full_adm}, and \ref{fig:triangle_full_nuadam} we include the full triangle plots (1D and 2D posterior distributions; $1\sigma$ and $2\sigma$ contours) of the most important parameters of the \lcdm, ADM, and \nuadam models, when analyzed with the $\mD$, $\mDH$, $\mDF$, and $\mDHF$ datasets.

\section{Lyman-$\alpha$}
\label{appendix_lyman}

\begin{figure}[h!]
    \centering
    \includegraphics[width=.6\linewidth]{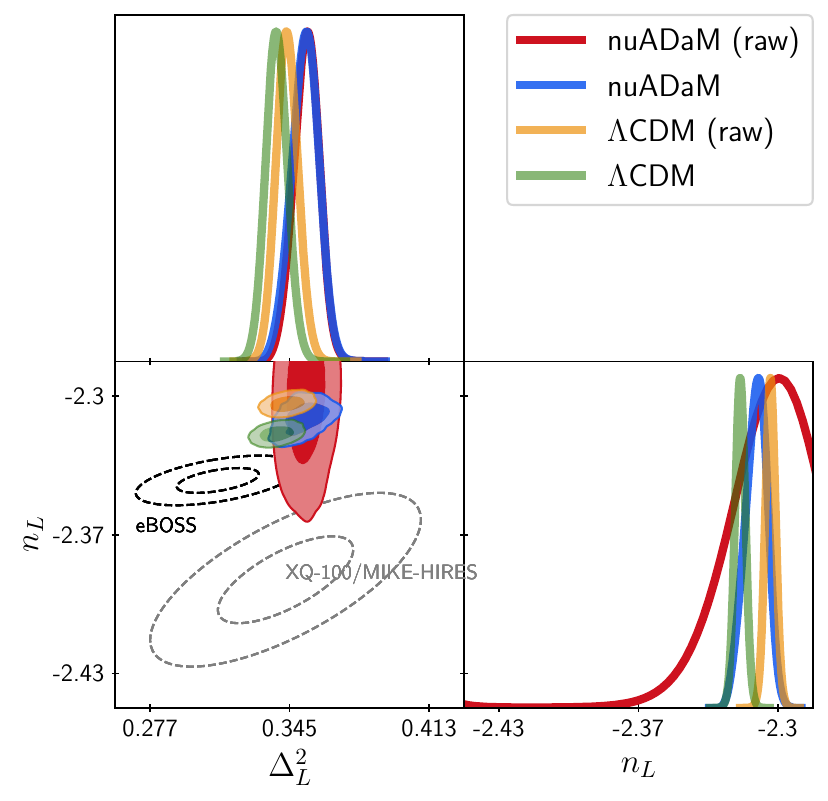}
    \includegraphics[width=.49\linewidth]{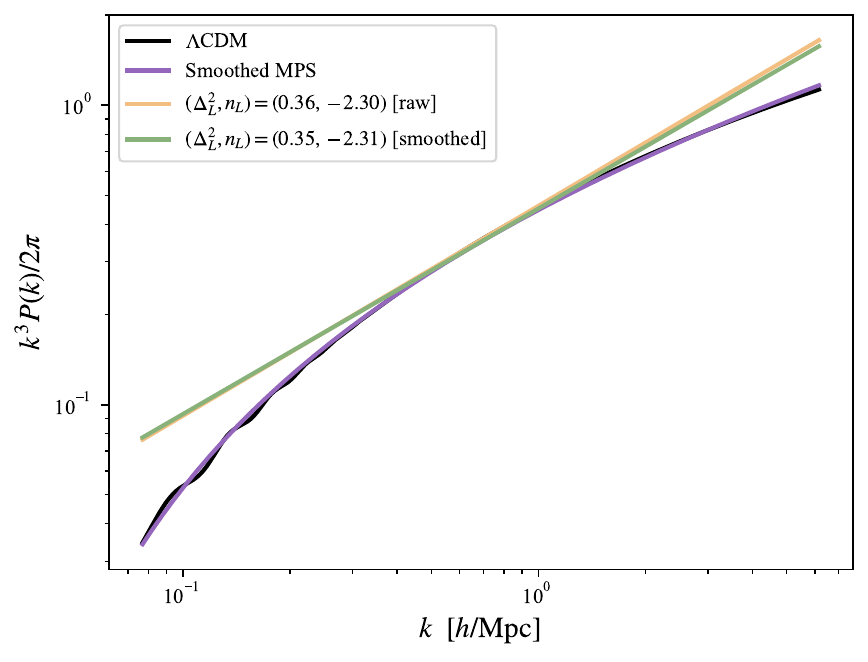}
    \includegraphics[width=.49\linewidth]{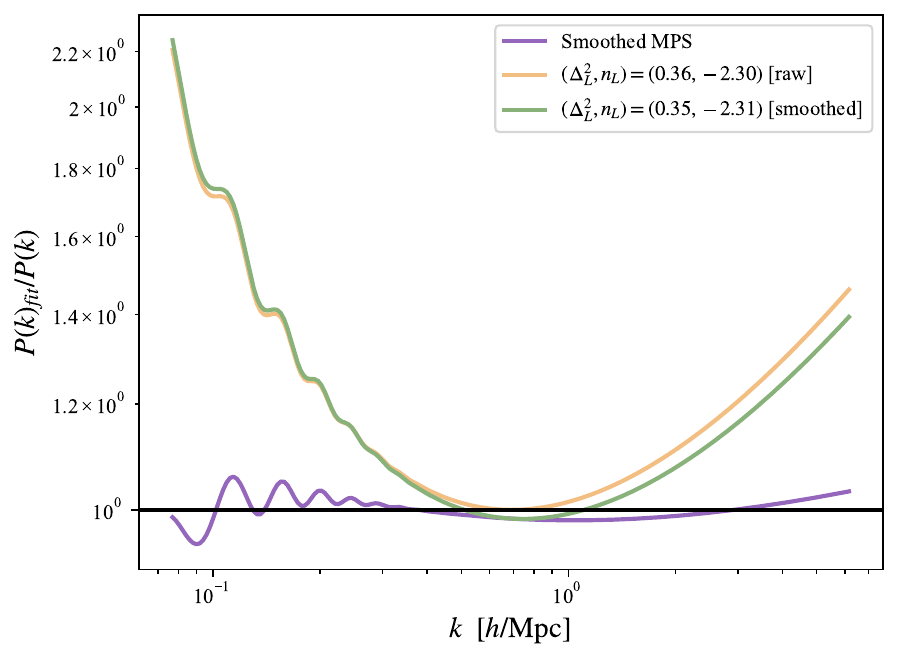}
	\caption{Comparison between the procedure described in \Subsec{subsubsec:lya} and the one described in \App{appendix_lyman} (denoted by `raw') to obtain the Lyman-$\alpha$ parameters $\Delta_L^2$ and $n_L$, for the \lcdm (in green and yellow, respectively) and \nuadam (in blue and red, respectively) models. {\it Top:} 1D and 2D $\Delta_L^2$ and $n_L$ posteriors. {\it Bottom:} The power-law MPS based on the deduced values of $\Delta_L^2$ and $n_L$ (with best fit \lcdm parameters), compared to the exact \lcdm linear MPS (black) and its smoothed version, given by \Eq{eq:template_mpk} (purple).}
	\label{fig:triangle_lya-comparison}
\end{figure}

In \Subsec{subsubsec:lya} we described our procedure to extract the Lyman-$\alpha$ parameters $\Delta_L^2$ and $n_L$ from a model's linear MPS (at a redshift $z_* = 3$) by fitting to it the template curve in \Eq{eq:template_mpk} over the observed wavenumber range, \ie, $0.001 \, \seg/\km \leq k \leq 0.08\,\seg/\km$.
Then, we evaluate the fitted curve at $k_* = 0.009\,\seg/\km$ and extract the amplitude $\Delta_L^2$ and tilt $n_L$.
The main advantage of this procedure is that it will average out any oscillatory features present in the MPS of the model in question (be that BAO and DAO), effectively allowing the MPS to be more accurately modeled by the power-law behavior which was used in creating the Lyman-$\alpha$ likelihoods to begin with.
In particular, we expect that once the DAO modes become non-linear, some degree of ``smoothing'' of these features will take place when the various modes mix.
However, we recognize the somewhat {\it ad hoc} nature of this procedure, and therefore devote the present appendix to a comparison with a more naive method consisting in simply ``reading'' $\Delta_L^2$ and $n_L$ from the linear MPS at $z_*=3$ and the single point $k_* = 0.009\,\seg/\km$, without performing any fitting procedure over the range of $k$'s being measured.
We call the resulting $\Delta_L^2$ and $n_L$ parameters ``raw'', and plot them, along with those derived from our fitting procedure, in \Fig{fig:triangle_lya-comparison}.
In the bottom right plot, we can see that for \lcdm the ``raw'' method more accurately captures the precise amplitude and tilt of the linear MPS at the single point $k_*$, but that it misses the mark at other $k$'s still important for Lyman-$\alpha$ observations, reaching 100\% discrepancies at the extremes of the plotted region.
On the other hand, our fitting procedure yields slightly smaller ($\mathcal{O}(\text{few}\, \%)$) amplitude and tilt, but is able to match the exact linear MPS slightly more accurately over a broader range of $k$ values, although the overall differences are quite small.
For \nuadam the `raw' and fitting procedures yield essentially the same posterior for the $\Delta_L^2$, but the posterior for $n_L$ is much broader for the `raw' method than the fitting method.
This is because the oscillatory features of the DAO, which for the \nuadam parameters of considered lie in the $k$ range of interest, will in general induce a local change in the tilt that will bias the `raw` result, but will be averaged out in the fitting result.
Thus, in order to get a more accurate qualitative picture of the $\Delta_L^2$ and $n_L$ parameters describing the linear MPS at $z_* = 3$ and $0.001 \, \seg/\km \leq k \leq 0.08\,\seg/\km$, as relevant for the measurement of the non-linear Lyman-$\alpha$ scales, we use the fitting procedure.

\clearpage

\bibliographystyle{JHEP}
\bibliography{refs}

@article{Planck:2018vyg,
    author = {Aghanim, N. and others},
    collaboration = {Planck},
    title = {{Planck 2018 results. VI. Cosmological parameters}},
    eprint = {1807.06209},
    archiveprefix = {arXiv},
    primaryclass = {astro-ph.CO},
    doi = {10.1051/0004-6361/201833910},
    journal = {Astron. Astrophys.},
    volume = {641},
    pages = {A6},
    year = {2020},
    note = {[Erratum: Astron.Astrophys. 652, C4 (2021)]},
}

@article{Riess:2021jrx,
    author = {Riess, Adam G. and others},
    title = {{A Comprehensive Measurement of the Local Value of the Hubble Constant with 1 km s$^{-1}$ Mpc$^{-1}$ Uncertainty from the Hubble Space Telescope and the SH0ES Team}},
    eprint = {2112.04510},
    archiveprefix = {arXiv},
    primaryclass = {astro-ph.CO},
    doi = {10.3847/2041-8213/ac5c5b},
    journal = {Astrophys. J. Lett.},
    volume = {934},
    number = {1},
    pages = {L7},
    year = {2022},
}

@article{Kamionkowski:2022pkx,
    author = {Kamionkowski, Marc and Riess, Adam G.},
    title = {{The Hubble Tension and Early Dark Energy}},
    eprint = {2211.04492},
    archiveprefix = {arXiv},
    primaryclass = {astro-ph.CO},
    doi = {10.1146/annurev-nucl-111422-024107},
    journal = {Ann. Rev. Nucl. Part. Sci.},
    volume = {73},
    pages = {153--180},
    year = {2023},
}

@article{Freedman:2021ahq,
    author = {Freedman, Wendy L.},
    title = {{Measurements of the Hubble Constant: Tensions in Perspective}},
    eprint = {2106.15656},
    archiveprefix = {arXiv},
    primaryclass = {astro-ph.CO},
    doi = {10.3847/1538-4357/ac0e95},
    journal = {Astrophys. J.},
    volume = {919},
    number = {1},
    pages = {16},
    year = {2021},
}

@article{Freedman:2024eph,
    author = {Freedman, Wendy L. and Madore, Barry F. and Hoyt, Taylor J. and Jang, In Sung and Lee, Abigail J. and Owens, Kayla A.},
    title = {{Status Report on the Chicago-Carnegie Hubble Program (CCHP): Measurement of the Hubble Constant Using the Hubble and James Webb Space Telescopes}},
    eprint = {2408.06153},
    archiveprefix = {arXiv},
    primaryclass = {astro-ph.CO},
    year = {2025},
    doi = {10.3847/1538-4357/adce78},
    journal = {Astrophys. J.},
    volume = {985},
    number = {2},
    pages = {203},
}

@article{Buen-Abad:2015ova,
    author = {Buen-Abad, Manuel A. and Marques-Tavares, Gustavo and Schmaltz, Martin},
    title = {{Non-Abelian dark matter and dark radiation}},
    eprint = {1505.03542},
    archiveprefix = {arXiv},
    primaryclass = {hep-ph},
    doi = {10.1103/PhysRevD.92.023531},
    journal = {Phys. Rev. D},
    volume = {92},
    number = {2},
    pages = {023531},
    year = {2015},
}

@article{Lesgourgues:2015wza,
    author = {Lesgourgues, Julien and Marques-Tavares, Gustavo and Schmaltz, Martin},
    title = {{Evidence for dark matter interactions in cosmological precision data?}},
    eprint = {1507.04351},
    archiveprefix = {arXiv},
    primaryclass = {astro-ph.CO},
    doi = {10.1088/1475-7516/2016/02/037},
    journal = {JCAP},
    volume = {02},
    pages = {037},
    year = {2016},
}

@article{Buen-Abad:2017gxg,
    author = {Buen-Abad, Manuel A. and Schmaltz, Martin and Lesgourgues, Julien and Brinckmann, Thejs},
    title = {{Interacting Dark Sector and Precision Cosmology}},
    eprint = {1708.09406},
    archiveprefix = {arXiv},
    primaryclass = {astro-ph.CO},
    doi = {10.1088/1475-7516/2018/01/008},
    journal = {JCAP},
    volume = {01},
    pages = {008},
    year = {2018},
}

@article{Zhao:2017cud,
    author = {Zhao, Gong-Bo and others},
    title = {{Dynamical dark energy in light of the latest observations}},
    eprint = {1701.08165},
    archiveprefix = {arXiv},
    primaryclass = {astro-ph.CO},
    doi = {10.1038/s41550-017-0216-z},
    journal = {Nature Astron.},
    volume = {1},
    number = {9},
    pages = {627--632},
    year = {2017},
}

@article{DiValentino:2017gzb,
    author = {Di Valentino, Eleonora},
    title = {{Crack in the cosmological paradigm}},
    eprint = {1709.04046},
    archiveprefix = {arXiv},
    primaryclass = {physics.pop-ph},
    doi = {10.1038/s41550-017-0236-8},
    journal = {Nature Astron.},
    volume = {1},
    number = {9},
    pages = {569--570},
    year = {2017},
}

@article{Poulin:2018cxd,
    author = {Poulin, Vivian and Smith, Tristan L. and Karwal, Tanvi and Kamionkowski, Marc},
    title = {{Early Dark Energy Can Resolve The Hubble Tension}},
    eprint = {1811.04083},
    archiveprefix = {arXiv},
    primaryclass = {astro-ph.CO},
    doi = {10.1103/PhysRevLett.122.221301},
    journal = {Phys. Rev. Lett.},
    volume = {122},
    number = {22},
    pages = {221301},
    year = {2019},
}

@article{Smith:2019ihp,
    author = {Smith, Tristan L. and Poulin, Vivian and Amin, Mustafa A.},
    title = {{Oscillating scalar fields and the Hubble tension: a resolution with novel signatures}},
    eprint = {1908.06995},
    archiveprefix = {arXiv},
    primaryclass = {astro-ph.CO},
    doi = {10.1103/PhysRevD.101.063523},
    journal = {Phys. Rev. D},
    volume = {101},
    number = {6},
    pages = {063523},
    year = {2020},
}

@article{Lin:2019qug,
    author = {Lin, Meng-Xiang and Benevento, Giampaolo and Hu, Wayne and Raveri, Marco},
    title = {{Acoustic Dark Energy: Potential Conversion of the Hubble Tension}},
    eprint = {1905.12618},
    archiveprefix = {arXiv},
    primaryclass = {astro-ph.CO},
    doi = {10.1103/PhysRevD.100.063542},
    journal = {Phys. Rev. D},
    volume = {100},
    number = {6},
    pages = {063542},
    year = {2019},
}

@article{Alexander:2019rsc,
    author = {Alexander, Stephon and McDonough, Evan},
    title = {{Axion-Dilaton Destabilization and the Hubble Tension}},
    eprint = {1904.08912},
    archiveprefix = {arXiv},
    primaryclass = {astro-ph.CO},
    doi = {10.1016/j.physletb.2019.134830},
    journal = {Phys. Lett. B},
    volume = {797},
    pages = {134830},
    year = {2019},
}

@article{Agrawal:2019lmo,
    author = {Agrawal, Prateek and Cyr-Racine, Francis-Yan and Pinner, David and Randall, Lisa},
    title = {{Rock \textquoteleft{}n\textquoteright{} roll solutions to the Hubble tension}},
    eprint = {1904.01016},
    archiveprefix = {arXiv},
    primaryclass = {astro-ph.CO},
    doi = {10.1016/j.dark.2023.101347},
    journal = {Phys. Dark Univ.},
    volume = {42},
    pages = {101347},
    year = {2023},
}

@article{Escudero:2019gvw,
    author = {Escudero, Miguel and Witte, Samuel J.},
    title = {{A CMB search for the neutrino mass mechanism and its relation to the Hubble tension}},
    eprint = {1909.04044},
    archiveprefix = {arXiv},
    primaryclass = {astro-ph.CO},
    reportnumber = {KCL-2019-71},
    doi = {10.1140/epjc/s10052-020-7854-5},
    journal = {Eur. Phys. J. C},
    volume = {80},
    number = {4},
    pages = {294},
    year = {2020},
}

@article{Berghaus:2019cls,
    author = {Berghaus, Kim V. and Karwal, Tanvi},
    title = {{Thermal Friction as a Solution to the Hubble Tension}},
    eprint = {1911.06281},
    archiveprefix = {arXiv},
    primaryclass = {astro-ph.CO},
    doi = {10.1103/PhysRevD.101.083537},
    journal = {Phys. Rev. D},
    volume = {101},
    number = {8},
    pages = {083537},
    year = {2020},
}

@article{Vagnozzi:2019ezj,
    author = {Vagnozzi, Sunny},
    title = {{New physics in light of the $H_0$ tension: An alternative view}},
    eprint = {1907.07569},
    archiveprefix = {arXiv},
    primaryclass = {astro-ph.CO},
    doi = {10.1103/PhysRevD.102.023518},
    journal = {Phys. Rev. D},
    volume = {102},
    number = {2},
    pages = {023518},
    year = {2020},
}

@article{Ye:2020btb,
    author = {Ye, Gen and Piao, Yun-Song},
    title = {{Is the Hubble tension a hint of AdS phase around recombination?}},
    eprint = {2001.02451},
    archiveprefix = {arXiv},
    primaryclass = {astro-ph.CO},
    doi = {10.1103/PhysRevD.101.083507},
    journal = {Phys. Rev. D},
    volume = {101},
    number = {8},
    pages = {083507},
    year = {2020},
}

@article{Das:2020xke,
    author = {Das, Anirban and Ghosh, Subhajit},
    title = {{Flavor-specific interaction favors strong neutrino self-coupling in the early universe}},
    eprint = {2011.12315},
    archiveprefix = {arXiv},
    primaryclass = {astro-ph.CO},
    reportnumber = {SLAC-PUB-17547},
    doi = {10.1088/1475-7516/2021/07/038},
    journal = {JCAP},
    volume = {07},
    pages = {038},
    year = {2021},
}

@article{RoyChoudhury:2020dmd,
    author = {Roy Choudhury, Shouvik and Hannestad, Steen and Tram, Thomas},
    title = {{Updated constraints on massive neutrino self-interactions from cosmology in light of the $H_0$ tension}},
    eprint = {2012.07519},
    archiveprefix = {arXiv},
    primaryclass = {astro-ph.CO},
    doi = {10.1088/1475-7516/2021/03/084},
    journal = {JCAP},
    volume = {03},
    pages = {084},
    year = {2021},
}

@article{Brinckmann:2020bcn,
    author = {Brinckmann, Thejs and Chang, Jae Hyeok and LoVerde, Marilena},
    title = {{Self-interacting neutrinos, the Hubble parameter tension, and the cosmic microwave background}},
    eprint = {2012.11830},
    archiveprefix = {arXiv},
    primaryclass = {astro-ph.CO},
    reportnumber = {YITP-SB-2020-40},
    doi = {10.1103/PhysRevD.104.063523},
    journal = {Phys. Rev. D},
    volume = {104},
    number = {6},
    pages = {063523},
    year = {2021},
}

@article{Krishnan:2020vaf,
    author = {Krishnan, C. and Colg\'ain, E. \'O. and Sheikh-Jabbari, M. M. and Yang, Tao},
    title = {{Running Hubble Tension and a H0 Diagnostic}},
    eprint = {2011.02858},
    archiveprefix = {arXiv},
    primaryclass = {astro-ph.CO},
    doi = {10.1103/PhysRevD.103.103509},
    journal = {Phys. Rev. D},
    volume = {103},
    number = {10},
    pages = {103509},
    year = {2021},
}

@article{Seto:2021xua,
    author = {Seto, Osamu and Toda, Yo},
    title = {{Comparing early dark energy and extra radiation solutions to the Hubble tension with BBN}},
    eprint = {2101.03740},
    archiveprefix = {arXiv},
    primaryclass = {astro-ph.CO},
    reportnumber = {EPHOU-21-002},
    doi = {10.1103/PhysRevD.103.123501},
    journal = {Phys. Rev. D},
    volume = {103},
    number = {12},
    pages = {123501},
    year = {2021},
}

@article{Ye:2021iwa,
    author = {Ye, Gen and Zhang, Jun and Piao, Yun-Song},
    title = {{Alleviating both H0 and S8 tensions: Early dark energy lifts the CMB-lockdown on ultralight axion}},
    eprint = {2107.13391},
    archiveprefix = {arXiv},
    primaryclass = {astro-ph.CO},
    doi = {10.1016/j.physletb.2023.137770},
    journal = {Phys. Lett. B},
    volume = {839},
    pages = {137770},
    year = {2023},
}

@article{Niedermann:2021vgd,
    author = {Niedermann, Florian and Sloth, Martin S.},
    title = {{Hot new early dark energy}},
    eprint = {2112.00770},
    archiveprefix = {arXiv},
    primaryclass = {hep-ph},
    doi = {10.1103/PhysRevD.105.063509},
    journal = {Phys. Rev. D},
    volume = {105},
    number = {6},
    pages = {063509},
    year = {2022},
}

@article{Aloni:2021eaq,
    author = {Aloni, Daniel and Berlin, Asher and Joseph, Melissa and Schmaltz, Martin and Weiner, Neal},
    title = {{A Step in understanding the Hubble tension}},
    eprint = {2111.00014},
    archiveprefix = {arXiv},
    primaryclass = {astro-ph.CO},
    doi = {10.1103/PhysRevD.105.123516},
    journal = {Phys. Rev. D},
    volume = {105},
    number = {12},
    pages = {123516},
    year = {2022},
}

@article{Dainotti:2021pqg,
    author = {Dainotti, Maria Giovanna and De Simone, Biagio and Schiavone, Tiziano and Montani, Giovanni and Rinaldi, Enrico and Lambiase, Gaetano},
    title = {{On the Hubble constant tension in the SNe Ia Pantheon sample}},
    eprint = {2103.02117},
    archiveprefix = {arXiv},
    primaryclass = {astro-ph.CO},
    reportnumber = {AAS28095R4, RIKEN-iTHEMS-Report-21},
    doi = {10.3847/1538-4357/abeb73},
    journal = {Astrophys. J.},
    volume = {912},
    number = {2},
    pages = {150},
    year = {2021},
}

@article{Odintsov:2022eqm,
    author = {Odintsov, S. D. and Oikonomou, V. K.},
    title = {{Did the Universe experience a pressure non-crushing type cosmological singularity in the recent past?}},
    eprint = {2201.07647},
    archiveprefix = {arXiv},
    primaryclass = {gr-qc},
    doi = {10.1209/0295-5075/ac52dc},
    journal = {EPL},
    volume = {137},
    number = {3},
    pages = {39001},
    year = {2022},
}

@article{Berghaus:2022cwf,
    author = {Berghaus, Kim V. and Karwal, Tanvi},
    title = {{Thermal friction as a solution to the Hubble and large-scale structure tensions}},
    eprint = {2204.09133},
    archiveprefix = {arXiv},
    primaryclass = {astro-ph.CO},
    doi = {10.1103/PhysRevD.107.103515},
    journal = {Phys. Rev. D},
    volume = {107},
    number = {10},
    pages = {103515},
    year = {2023},
}

@article{Schoneberg:2022grr,
    author = {Sch\"oneberg, Nils and Franco Abell\'an, Guillermo},
    title = {{A step in the right direction? Analyzing the Wess Zumino Dark Radiation solution to the Hubble tension}},
    eprint = {2206.11276},
    archiveprefix = {arXiv},
    primaryclass = {astro-ph.CO},
    doi = {10.1088/1475-7516/2022/12/001},
    journal = {JCAP},
    volume = {12},
    pages = {001},
    year = {2022},
}

@article{Joseph:2022jsf,
    author = {Joseph, Melissa and Aloni, Daniel and Schmaltz, Martin and Sivarajan, Eashwar N. and Weiner, Neal},
    title = {{A Step in understanding the S8 tension}},
    eprint = {2207.03500},
    archiveprefix = {arXiv},
    primaryclass = {astro-ph.CO},
    doi = {10.1103/PhysRevD.108.023520},
    journal = {Phys. Rev. D},
    volume = {108},
    number = {2},
    pages = {023520},
    year = {2023},
}

@article{Buen-Abad:2022kgf,
    author = {Buen-Abad, Manuel A. and Chacko, Zackaria and Kilic, Can and Marques-Tavares, Gustavo and Youn, Taewook},
    title = {{Stepped partially acoustic dark matter, large scale structure, and the Hubble tension}},
    eprint = {2208.05984},
    archiveprefix = {arXiv},
    primaryclass = {hep-ph},
    reportnumber = {UTTG-10-2022},
    doi = {10.1007/JHEP06(2023)012},
    journal = {JHEP},
    volume = {06},
    pages = {012},
    year = {2023},
}

@article{Rezazadeh:2022lsf,
    author = {Rezazadeh, Kazem and Ashoorioon, Amjad and Grin, Daniel},
    title = {{Cascading Dark Energy}},
    eprint = {2208.07631},
    archiveprefix = {arXiv},
    primaryclass = {astro-ph.CO},
    reportnumber = {IPM/P-2022/60},
    doi = {10.3847/1538-4357/ad7b16},
    journal = {Astrophys. J.},
    volume = {975},
    number = {1},
    pages = {137},
    year = {2024},
}

@article{Brinckmann:2022ajr,
    author = {Brinckmann, Thejs and Chang, Jae Hyeok and Du, Peizhi and LoVerde, Marilena},
    title = {{Confronting interacting dark radiation scenarios with cosmological data}},
    eprint = {2212.13264},
    archiveprefix = {arXiv},
    primaryclass = {astro-ph.CO},
    doi = {10.1103/PhysRevD.107.123517},
    journal = {Phys. Rev. D},
    volume = {107},
    number = {12},
    pages = {123517},
    year = {2023},
}

@article{Wang:2022bmk,
    author = {Wang, Hao and Piao, Yun-Song},
    title = {{Does the fraction of dark matter diminish with early dark energy?}},
    eprint = {2209.09685},
    archiveprefix = {arXiv},
    primaryclass = {astro-ph.CO},
    doi = {10.1103/PhysRevD.108.083516},
    journal = {Phys. Rev. D},
    volume = {108},
    number = {8},
    pages = {083516},
    year = {2023},
}

@article{Bansal:2022qbi,
    author = {Bansal, Saurabh and Barron, Jared and Curtin, David and Tsai, Yuhsin},
    title = {{Precision cosmological constraints on atomic dark matter}},
    eprint = {2212.02487},
    archiveprefix = {arXiv},
    primaryclass = {hep-ph},
    doi = {10.1007/JHEP10(2023)095},
    journal = {JHEP},
    volume = {10},
    pages = {095},
    year = {2023},
}

@article{Buen-Abad:2023uva,
    author = {Buen-Abad, Manuel A. and Chacko, Zackaria and Kilic, Can and Marques-Tavares, Gustavo and Youn, Taewook},
    title = {{Stepped partially acoustic dark matter: likelihood analysis and cosmological tensions}},
    eprint = {2306.01844},
    archiveprefix = {arXiv},
    primaryclass = {astro-ph.CO},
    reportnumber = {UTWI-19-2023},
    doi = {10.1088/1475-7516/2023/11/005},
    journal = {JCAP},
    volume = {11},
    pages = {005},
    year = {2023},
}

@article{Sandner:2023ptm,
    author = {Sandner, Stefan and Escudero, Miguel and Witte, Samuel J.},
    title = {{Precision CMB constraints on eV-scale bosons coupled to neutrinos}},
    eprint = {2305.01692},
    archiveprefix = {arXiv},
    primaryclass = {hep-ph},
    reportnumber = {IFIC/23-13, FTUV-23-0413.0599, CERN-TH-2023-073},
    doi = {10.1140/epjc/s10052-023-11864-6},
    journal = {Eur. Phys. J. C},
    volume = {83},
    number = {8},
    pages = {709},
    year = {2023},
}

@article{Zu:2023rmc,
    author = {Zu, Lei and Zhang, Chi and Chen, Hou-Zun and Wang, Wei and Tsai, Yue-Lin Sming and Tsai, Yuhsin and Luo, Wentao and Fan, Yi-Zhong},
    title = {{Exploring mirror twin Higgs cosmology with present and future weak lensing surveys}},
    eprint = {2304.06308},
    archiveprefix = {arXiv},
    primaryclass = {astro-ph.CO},
    doi = {10.1088/1475-7516/2023/08/023},
    journal = {JCAP},
    volume = {08},
    pages = {023},
    year = {2023},
}

@article{Niedermann:2023ssr,
    author = {Niedermann, Florian and Sloth, Martin S.},
    title = {{New Early Dark Energy as a solution to the $H_0$ and $S_8$ tensions}},
    eprint = {2307.03481},
    archiveprefix = {arXiv},
    primaryclass = {hep-ph},
    doi = {10.1007/978-981-99-0177-7_23},
    month = {7},
    year = {2023},
}

@article{Hughes:2023tcn,
    author = {Hughes, Ellie and Ge, Fei and Cyr-Racine, Francis-Yan and Knox, Lloyd and Raghunathan, Srinivasan},
    title = {{Cool dark sector, concordance, and a low \ensuremath{\sigma}8}},
    eprint = {2311.05678},
    archiveprefix = {arXiv},
    primaryclass = {astro-ph.CO},
    doi = {10.1103/PhysRevD.109.103516},
    journal = {Phys. Rev. D},
    volume = {109},
    number = {10},
    pages = {103516},
    year = {2024},
}

@article{Greene:2024qis,
    author = {Greene, Kylar and Cyr-Racine, Francis-Yan},
    title = {{Ratio-preserving approach to cosmological concordance}},
    eprint = {2403.05619},
    archiveprefix = {arXiv},
    primaryclass = {astro-ph.CO},
    reportnumber = {FERMILAB-PUB-24-0721-V},
    doi = {10.1103/PhysRevD.110.043524},
    journal = {Phys. Rev. D},
    volume = {110},
    number = {4},
    pages = {043524},
    year = {2024},
}

@article{Allali:2024anb,
    author = {Allali, Itamar J. and Aloni, Daniel and Sch\"oneberg, Nils},
    title = {{Cosmological probes of Dark Radiation from Neutrino Mixing}},
    eprint = {2404.16822},
    archiveprefix = {arXiv},
    primaryclass = {astro-ph.CO},
    doi = {10.1088/1475-7516/2024/09/019},
    journal = {JCAP},
    volume = {09},
    pages = {019},
    year = {2024},
}

@article{Montani:2024pou,
    author = {Montani, Giovanni and Carlevaro, Nakia and Escamilla, Luis A. and Di Valentino, Eleonora},
    title = {{Kinetic model for dark energy{\textemdash}dark matter interaction: Scenario for the hubble tension}},
    eprint = {2404.15977},
    archiveprefix = {arXiv},
    primaryclass = {gr-qc},
    year = {2025},
    doi = {10.1016/j.dark.2025.101848},
    journal = {Phys. Dark Univ.},
    volume = {48},
    pages = {101848},
}

@article{Co:2024oek,
    author = {Co, Raymond T. and Fernandez, Nicolas and Ghalsasi, Akshay and Harigaya, Keisuke and Shelton, Jessie},
    title = {{Axion baryogenesis puts a new spin on the Hubble tension}},
    eprint = {2405.12268},
    archiveprefix = {arXiv},
    primaryclass = {hep-ph},
    reportnumber = {PITT-PACC-2402},
    doi = {10.1103/PhysRevD.110.083534},
    journal = {Phys. Rev. D},
    volume = {110},
    number = {8},
    pages = {083534},
    year = {2024},
}

@article{Cho:2024lhp,
    author = {Cho, Wonsub and Choi, Ki-Young and Mahapatra, Satyabrata},
    title = {{Reconciling cosmological tensions with inelastic dark matter and dark radiation in a U(1)D framework}},
    eprint = {2408.03004},
    archiveprefix = {arXiv},
    primaryclass = {hep-ph},
    doi = {10.1088/1475-7516/2024/09/065},
    journal = {JCAP},
    volume = {09},
    pages = {065},
    year = {2024},
}

@article{Simon:2024jmu,
    author = {Simon, Th{\'e}o and Adi, Tal and Bernal, Jos{\'e} Luis and Kovetz, Ely D. and Poulin, Vivian and Smith, Tristan L.},
    title = {{Toward alleviating the $H_0$ and $S_8$ tensions with early dark energy-dark matter drag}},
    eprint = {2410.21459},
    archiveprefix = {arXiv},
    primaryclass = {astro-ph.CO},
    year = {2025},
    doi = {10.1103/PhysRevD.111.023523},
    journal = {Phys. Rev. D},
    volume = {111},
    number = {2},
    pages = {023523},
}

@article{DeSimone:2024lvy,
    author = {De Simone, B. and van Putten, M. H. P. M. and Dainotti, M. G. and Lambiase, G.},
    title = {{A doublet of cosmological models to challenge the H0 tension in the Pantheon Supernovae Ia catalog}},
    eprint = {2411.05744},
    archiveprefix = {arXiv},
    primaryclass = {astro-ph.CO},
    year = {2025},
    doi = {10.1016/j.jheap.2024.12.003},
    journal = {JHEAp},
    volume = {45},
    pages = {290--298},
}

@article{DiValentino:2021izs,
    author = {Di Valentino, Eleonora and Mena, Olga and Pan, Supriya and Visinelli, Luca and Yang, Weiqiang and Melchiorri, Alessandro and Mota, David F. and Riess, Adam G. and Silk, Joseph},
    title = {{In the realm of the Hubble tension\textemdash{}a review of solutions}},
    eprint = {2103.01183},
    archiveprefix = {arXiv},
    primaryclass = {astro-ph.CO},
    reportnumber = {IPPP/20/108},
    doi = {10.1088/1361-6382/ac086d},
    journal = {Class. Quant. Grav.},
    volume = {38},
    number = {15},
    pages = {153001},
    year = {2021},
}

@article{Schoneberg:2021qvd,
    author = {Sch\"oneberg, Nils and Franco Abell\'an, Guillermo and P\'erez S\'anchez, Andrea and Witte, Samuel J. and Poulin, Vivian and Lesgourgues, Julien},
    title = {{The H0 Olympics: A fair ranking of proposed models}},
    eprint = {2107.10291},
    archiveprefix = {arXiv},
    primaryclass = {astro-ph.CO},
    doi = {10.1016/j.physrep.2022.07.001},
    journal = {Phys. Rept.},
    volume = {984},
    pages = {1--55},
    year = {2022},
}

@article{Abdalla:2022yfr,
    author = {Abdalla, Elcio and others},
    title = {{Cosmology intertwined: A review of the particle physics, astrophysics, and cosmology associated with the cosmological tensions and anomalies}},
    eprint = {2203.06142},
    archiveprefix = {arXiv},
    primaryclass = {astro-ph.CO},
    reportnumber = {FERMILAB-CONF-22-192-SCD},
    doi = {10.1016/j.jheap.2022.04.002},
    journal = {JHEAp},
    volume = {34},
    pages = {49--211},
    year = {2022},
}

@article{Poulin:2023lkg,
    author = {Poulin, Vivian and Smith, Tristan L. and Karwal, Tanvi},
    title = {{The Ups and Downs of Early Dark Energy solutions to the Hubble tension: A review of models, hints and constraints circa 2023}},
    eprint = {2302.09032},
    archiveprefix = {arXiv},
    primaryclass = {astro-ph.CO},
    doi = {10.1016/j.dark.2023.101348},
    journal = {Phys. Dark Univ.},
    volume = {42},
    pages = {101348},
    year = {2023},
}

@article{Khalife:2023qbu,
    author = {Khalife, Ali Rida and Zanjani, Maryam Bahrami and Galli, Silvia and G\"unther, Sven and Lesgourgues, Julien and Benabed, Karim},
    title = {{Review of Hubble tension solutions with new SH0ES and SPT-3G data}},
    eprint = {2312.09814},
    archiveprefix = {arXiv},
    primaryclass = {astro-ph.CO},
    reportnumber = {TTK-23-36},
    doi = {10.1088/1475-7516/2024/04/059},
    journal = {JCAP},
    volume = {04},
    pages = {059},
    year = {2024},
}

@article{Aylor:2018drw,
    author = {Aylor, Kevin and Joy, MacKenzie and Knox, Lloyd and Millea, Marius and Raghunathan, Srinivasan and Wu, W. L. Kimmy},
    title = {{Sounds Discordant: Classical Distance Ladder \& $\Lambda$CDM -based Determinations of the Cosmological Sound Horizon}},
    eprint = {1811.00537},
    archiveprefix = {arXiv},
    primaryclass = {astro-ph.CO},
    doi = {10.3847/1538-4357/ab0898},
    journal = {Astrophys. J.},
    volume = {874},
    number = {1},
    pages = {4},
    year = {2019},
}

@article{Philcox:2022sgj,
    author = {Philcox, Oliver H. E. and Farren, Gerrit S. and Sherwin, Blake D. and Baxter, Eric J. and Brout, Dillon J.},
    title = {{Determining the Hubble constant without the sound horizon: A 3.6\% constraint on H0 from galaxy surveys, CMB lensing, and supernovae}},
    eprint = {2204.02984},
    archiveprefix = {arXiv},
    primaryclass = {astro-ph.CO},
    doi = {10.1103/PhysRevD.106.063530},
    journal = {Phys. Rev. D},
    volume = {106},
    number = {6},
    pages = {063530},
    year = {2022},
}

@article{Zaborowski:2024car,
    author = {Zaborowski, E. A. and others},
    title = {{A Sound Horizon-Free Measurement of $H_0$ in DESI 2024}},
    eprint = {2411.16677},
    archiveprefix = {arXiv},
    primaryclass = {astro-ph.CO},
    reportnumber = {FERMILAB-PUB-24-0885-PPD},
    month = {11},
    year = {2024},
    doi = {10.1088/1475-7516/2025/06/020},
}

@article{Blinov:2020hmc,
    author = {Blinov, Nikita and Marques-Tavares, Gustavo},
    title = {{Interacting radiation after Planck and its implications for the Hubble Tension}},
    eprint = {2003.08387},
    archiveprefix = {arXiv},
    primaryclass = {astro-ph.CO},
    reportnumber = {FERMILAB-PUB-20-114-A-T},
    doi = {10.1088/1475-7516/2020/09/029},
    journal = {JCAP},
    volume = {09},
    pages = {029},
    year = {2020},
}

@article{Hou:2011ec,
    author = {Hou, Zhen and Keisler, Ryan and Knox, Lloyd and Millea, Marius and Reichardt, Christian},
    title = {{How Massless Neutrinos Affect the Cosmic Microwave Background Damping Tail}},
    eprint = {1104.2333},
    archiveprefix = {arXiv},
    primaryclass = {astro-ph.CO},
    doi = {10.1103/PhysRevD.87.083008},
    journal = {Phys. Rev. D},
    volume = {87},
    pages = {083008},
    year = {2013},
}

@article{Goldstein:2023gnw,
    author = {Goldstein, Samuel and Hill, J. Colin and Ir\v{s}i\v{c}, Vid and Sherwin, Blake D.},
    title = {{Canonical Hubble-Tension-Resolving Early Dark Energy Cosmologies Are Inconsistent with the Lyman-\ensuremath{\alpha} Forest}},
    eprint = {2303.00746},
    archiveprefix = {arXiv},
    primaryclass = {astro-ph.CO},
    doi = {10.1103/PhysRevLett.131.201001},
    journal = {Phys. Rev. Lett.},
    volume = {131},
    number = {20},
    pages = {201001},
    year = {2023},
}

@article{Rogers:2023upm,
    author = {Rogers, Keir K. and Poulin, Vivian},
    title = {{5{\ensuremath{\sigma}} tension between Planck cosmic microwave background and eBOSS Lyman-alpha forest and constraints on physics beyond {\ensuremath{\Lambda}}CDM}},
    eprint = {2311.16377},
    archiveprefix = {arXiv},
    primaryclass = {astro-ph.CO},
    year = {2025},
    doi = {10.1103/PhysRevResearch.7.L012018},
    journal = {Phys. Rev. Res.},
    volume = {7},
    number = {1},
    pages = {L012018},
}

@article{Allali:2023zbi,
    author = {Allali, Itamar J. and Rompineve, Fabrizio and Hertzberg, Mark P.},
    title = {{Dark sectors with mass thresholds face cosmological datasets}},
    eprint = {2305.14166},
    archiveprefix = {arXiv},
    primaryclass = {astro-ph.CO},
    reportnumber = {CERN-TH-2023-084},
    doi = {10.1103/PhysRevD.108.023527},
    journal = {Phys. Rev. D},
    volume = {108},
    number = {2},
    pages = {023527},
    year = {2023},
}

@article{Schoneberg:2023rnx,
    author = {Sch\"oneberg, Nils and Franco Abell\'an, Guillermo and Simon, Th\'eo and Bartlett, Alexa and Patel, Yashvi and Smith, Tristan L.},
    title = {{Comparative analysis of interacting stepped dark radiation}},
    eprint = {2306.12469},
    archiveprefix = {arXiv},
    primaryclass = {astro-ph.CO},
    doi = {10.1103/PhysRevD.108.123513},
    journal = {Phys. Rev. D},
    volume = {108},
    number = {12},
    pages = {123513},
    year = {2023},
}

@article{Bagherian:2024obh,
    author = {Bagherian, Hengameh and Joseph, Melissa and Schmaltz, Martin and Sivarajan, Eashwar N.},
    title = {{Confronting interacting radiation models for the Hubble tension with Lyman-{\ensuremath{\alpha}} data}},
    eprint = {2405.17554},
    archiveprefix = {arXiv},
    primaryclass = {astro-ph.CO},
    year = {2025},
    doi = {10.1103/PhysRevD.111.043513},
    journal = {Phys. Rev. D},
    volume = {111},
    number = {4},
    pages = {043513},
}

@article{Goldberg:1986nk,
    author = {Goldberg, Haim and Hall, Lawrence J.},
    editor = {Tran Thanh Van, J.},
    title = {{A New Candidate for Dark Matter}},
    reportnumber = {HUTP-86/A029},
    doi = {10.1016/0370-2693(86)90731-8},
    journal = {Phys. Lett. B},
    volume = {174},
    pages = {151},
    year = {1986},
}

@article{Kaplan:2009de,
    author = {Kaplan, David E. and Krnjaic, Gordan Z. and Rehermann, Keith R. and Wells, Christopher M.},
    title = {{Atomic Dark Matter}},
    eprint = {0909.0753},
    archiveprefix = {arXiv},
    primaryclass = {hep-ph},
    doi = {10.1088/1475-7516/2010/05/021},
    journal = {JCAP},
    volume = {05},
    pages = {021},
    year = {2010},
}

@article{Kaplan:2011yj,
    author = {Kaplan, David E. and Krnjaic, Gordan Z. and Rehermann, Keith R. and Wells, Christopher M.},
    title = {{Dark Atoms: Asymmetry and Direct Detection}},
    eprint = {1105.2073},
    archiveprefix = {arXiv},
    primaryclass = {hep-ph},
    reportnumber = {FERMILAB-PUB-11-207-T},
    doi = {10.1088/1475-7516/2011/10/011},
    journal = {JCAP},
    volume = {10},
    pages = {011},
    year = {2011},
}

@article{Cyr-Racine:2013fsa,
    author = {Cyr-Racine, Francis-Yan and de Putter, Roland and Raccanelli, Alvise and Sigurdson, Kris},
    title = {{Constraints on Large-Scale Dark Acoustic Oscillations from Cosmology}},
    eprint = {1310.3278},
    archiveprefix = {arXiv},
    primaryclass = {astro-ph.CO},
    doi = {10.1103/PhysRevD.89.063517},
    journal = {Phys. Rev. D},
    volume = {89},
    number = {6},
    pages = {063517},
    year = {2014},
}

@article{Cyr-Racine:2012tfp,
    author = {Cyr-Racine, Francis-Yan and Sigurdson, Kris},
    title = {{Cosmology of atomic dark matter}},
    eprint = {1209.5752},
    archiveprefix = {arXiv},
    primaryclass = {astro-ph.CO},
    doi = {10.1103/PhysRevD.87.103515},
    journal = {Phys. Rev. D},
    volume = {87},
    number = {10},
    pages = {103515},
    year = {2013},
}

@article{Bansal:2021dfh,
    author = {Bansal, Saurabh and Kim, Jeong Han and Kolda, Christopher and Low, Matthew and Tsai, Yuhsin},
    title = {{Mirror twin Higgs cosmology: constraints and a possible resolution to the H$_{0}$ and S$_{8}$ tensions}},
    eprint = {2110.04317},
    archiveprefix = {arXiv},
    primaryclass = {hep-ph},
    reportnumber = {FERMILAB-PUB-21-371-T},
    doi = {10.1007/JHEP05(2022)050},
    journal = {JHEP},
    volume = {05},
    pages = {050},
    year = {2022},
}

@article{Blinnikov:1983gh,
    author = {Blinnikov, S. I. and Khlopov, M.},
    title = {{Possible astronomical effects of mirror particles}},
    journal = {Sov. Astron.},
    volume = {27},
    pages = {371--375},
    year = {1983},
}

@article{Hodges:1993yb,
    author = {Hodges, H. M.},
    title = {{Mirror baryons as the dark matter}},
    doi = {10.1103/PhysRevD.47.456},
    journal = {Phys. Rev. D},
    volume = {47},
    pages = {456--459},
    year = {1993},
}

@article{Berezhiani:1995am,
    author = {Berezhiani, Z. G. and Dolgov, A. D. and Mohapatra, R. N.},
    title = {{Asymmetric inflationary reheating and the nature of mirror universe}},
    eprint = {hep-ph/9511221},
    archiveprefix = {arXiv},
    reportnumber = {INFN-FE-16-95, UMD-PP-96-15, IFIC-95-64, FTUV-95-61},
    doi = {10.1016/0370-2693(96)00219-5},
    journal = {Phys. Lett. B},
    volume = {375},
    pages = {26--36},
    year = {1996},
}

@inproceedings{Mohapatra:2000rk,
    author = {Mohapatra, R. N. and Teplitz, V. L.},
    title = {{Mirror dark matter}},
    booktitle = {{7th International Symposium on Particles, Strings and Cosmology}},
    eprint = {astro-ph/0004046},
    archiveprefix = {arXiv},
    reportnumber = {UMD-PP-00-066},
    pages = {211--221},
    month = {3},
    year = {2000},
}

@article{Berezhiani:2000gw,
    author = {Berezhiani, Zurab and Comelli, Denis and Villante, Francesco L.},
    title = {{The Early mirror universe: Inflation, baryogenesis, nucleosynthesis and dark matter}},
    eprint = {hep-ph/0008105},
    archiveprefix = {arXiv},
    reportnumber = {DFAQ-TH-2000-03, INFN-FE-04-00},
    doi = {10.1016/S0370-2693(01)00217-9},
    journal = {Phys. Lett. B},
    volume = {503},
    pages = {362--375},
    year = {2001},
}

@article{Foot:2003jt,
    author = {Foot, Robert and Volkas, R. R.},
    title = {{Was ordinary matter synthesized from mirror matter? An Attempt to explain why Omega(Baryon) approximately equal to 0.2 Omega(Dark)}},
    eprint = {hep-ph/0304261},
    archiveprefix = {arXiv},
    doi = {10.1103/PhysRevD.68.021304},
    journal = {Phys. Rev. D},
    volume = {68},
    pages = {021304},
    year = {2003},
}

@inproceedings{Berezhiani:2005ek,
    author = {Berezhiani, Zurab},
    title = {{Through the looking-glass: Alice's adventures in mirror world}},
    booktitle = {{From Fields to Strings: Circumnavigating Theoretical Physics: A Conference in Tribute to Ian Kogan}},
    eprint = {hep-ph/0508233},
    archiveprefix = {arXiv},
    doi = {10.1142/9789812775344_0055},
    pages = {2147--2195},
    year = {2005},
}

@article{An:2009vq,
    author = {An, Haipeng and Chen, Shao-Long and Mohapatra, Rabindra N. and Zhang, Yue},
    title = {{Leptogenesis as a Common Origin for Matter and Dark Matter}},
    eprint = {0911.4463},
    archiveprefix = {arXiv},
    primaryclass = {hep-ph},
    reportnumber = {UMD-40762-471, UMD-PP-09-062, IC-2009-090},
    doi = {10.1007/JHEP03(2010)124},
    journal = {JHEP},
    volume = {03},
    pages = {124},
    year = {2010},
}

@article{Lonsdale:2018xwd,
    author = {Lonsdale, Stephen J. and Volkas, Raymond R.},
    title = {{Comprehensive asymmetric dark matter model}},
    eprint = {1801.05561},
    archiveprefix = {arXiv},
    primaryclass = {hep-ph},
    doi = {10.1103/PhysRevD.97.103510},
    journal = {Phys. Rev. D},
    volume = {97},
    number = {10},
    pages = {103510},
    year = {2018},
}

@article{Baumann:2015rya,
    author = {Baumann, Daniel and Green, Daniel and Meyers, Joel and Wallisch, Benjamin},
    title = {{Phases of New Physics in the CMB}},
    eprint = {1508.06342},
    archiveprefix = {arXiv},
    primaryclass = {astro-ph.CO},
    doi = {10.1088/1475-7516/2016/01/007},
    journal = {JCAP},
    volume = {01},
    pages = {007},
    year = {2016},
}

@article{Brust:2017nmv,
    author = {Brust, Christopher and Cui, Yanou and Sigurdson, Kris},
    title = {{Cosmological Constraints on Interacting Light Particles}},
    eprint = {1703.10732},
    archiveprefix = {arXiv},
    primaryclass = {astro-ph.CO},
    doi = {10.1088/1475-7516/2017/08/020},
    journal = {JCAP},
    volume = {08},
    pages = {020},
    year = {2017},
}

@article{Dvorkin:2013cea,
    author = {Dvorkin, Cora and Blum, Kfir and Kamionkowski, Marc},
    title = {{Constraining Dark Matter-Baryon Scattering with Linear Cosmology}},
    eprint = {1311.2937},
    archiveprefix = {arXiv},
    primaryclass = {astro-ph.CO},
    doi = {10.1103/PhysRevD.89.023519},
    journal = {Phys. Rev. D},
    volume = {89},
    number = {2},
    pages = {023519},
    year = {2014},
}

@article{Xu:2018efh,
    author = {Xu, Weishuang Linda and Dvorkin, Cora and Chael, Andrew},
    title = {{Probing sub-GeV Dark Matter-Baryon Scattering with Cosmological Observables}},
    eprint = {1802.06788},
    archiveprefix = {arXiv},
    primaryclass = {astro-ph.CO},
    doi = {10.1103/PhysRevD.97.103530},
    journal = {Phys. Rev. D},
    volume = {97},
    number = {10},
    pages = {103530},
    year = {2018},
}

@article{Boddy:2018wzy,
    author = {Boddy, Kimberly K. and Gluscevic, Vera and Poulin, Vivian and Kovetz, Ely D. and Kamionkowski, Marc and Barkana, Rennan},
    title = {{Critical assessment of CMB limits on dark matter-baryon scattering: New treatment of the relative bulk velocity}},
    eprint = {1808.00001},
    archiveprefix = {arXiv},
    primaryclass = {astro-ph.CO},
    doi = {10.1103/PhysRevD.98.123506},
    journal = {Phys. Rev. D},
    volume = {98},
    number = {12},
    pages = {123506},
    year = {2018},
}

@article{Dvorkin:2020xga,
    author = {Dvorkin, Cora and Lin, Tongyan and Schutz, Katelin},
    title = {{Cosmology of Sub-MeV Dark Matter Freeze-In}},
    eprint = {2011.08186},
    archiveprefix = {arXiv},
    primaryclass = {astro-ph.CO},
    reportnumber = {MIT-CTP/5256},
    doi = {10.1103/PhysRevLett.127.111301},
    journal = {Phys. Rev. Lett.},
    volume = {127},
    number = {11},
    pages = {111301},
    year = {2021},
}

@book{Gottfried:2003aaa,
    author = {Gottfried, Kurt and Yan, Tung-Mow},
    title = {{Quantum Mechanics: Fundamentals}},
    doi = {10.1007/978-0-387-21623-2},
    publisher = {Springer},
    address = {New York},
    year = {2003},
}

@article{Karzas:1961aaa,
    author = {Karzas, W. J and Latter, R},
    title = {{Electron Radiative Transitions in a Coulomb Field}},
    doi = {10.1086/190063},
    journal = {Astrophys. J. Suppl. Ser.},
    volume = {6},
    pages = {167},
    year = {1961},
}

@book{Spitzer:1978aaa,
    author = {Spitzer, Lyman},
    title = {{Physical processes in the interstellar medium}},
    doi = {10.1002/9783527617722},
    publisher = {Wiley},
    address = {New York},
    year = {1978},
}

@article{Peebles:1968ja,
    author = {Peebles, P. J. E.},
    title = {{Recombination of the Primeval Plasma}},
    doi = {10.1086/149628},
    journal = {Astrophys. J.},
    volume = {153},
    pages = {1},
    year = {1968},
}

@article{Pequignot:1991,
    author = {Pequignot, D. and Petitjean, P. and Boisson, C.},
    title = {{Total and effective radiative recombination coefficients}},
    journal = {Astron. Astrophys.},
    volume = {251},
    pages = {680-688},
    year = {1991},
}

@article{Seager:1999km,
    author = {Seager, Sara and Sasselov, Dimitar D. and Scott, Douglas},
    title = {{How exactly did the universe become neutral?}},
    eprint = {astro-ph/9912182},
    archiveprefix = {arXiv},
    reportnumber = {UBC-COS-99-07},
    doi = {10.1086/313388},
    journal = {Astrophys. J. Suppl.},
    volume = {128},
    pages = {407--430},
    year = {2000},
}

@article{Switzer:2007sq,
    author = {Switzer, Eric R. and Hirata, Christopher M.},
    title = {{Primordial helium recombination. 3. Thomson scattering, isotope shifts, and cumulative results}},
    eprint = {astro-ph/0702145},
    archiveprefix = {arXiv},
    doi = {10.1103/PhysRevD.77.083008},
    journal = {Phys. Rev. D},
    volume = {77},
    pages = {083008},
    year = {2008},
}

@article{Hirata:2008ny,
    author = {Hirata, Christopher M.},
    title = {{Two-photon transitions in primordial hydrogen recombination}},
    eprint = {0803.0808},
    archiveprefix = {arXiv},
    primaryclass = {astro-ph},
    doi = {10.1103/PhysRevD.78.023001},
    journal = {Phys. Rev. D},
    volume = {78},
    pages = {023001},
    year = {2008},
}

@article{Ali-Haimoud:2010tlj,
    author = {Ali-Haimoud, Yacine and Hirata, Christopher M.},
    title = {{Ultrafast effective multi-level atom method for primordial hydrogen recombination}},
    eprint = {1006.1355},
    archiveprefix = {arXiv},
    primaryclass = {astro-ph.CO},
    doi = {10.1103/PhysRevD.82.063521},
    journal = {Phys. Rev. D},
    volume = {82},
    pages = {063521},
    year = {2010},
}

@article{Ali-Haimoud:2010hou,
    author = {Ali-Haimoud, Yacine and Hirata, Christopher M.},
    title = {{HyRec: A fast and highly accurate primordial hydrogen and helium recombination code}},
    eprint = {1011.3758},
    archiveprefix = {arXiv},
    primaryclass = {astro-ph.CO},
    doi = {10.1103/PhysRevD.83.043513},
    journal = {Phys. Rev. D},
    volume = {83},
    pages = {043513},
    year = {2011},
}

@article{Lee:2020obi,
    author = {Lee, Nanoom and Ali-Ha\"\i{}moud, Yacine},
    title = {{HYREC-2: a highly accurate sub-millisecond recombination code}},
    eprint = {2007.14114},
    archiveprefix = {arXiv},
    primaryclass = {astro-ph.CO},
    doi = {10.1103/PhysRevD.102.083517},
    journal = {Phys. Rev. D},
    volume = {102},
    number = {8},
    pages = {083517},
    year = {2020},
}

@article{STEPANEK200399,
    title = {Electron and positron atomic elastic scattering cross sections},
    journal = {Radiation Physics and Chemistry},
    volume = {66},
    number = {2},
    pages = {99-116},
    year = {2003},
    issn = {0969-806X},
    doi = {10.1016/S0969-806X(02)00386-9},
    url = {https://www.sciencedirect.com/science/article/pii/S0969806X02003869},
    author = {Jiri Stepanek},
}

@article{Chacko:2016kgg,
    author = {Chacko, Zackaria and Cui, Yanou and Hong, Sungwoo and Okui, Takemichi and Tsai, Yuhsin},
    title = {{Partially Acoustic Dark Matter, Interacting Dark Radiation, and Large Scale Structure}},
    eprint = {1609.03569},
    archiveprefix = {arXiv},
    primaryclass = {astro-ph.CO},
    doi = {10.1007/JHEP12(2016)108},
    journal = {JHEP},
    volume = {12},
    pages = {108},
    year = {2016},
}

@article{Ma:1995ey,
    author = {Ma, Chung-Pei and Bertschinger, Edmund},
    title = {{Cosmological perturbation theory in the synchronous and conformal Newtonian gauges}},
    eprint = {astro-ph/9506072},
    archiveprefix = {arXiv},
    doi = {10.1086/176550},
    journal = {Astrophys. J.},
    volume = {455},
    pages = {7--25},
    year = {1995},
}

@article{Chacko:2003dt,
    author = {Chacko, Z. and Hall, Lawrence J. and Okui, Takemichi and Oliver, Steven J.},
    title = {{CMB signals of neutrino mass generation}},
    eprint = {hep-ph/0312267},
    archiveprefix = {arXiv},
    reportnumber = {UCB-PTH-03-34, LBNL-54211, BUHEP-03-22},
    doi = {10.1103/PhysRevD.70.085008},
    journal = {Phys. Rev. D},
    volume = {70},
    pages = {085008},
    year = {2004},
}

@article{Friedland:2007vv,
    author = {Friedland, Alexander and Zurek, Kathryn M. and Bashinsky, Sergei},
    title = {{Constraining Models of Neutrino Mass and Neutrino Interactions with the Planck Satellite}},
    eprint = {0704.3271},
    archiveprefix = {arXiv},
    primaryclass = {astro-ph},
    reportnumber = {LA-UR-07-2135, MADPH-07-1483},
    month = {4},
    year = {2007},
}

@article{Ghosh:2024wva,
    author = {Ghosh, Subhajit and Ho, Daven Wei Ren and Tsai, Yuhsin},
    title = {{Dark matter-radiation scattering enhances CMB phase shift through dark matter-loading}},
    eprint = {2405.08064},
    archiveprefix = {arXiv},
    primaryclass = {astro-ph.CO},
    reportnumber = {UT-WI-17-2024},
    year = {2025},
    doi = {10.1088/1475-7516/2025/01/058},
    journal = {JCAP},
    volume = {01},
    pages = {058},
}

@article{Bashinsky:2003tk,
    author = {Bashinsky, Sergei and Seljak, Uros},
    title = {{Neutrino perturbations in CMB anisotropy and matter clustering}},
    eprint = {astro-ph/0310198},
    archiveprefix = {arXiv},
    doi = {10.1103/PhysRevD.69.083002},
    journal = {Phys. Rev. D},
    volume = {69},
    pages = {083002},
    year = {2004},
}

@article{Lesgourgues:2011re,
    author = {Lesgourgues, Julien},
    title = {{The Cosmic Linear Anisotropy Solving System (CLASS) I: Overview}},
    eprint = {1104.2932},
    archiveprefix = {arXiv},
    primaryclass = {astro-ph.IM},
    month = {4},
    year = {2011},
}

@article{Blas:2011rf,
    author = {Blas, Diego and Lesgourgues, Julien and Tram, Thomas},
    title = {{The Cosmic Linear Anisotropy Solving System (CLASS) II: Approximation schemes}},
    eprint = {1104.2933},
    archiveprefix = {arXiv},
    primaryclass = {astro-ph.CO},
    reportnumber = {CERN-PH-TH-2011-082, LAPTH-010-11},
    doi = {10.1088/1475-7516/2011/07/034},
    journal = {JCAP},
    volume = {07},
    pages = {034},
    year = {2011},
}

@article{Lesgourgues:2011rg,
    author = {Lesgourgues, Julien},
    title = {{The Cosmic Linear Anisotropy Solving System (CLASS) III: Comparision with CAMB for LambdaCDM}},
    eprint = {1104.2934},
    archiveprefix = {arXiv},
    primaryclass = {astro-ph.CO},
    reportnumber = {CERN-PH-TH-2011-083, LAPTH-011-11},
    month = {4},
    year = {2011},
}

@article{Lesgourgues:2011rh,
    author = {Lesgourgues, Julien and Tram, Thomas},
    title = {{The Cosmic Linear Anisotropy Solving System (CLASS) IV: efficient implementation of non-cold relics}},
    eprint = {1104.2935},
    archiveprefix = {arXiv},
    primaryclass = {astro-ph.CO},
    reportnumber = {CERN-PH-TH-2011-084, LAPTH-012-11},
    doi = {10.1088/1475-7516/2011/09/032},
    journal = {JCAP},
    volume = {09},
    pages = {032},
    year = {2011},
}

@article{Audren:2012wb,
    author = {Audren, Benjamin and Lesgourgues, Julien and Benabed,
                        Karim and Prunet, Simon},
    title = {{Conservative Constraints on Early Cosmology: an
                        illustration of the Monte Python cosmological parameter
                        inference code}},
    journal = {JCAP},
    volume = {02},
    pages = {001},
    doi = {10.1088/1475-7516/2013/02/001},
    year = {2013},
    eprint = {1210.7183},
    archiveprefix = {arXiv},
    primaryclass = {astro-ph.CO},
    reportnumber = {CERN-PH-TH-2012-290, LAPTH-048-12},
    slaccitation = {%%CITATION = ARXIV:1210.7183;%%},
}

@article{Brinckmann:2018cvx,
    author = {Brinckmann, Thejs and Lesgourgues, Julien},
    title = {{MontePython 3: boosted MCMC sampler and other features}},
    year = {2019},
    eprint = {1804.07261},
    archiveprefix = {arXiv},
    primaryclass = {astro-ph.CO},
    slaccitation = {%%CITATION = ARXIV:1804.07261;%%},
    doi = {10.1016/j.dark.2018.100260},
    journal = {Phys. Dark Univ.},
    volume = {24},
    pages = {100260},
}

@article{Smith:2002dz,
    author = {Smith, R. E. and Peacock, J. A. and Jenkins, A. and White, S. D. M. and Frenk, C. S. and Pearce, F. R. and Thomas, P. A. and Efstathiou, G. and Couchmann, H. M. P.},
    collaboration = {VIRGO Consortium},
    title = {{Stable clustering, the halo model and nonlinear cosmological power spectra}},
    eprint = {astro-ph/0207664},
    archiveprefix = {arXiv},
    doi = {10.1046/j.1365-8711.2003.06503.x},
    journal = {Mon. Not. Roy. Astron. Soc.},
    volume = {341},
    pages = {1311},
    year = {2003},
}

@article{Takahashi:2012em,
    author = {Takahashi, Ryuichi and Sato, Masanori and Nishimichi, Takahiro and Taruya, Atsushi and Oguri, Masamune},
    title = {{Revising the Halofit Model for the Nonlinear Matter Power Spectrum}},
    eprint = {1208.2701},
    archiveprefix = {arXiv},
    primaryclass = {astro-ph.CO},
    doi = {10.1088/0004-637X/761/2/152},
    journal = {Astrophys. J.},
    volume = {761},
    pages = {152},
    year = {2012},
}

@article{Gelman:1992zz,
    author = {Gelman, Andrew and Rubin, Donald B.},
    title = {{Inference from Iterative Simulation Using Multiple Sequences}},
    doi = {10.1214/ss/1177011136},
    journal = {Statist. Sci.},
    volume = {7},
    pages = {457--472},
    year = {1992},
}

@article{Beutler:2011hx,
    author = {Beutler, Florian and Blake, Chris and Colless, Matthew and Jones, D. Heath and Staveley-Smith, Lister and Campbell, Lachlan and Parker, Quentin and Saunders, Will and Watson, Fred},
    title = {{The 6dF Galaxy Survey: Baryon Acoustic Oscillations and the Local Hubble Constant}},
    eprint = {1106.3366},
    archiveprefix = {arXiv},
    primaryclass = {astro-ph.CO},
    doi = {10.1111/j.1365-2966.2011.19250.x},
    journal = {Mon. Not. Roy. Astron. Soc.},
    volume = {416},
    pages = {3017--3032},
    year = {2011},
}

@article{Ross:2014qpa,
    author = {Ross, Ashley J. and Samushia, Lado and Howlett, Cullan and Percival, Will J. and Burden, Angela and Manera, Marc},
    title = {{The clustering of the SDSS DR7 main Galaxy sample \textendash{} I. A 4 per cent distance measure at $z = 0.15$}},
    eprint = {1409.3242},
    archiveprefix = {arXiv},
    primaryclass = {astro-ph.CO},
    doi = {10.1093/mnras/stv154},
    journal = {Mon. Not. Roy. Astron. Soc.},
    volume = {449},
    number = {1},
    pages = {835--847},
    year = {2015},
}

@article{eBOSS:2020yzd,
    author = {Alam, Shadab and others},
    collaboration = {eBOSS},
    title = {{Completed SDSS-IV extended Baryon Oscillation Spectroscopic Survey: Cosmological implications from two decades of spectroscopic surveys at the Apache Point Observatory}},
    eprint = {2007.08991},
    archiveprefix = {arXiv},
    primaryclass = {astro-ph.CO},
    doi = {10.1103/PhysRevD.103.083533},
    journal = {Phys. Rev. D},
    volume = {103},
    number = {8},
    pages = {083533},
    year = {2021},
}

@article{eBOSS:2020qek,
    author = {Tamone, Am\'elie and others},
    collaboration = {eBOSS},
    title = {{The Completed SDSS-IV extended Baryon Oscillation Spectroscopic Survey: Growth rate of structure measurement from anisotropic clustering analysis in configuration space between redshift 0.6 and 1.1 for the Emission Line Galaxy sample}},
    eprint = {2007.09009},
    archiveprefix = {arXiv},
    primaryclass = {astro-ph.CO},
    doi = {10.1093/mnras/staa3050},
    journal = {Mon. Not. Roy. Astron. Soc.},
    volume = {499},
    number = {4},
    pages = {5527--5546},
    year = {2020},
}

@article{eBOSS:2020fvk,
    author = {de Mattia, Arnaud and others},
    collaboration = {eBOSS},
    title = {{The Completed SDSS-IV extended Baryon Oscillation Spectroscopic Survey: measurement of the BAO and growth rate of structure of the emission line galaxy sample from the anisotropic power spectrum between redshift 0.6 and 1.1}},
    eprint = {2007.09008},
    archiveprefix = {arXiv},
    primaryclass = {astro-ph.CO},
    doi = {10.1093/mnras/staa3891},
    journal = {Mon. Not. Roy. Astron. Soc.},
    volume = {501},
    number = {4},
    pages = {5616--5645},
    year = {2021},
}

@article{eBOSS:2020gbb,
    author = {Hou, Jiamin and others},
    collaboration = {eBOSS},
    title = {{The Completed SDSS-IV extended Baryon Oscillation Spectroscopic Survey: BAO and RSD measurements from anisotropic clustering analysis of the Quasar Sample in configuration space between redshift 0.8 and 2.2}},
    eprint = {2007.08998},
    archiveprefix = {arXiv},
    primaryclass = {astro-ph.CO},
    doi = {10.1093/mnras/staa3234},
    journal = {Mon. Not. Roy. Astron. Soc.},
    volume = {500},
    number = {1},
    pages = {1201--1221},
    year = {2020},
}

@article{eBOSS:2020uxp,
    author = {Neveux, Richard and others},
    collaboration = {eBOSS},
    title = {{The completed SDSS-IV extended Baryon Oscillation Spectroscopic Survey: BAO and RSD measurements from the anisotropic power spectrum of the quasar sample between redshift 0.8 and 2.2}},
    eprint = {2007.08999},
    archiveprefix = {arXiv},
    primaryclass = {astro-ph.CO},
    doi = {10.1093/mnras/staa2780},
    journal = {Mon. Not. Roy. Astron. Soc.},
    volume = {499},
    number = {1},
    pages = {210--229},
    year = {2020},
}

@article{eBOSS:2020tmo,
    author = {du Mas des Bourboux, Helion and others},
    collaboration = {eBOSS},
    title = {{The Completed SDSS-IV Extended Baryon Oscillation Spectroscopic Survey: Baryon Acoustic Oscillations with Ly\ensuremath{\alpha} Forests}},
    eprint = {2007.08995},
    archiveprefix = {arXiv},
    primaryclass = {astro-ph.CO},
    doi = {10.3847/1538-4357/abb085},
    journal = {Astrophys. J.},
    volume = {901},
    number = {2},
    pages = {153},
    year = {2020},
}

@article{Scolnic:2021amr,
    author = {Scolnic, Dan and others},
    title = {{The Pantheon+ Analysis: The Full Data Set and Light-curve Release}},
    eprint = {2112.03863},
    archiveprefix = {arXiv},
    primaryclass = {astro-ph.CO},
    doi = {10.3847/1538-4357/ac8b7a},
    journal = {Astrophys. J.},
    volume = {938},
    number = {2},
    pages = {113},
    year = {2022},
}

@article{Brout:2022vxf,
    author = {Brout, Dillon and others},
    title = {{The Pantheon+ Analysis: Cosmological Constraints}},
    eprint = {2202.04077},
    archiveprefix = {arXiv},
    primaryclass = {astro-ph.CO},
    doi = {10.3847/1538-4357/ac8e04},
    journal = {Astrophys. J.},
    volume = {938},
    number = {2},
    pages = {110},
    year = {2022},
}

@article{Hooper:2022byl,
    author = {Hooper, Deanna C. and Sch\"oneberg, Nils and Murgia, Riccardo and Archidiacono, Maria and Lesgourgues, Julien and Viel, Matteo},
    title = {{One likelihood to bind them all: Lyman-\ensuremath{\alpha} constraints on non-standard dark matter}},
    eprint = {2206.08188},
    archiveprefix = {arXiv},
    primaryclass = {astro-ph.CO},
    reportnumber = {HIP-2022-17/TH, TTK-22-21},
    doi = {10.1088/1475-7516/2022/10/032},
    journal = {JCAP},
    volume = {10},
    pages = {032},
    year = {2022},
}

@article{Pedersen:2022anu,
    author = {Pedersen, Christian and Font-Ribera, Andreu and Gnedin, Nickolay Y.},
    title = {{Compressing the Cosmological Information in One-dimensional Correlations of the Ly\ensuremath{\alpha} Forest}},
    eprint = {2209.09895},
    archiveprefix = {arXiv},
    primaryclass = {astro-ph.CO},
    reportnumber = {FERMILAB-PUB-22-713-T},
    doi = {10.3847/1538-4357/acb433},
    journal = {Astrophys. J.},
    volume = {944},
    number = {2},
    pages = {223},
    year = {2023},
}

@article{eBOSS:2018qyj,
    author = {Chabanier, Sol\`ene and others},
    collaboration = {eBOSS},
    title = {{The one-dimensional power spectrum from the SDSS DR14 Ly$\alpha$ forests}},
    eprint = {1812.03554},
    archiveprefix = {arXiv},
    primaryclass = {astro-ph.CO},
    doi = {10.1088/1475-7516/2019/07/017},
    journal = {JCAP},
    volume = {07},
    pages = {017},
    year = {2019},
}

@book{van1995python,
    title = {Python tutorial},
    author = {Van Rossum, Guido and Drake Jr, Fred L},
    year = {1995},
    publisher = {Centrum voor Wiskunde en Informatica Amsterdam, The Netherlands},
}

@book{10.5555/1593511,
    author = {Van Rossum, Guido and Drake, Fred L.},
    title = {Python 3 Reference Manual},
    year = {2009},
    isbn = {1441412697},
    publisher = {CreateSpace},
    address = {Scotts Valley, CA},
}

@article{Virtanen:2019joe,
    author = {Virtanen, Pauli and others},
    title = {{SciPy 1.0--Fundamental Algorithms for Scientific Computing in Python}},
    eprint = {1907.10121},
    archiveprefix = {arXiv},
    primaryclass = {cs.MS},
    doi = {10.1038/s41592-019-0686-2},
    journal = {Nature Meth.},
    volume = {17},
    pages = {261},
    year = {2020},
}

@article{Irsic:2017sop,
    author = {Ir\v{s}i\v{c}, Vid and others},
    title = {{The Lyman \ensuremath{\alpha} forest power spectrum from the XQ-100 Legacy Survey}},
    eprint = {1702.01761},
    archiveprefix = {arXiv},
    primaryclass = {astro-ph.CO},
    doi = {10.1093/mnras/stw3372},
    journal = {Mon. Not. Roy. Astron. Soc.},
    volume = {466},
    number = {4},
    pages = {4332--4345},
    year = {2017},
}

@article{Viel:2013fqw,
    author = {Viel, Matteo and Becker, George D. and Bolton, James S. and Haehnelt, Martin G.},
    title = {{Warm dark matter as a solution to the small scale crisis: New constraints from high redshift Lyman-\ensuremath{\alpha} forest data}},
    eprint = {1306.2314},
    archiveprefix = {arXiv},
    primaryclass = {astro-ph.CO},
    doi = {10.1103/PhysRevD.88.043502},
    journal = {Phys. Rev. D},
    volume = {88},
    pages = {043502},
    year = {2013},
}

@article{eBOSS:2020jck,
    author = {Lyke, Brad W. and others},
    collaboration = {eBOSS},
    title = {{The Sloan Digital Sky Survey Quasar Catalog: Sixteenth Data Release}},
    eprint = {2007.09001},
    archiveprefix = {arXiv},
    primaryclass = {astro-ph.GA},
    doi = {10.3847/1538-4365/aba623},
    journal = {Astrophys. J. Suppl.},
    volume = {250},
    number = {1},
    pages = {8},
    year = {2020},
}

@article{Heymans:2020gsg,
    author = {Heymans, Catherine and others},
    title = {{KiDS-1000 Cosmology: Multi-probe weak gravitational lensing and spectroscopic galaxy clustering constraints}},
    eprint = {2007.15632},
    archiveprefix = {arXiv},
    primaryclass = {astro-ph.CO},
    doi = {10.1051/0004-6361/202039063},
    journal = {Astron. Astrophys.},
    volume = {646},
    pages = {A140},
    year = {2021},
}

@article{DES:2021wwk,
    author = {Abbott, T. M. C. and others},
    collaboration = {DES},
    title = {{Dark Energy Survey Year 3 results: Cosmological constraints from galaxy clustering and weak lensing}},
    eprint = {2105.13549},
    archiveprefix = {arXiv},
    primaryclass = {astro-ph.CO},
    reportnumber = {FERMILAB-PUB-21-221-AE, DES-2020-0617},
    doi = {10.1103/PhysRevD.105.023520},
    journal = {Phys. Rev. D},
    volume = {105},
    number = {2},
    pages = {023520},
    year = {2022},
}

@article{Kilo-DegreeSurvey:2023gfr,
    author = {Abbott, T. M. C. and others},
    collaboration = {Kilo-Degree Survey, Dark Energy Survey},
    title = {{DES Y3 + KiDS-1000: Consistent cosmology combining cosmic shear surveys}},
    eprint = {2305.17173},
    archiveprefix = {arXiv},
    primaryclass = {astro-ph.CO},
    reportnumber = {FERMILAB-PUB-23-267-PPD, DES-2023-0769},
    doi = {10.21105/astro.2305.17173},
    journal = {Open J. Astrophys.},
    volume = {6},
    pages = {2305.17173},
    year = {2023},
}

@article{Ivanov:2024xgb,
    author = {Ivanov, Mikhail M. and Obuljen, Andrej and Cuesta-Lazaro, Carolina and Toomey, Michael W.},
    title = {{Full-shape analysis with simulation-based priors: Cosmological parameters and the structure growth anomaly}},
    eprint = {2409.10609},
    archiveprefix = {arXiv},
    primaryclass = {astro-ph.CO},
    reportnumber = {MIT-CTP/5762},
    year = {2025},
    doi = {10.1103/PhysRevD.111.063548},
    journal = {Phys. Rev. D},
    volume = {111},
    number = {6},
    pages = {063548},
}

@article{Battye:2014qga,
    author = {Battye, Richard A. and Charnock, Tom and Moss, Adam},
    title = {{Tension between the power spectrum of density perturbations measured on large and small scales}},
    eprint = {1409.2769},
    archiveprefix = {arXiv},
    primaryclass = {astro-ph.CO},
    doi = {10.1103/PhysRevD.91.103508},
    journal = {Phys. Rev. D},
    volume = {91},
    number = {10},
    pages = {103508},
    year = {2015},
}

@article{Enqvist:2015ara,
    author = {Enqvist, Kari and Nadathur, Seshadri and Sekiguchi, Toyokazu and Takahashi, Tomo},
    title = {{Decaying dark matter and the tension in $\sigma_8$}},
    eprint = {1505.05511},
    archiveprefix = {arXiv},
    primaryclass = {astro-ph.CO},
    reportnumber = {HIP-2015-16-TH},
    doi = {10.1088/1475-7516/2015/09/067},
    journal = {JCAP},
    volume = {09},
    pages = {067},
    year = {2015},
}

@article{Murgia:2016ccp,
    author = {Murgia, Riccardo and Gariazzo, Stefano and Fornengo, Nicolao},
    title = {{Constraints on the Coupling between Dark Energy and Dark Matter from CMB data}},
    eprint = {1602.01765},
    archiveprefix = {arXiv},
    primaryclass = {astro-ph.CO},
    doi = {10.1088/1475-7516/2016/04/014},
    journal = {JCAP},
    volume = {04},
    pages = {014},
    year = {2016},
}

@article{Kumar:2016zpg,
    author = {Kumar, Suresh and Nunes, Rafael C.},
    title = {{Probing the interaction between dark matter and dark energy in the presence of massive neutrinos}},
    eprint = {1608.02454},
    archiveprefix = {arXiv},
    primaryclass = {astro-ph.CO},
    doi = {10.1103/PhysRevD.94.123511},
    journal = {Phys. Rev. D},
    volume = {94},
    number = {12},
    pages = {123511},
    year = {2016},
}

@article{Poulin:2016nat,
    author = {Poulin, Vivian and Serpico, Pasquale D. and Lesgourgues, Julien},
    title = {{A fresh look at linear cosmological constraints on a decaying dark matter component}},
    eprint = {1606.02073},
    archiveprefix = {arXiv},
    primaryclass = {astro-ph.CO},
    reportnumber = {LAPTH-027-16},
    doi = {10.1088/1475-7516/2016/08/036},
    journal = {JCAP},
    volume = {08},
    pages = {036},
    year = {2016},
}

@article{Buen-Abad:2018mas,
    author = {Buen-Abad, Manuel A. and Emami, Razieh and Schmaltz, Martin},
    title = {{Cannibal Dark Matter and Large Scale Structure}},
    eprint = {1803.08062},
    archiveprefix = {arXiv},
    primaryclass = {hep-ph},
    doi = {10.1103/PhysRevD.98.083517},
    journal = {Phys. Rev. D},
    volume = {98},
    number = {8},
    pages = {083517},
    year = {2018},
}

@article{Dessert:2018khu,
    author = {Dessert, Christopher and Kilic, Can and Trendafilova, Cynthia and Tsai, Yuhsin},
    title = {{Addressing Astrophysical and Cosmological Problems With Secretly Asymmetric Dark Matter}},
    eprint = {1811.05534},
    archiveprefix = {arXiv},
    primaryclass = {hep-ph},
    reportnumber = {UTTG-23-18, LCTP-18-24},
    doi = {10.1103/PhysRevD.100.015029},
    journal = {Phys. Rev. D},
    volume = {100},
    number = {1},
    pages = {015029},
    year = {2019},
}

@article{Archidiacono:2019wdp,
    author = {Archidiacono, Maria and Hooper, Deanna C. and Murgia, Riccardo and Bohr, Sebastian and Lesgourgues, Julien and Viel, Matteo},
    title = {{Constraining Dark Matter-Dark Radiation interactions with CMB, BAO, and Lyman-$\alpha$}},
    eprint = {1907.01496},
    archiveprefix = {arXiv},
    primaryclass = {astro-ph.CO},
    doi = {10.1088/1475-7516/2019/10/055},
    journal = {JCAP},
    volume = {10},
    pages = {055},
    year = {2019},
}

@article{Heimersheim:2020aoc,
    author = {Heimersheim, Stefan and Sch\"oneberg, Nils and Hooper, Deanna C. and Lesgourgues, Julien},
    title = {{Cannibalism hinders growth: Cannibal Dark Matter and the $S_8$ tension}},
    eprint = {2008.08486},
    archiveprefix = {arXiv},
    primaryclass = {astro-ph.CO},
    reportnumber = {TTK-20-26, ULB-TH/20-11},
    doi = {10.1088/1475-7516/2020/12/016},
    journal = {JCAP},
    volume = {12},
    pages = {016},
    year = {2020},
}

@article{Clark:2020miy,
    author = {Clark, Steven J. and Vattis, Kyriakos and Koushiappas, Savvas M.},
    title = {{Cosmological constraints on late-universe decaying dark matter as a solution to the $H_0$ tension}},
    eprint = {2006.03678},
    archiveprefix = {arXiv},
    primaryclass = {astro-ph.CO},
    doi = {10.1103/PhysRevD.103.043014},
    journal = {Phys. Rev. D},
    volume = {103},
    number = {4},
    pages = {043014},
    year = {2021},
}

@article{FrancoAbellan:2021sxk,
    author = {Franco Abell\'an, Guillermo and Murgia, Riccardo and Poulin, Vivian},
    title = {{Linear cosmological constraints on two-body decaying dark matter scenarios and the S8 tension}},
    eprint = {2102.12498},
    archiveprefix = {arXiv},
    primaryclass = {astro-ph.CO},
    doi = {10.1103/PhysRevD.104.123533},
    journal = {Phys. Rev. D},
    volume = {104},
    number = {12},
    pages = {123533},
    year = {2021},
}

@article{Cruz:2023lmn,
    author = {Cruz, Juan S. and Niedermann, Florian and Sloth, Martin S.},
    title = {{Cold New Early Dark Energy pulls the trigger on the H $_{0}$ and S $_{8}$ tensions: a simultaneous solution to both tensions without new ingredients}},
    eprint = {2305.08895},
    archiveprefix = {arXiv},
    primaryclass = {astro-ph.CO},
    doi = {10.1088/1475-7516/2023/11/033},
    journal = {JCAP},
    volume = {11},
    pages = {033},
    year = {2023},
}

@book{Weinberg:2008zzc,
    author = {Weinberg, Steven},
    title = {{Cosmology}},
    isbn = {978-0-19-852682-7},
    year = {2008},
    publisher = {Oxford University Press},
    url = {https://global.oup.com/academic/product/cosmology-9780198526827},
}

@article{DES:2022qpf,
    author = {Doux, C. and others},
    collaboration = {DES},
    title = {{Dark energy survey year 3 results: cosmological constraints from the analysis of cosmic shear in harmonic space}},
    eprint = {2203.07128},
    archiveprefix = {arXiv},
    primaryclass = {astro-ph.CO},
    reportnumber = {FERMILAB-PUB-22-042-PPD-SCD},
    doi = {10.1093/mnras/stac1826},
    journal = {Mon. Not. Roy. Astron. Soc.},
    volume = {515},
    number = {2},
    pages = {1942--1972},
    year = {2022},
}

@article{Zhou:2024igb,
    author = {Zhou, Zilu and Weiner, Neal},
    title = {{Searching for Dark Matter Interactions with ACT, SPT and DES}},
    eprint = {2409.06771},
    archiveprefix = {arXiv},
    primaryclass = {hep-ph},
    month = {9},
    year = {2024},
}

@article{Akaike:1974vps,
    author = {Akaike, H.},
    journal = {IEEE Transactions on Automatic Control},
    title = {A new look at the statistical model identification},
    year = {1974},
    volume = {19},
    number = {6},
    pages = {716-723},
    doi = {10.1109/TAC.1974.1100705},
}

@article{Simon:2022lde,
    author = {Simon, Th\'eo and Zhang, Pierre and Poulin, Vivian and Smith, Tristan L.},
    title = {{Consistency of effective field theory analyses of the BOSS power spectrum}},
    eprint = {2208.05929},
    archiveprefix = {arXiv},
    primaryclass = {astro-ph.CO},
    doi = {10.1103/PhysRevD.107.123530},
    journal = {Phys. Rev. D},
    volume = {107},
    number = {12},
    pages = {123530},
    year = {2023},
}

@article{Holm:2023laa,
    author = {Holm, Emil Brinch and Herold, Laura and Simon, Th\'eo and Ferreira, Elisa G. M. and Hannestad, Steen and Poulin, Vivian and Tram, Thomas},
    title = {{Bayesian and frequentist investigation of prior effects in EFT of LSS analyses of full-shape BOSS and eBOSS data}},
    eprint = {2309.04468},
    archiveprefix = {arXiv},
    primaryclass = {astro-ph.CO},
    doi = {10.1103/PhysRevD.108.123514},
    journal = {Phys. Rev. D},
    volume = {108},
    number = {12},
    pages = {123514},
    year = {2023},
}

@article{Chudaykin:2024wlw,
    author = {Chudaykin, Anton and Ivanov, Mikhail M. and Nishimichi, Takahiro},
    title = {{On priors and scale cuts in EFT-based full-shape analyses}},
    eprint = {2410.16358},
    archiveprefix = {arXiv},
    primaryclass = {astro-ph.CO},
    month = {10},
    year = {2024},
}

@article{McDonough:2023qcu,
    author = {McDonough, Evan and Hill, J. Colin and Ivanov, Mikhail M. and La Posta, Adrien and Toomey, Michael W.},
    title = {{Observational constraints on early dark energy}},
    eprint = {2310.19899},
    archiveprefix = {arXiv},
    primaryclass = {astro-ph.CO},
    reportnumber = {MIT-CTP/5618},
    doi = {10.1142/S0218271824300039},
    journal = {Int. J. Mod. Phys. D},
    volume = {33},
    number = {11},
    pages = {2430003},
    year = {2024},
}

\end{document}